\documentclass[11pt, twoside]{article}
\usepackage[T1]{fontenc}
\usepackage[latin9]{inputenc}
\usepackage{geometry}
\usepackage{amsmath}
\usepackage{amsthm}
\usepackage{amssymb}
\usepackage{amstext}
\usepackage{bm}
\usepackage{booktabs} 
\usepackage{color}
\usepackage{enumerate} 
\usepackage[shortlabels]{enumitem}
\usepackage{epsfig,epsf,psfrag}
\usepackage{epstopdf}
\usepackage{float}
\usepackage{graphics}
\usepackage{graphicx}
\usepackage{latexsym}
\usepackage{lscape}
\usepackage{mathabx} 
\usepackage{mathtools}
\usepackage{mathrsfs}
\usepackage{multirow}
\usepackage{natbib}
\usepackage{rotate}
\usepackage{setspace}

\usepackage{sectsty}

\usepackage{subcaption}
\usepackage{xargs}[2008/03/08]
\usepackage{xcolor}

\def\rvone{}

\usepackage{ifthen}
\usepackage{fancyhdr}
\usepackage[hidelinks]{hyperref}
\hypersetup{
	colorlinks,
	linkcolor={red!50!black},
	citecolor={blue!50!black},
	urlcolor={blue!80!black}
}

\renewcommand\thesection{\arabic{section}}
\renewcommand{\thesubsection}{\thesection.\arabic{subsection}}

\sectionfont{\centering\bf\large}
\subsectionfont{\normalsize}

\usepackage{titlesec}

\renewenvironment{thebibliography}[1]{
	\begin{oldthebibliography}{#1}
		\setlength{\parskip}{0ex}
		\setlength{\itemsep}{0ex}
	}
	{
	\end{oldthebibliography}
}

\newtheorem{thm}{Theorem}
\newtheorem{lem}{Lemma}
\newtheorem{cor}{Corollary}
{
	\theoremstyle{remark}
	
	\newtheorem{exm}{Example}
	
	\newtheorem{assumption}{Assumption}
}

\newtheorem*{lem*}{Lemma}

\newcommand{\bea}{\begin{eqnarray*}}
\newcommand{\eea}{\end{eqnarray*}}
\newcommand{\be}{\begin{eqnarray}}
\newcommand{\ee}{\end{eqnarray}}
\newcommand{\beq}{\begin{equation}}
\newcommand{\eeq}{\end{equation}}

\newcommand{\als}[1]{\begin{align*}#1\end{align*}}
\newcommand{\bal}{\begin{equation}\aligned}
\newcommand{\eal}{\endaligned\end{equation}}
\newcommand{\bgt}{\begin{equation}\begin{gathered}}
\newcommand{\egt}{\end{gathered}\end{equation}}
\newcommand{\ed}{\end{document}}

\newcommand{\btab}{\begin{tabular}}
\newcommand{\etab}{\end{tabular}}

\newcommand{\bi}{\begin{itemize}}
\newcommand{\ei}{\end{itemize}}
\newcommand{\bfi}{\begin{figure}}
\newcommand{\efi}{\end{figure}}
\newcommand{\ben}{\begin{enumerate}}
\newcommand{\een}{\end{enumerate}}
\newcommand{\bay}{\begin{array}}
\newcommand{\eay}{\end{array}}

\newcommand{\nn}{\nonumber}

\def\bco{\iffalse}

\newcommand{\no}{\noindent}
\newcommand{\bc}{\begin{center}}
\newcommand{\ec}{\end{center}}

\definecolor{DarkBlue}{rgb}{0,.08,.45}
\definecolor{DarkRed}{rgb}{.7,0,.4}
\definecolor{DarkGreen}{rgb}{0,0.4,0}

\newcommand{\bpmt}{\begin{pmatrix}}
	\newcommand{\epmt}{\end{pmatrix}}

\newcommand{\bsp}{\begin{split}}
	\newcommand{\esp}{\end{split}}

\newcommand{\bdes}{\begin{description}}
	\newcommand{\edes}{\end{description}}

\newcommand{\bass}{\begin{assumption}}
	\newcommand{\eass}{\end{assumption}}
\newcommand{\bthm}{\begin{thm}}
	\newcommand{\ethm}{\end{thm}}
\newcommand{\blem}{\begin{lem}}
	\newcommand{\elem}{\end{lem}}
\newcommand{\bcor}{\begin{cor}}
	\newcommand{\ecor}{\end{cor}}
\newcommand{\bpf}{\begin{proof}}
	\newcommand{\epf}{\end{proof}}

\def\i1{_{i1}}

\def\mhf{^{-1/2}}
\def\mtwo{^{-2}}
\def\half{^{1/2}}
\def\inv{^{-1}}

\def\diag{{\rm diag}}

\def\diam{{\rm diam}}

\def\id{{\rm id}}

\def\diffop{\mathrm{d}}

\def\wh{\widehat}
\def\wt{\widetilde}

\def\ra{\rightarrow}

\def \mbb {\mathbb}
\def \mbf {\mathbf}
\def \mc {\mathcal}
\def \msr{\mathscr}

\DeclareMathOperator*{\argmin}{argmin}

\def\expect{\mbb{E}}
\def\prob{P}
\def\real{\mbb{R}}
\def\ntnum{\mbb{N}}
\def\pint{\ntnum_+}
\def\integer{\mbb{Z}}
\def\O{O}
\def\o{o}
\def\Op{O_P}
\def\op{o_P}
\def\gify{\rightarrow\infty}
\def\indicator#1{\mbb I\left(#1\right)}
\def\ball#1#2{B_{#2}(#1)}

\usepackage{ifthen}
\newcommand{\hilbert}[1][]{%
	\ifthenelse{ \equal{#1}{} }
	{\ensuremath{\mathcal{L}^2}}
	{\ensuremath{\mathcal{L}^2_{#1}}}
} 
\newcommand{\ltwoNorm}[2][]{%
	\ifthenelse{ \equal{#1}{} }
	{\ensuremath{\|#2\|}}
	{\ensuremath{\|#2\|_{#1}}}
} 
\newcommand{\hsSpace}[2][]{
	\ifthenelse{\equal{#1}{}}
	{\ensuremath{\mathcal H_{#2}}}
	{\ensuremath{\mathcal H_{#1,#2}}}
} 

\newcommand{\innerprod}[3][]{%
	\ifthenelse{ \equal{#1}{} }
	{\ensuremath{\langle#2,#3\rangle}}
	{\ensuremath{\langle#2,#3\rangle_{#1}}}
}

\def\qaq{\quad\text{and}\quad}

\def\bprts#1{\bpmt{#1}\epmt}

\def\dcor{r}
\def\tsidx{s}
\def\nts{S}
\def\hubidx{k}
\def\vhubidx{l}
\def\sgnl{y}

\def\smap{\Gamma}
\def\smrCm{\Lambda}
\def\smrEst{\wh\Lambda}
\def\pmCm{\apdt_{\min}}
\def\pmEst{\wh{\apdt}_{\min}}

\def\pwSmap{\alpha_1}
\def\constSmap{C_1}
\def\pwSmr{\alpha_2}
\def\constSmr{C_2}
\def\radSmr{r}

\def\corm{\mbf{R}}

\def\idm{\mbf{I}}
\def\amat{\mbf{B}}
\def\amij{B_{\hubidx\vhubidx}}
\def\lapf{L}
\def\adjf{A}
\def\dgrf{D}
\def\ones{\mbf{1}}

\def\csm{\corm_{\dptidx}}
\def\csij{R_{\dptidx,\hubidx\vhubidx}}

\newcommandx*\metric[1][1=\msp, usedefault]{d_{#1}}
\def\msp{\mathcal{M}} 
\def\nobj{n} 
\def\vnobj{n'}
\def\objidx{i}
\def\varobjidx{i'}

\def\wp{h}

\def\wpcand{g}
\def\vwpcand{g^*}
\newcommand\wpi[1][\objidx]{\wp_{#1}}

\newcommand\wpiEst[1][\objidx]{\wh \wp_{#1}}
\def\wfsp{\mc{W}}

\def\anyObj{z}
\def\varAnyObj{z'}

\def\sumn{\sum_{\objidx=1}^{\nobj}}

\newcommandx*\pairwp[2][1=\varobjidx,2=\objidx]{g_{#1#2}}
\newcommandx*\pairwpOrc[2][1=\varobjidx,2=\objidx]{\wt g_{#1#2}}
\newcommandx*\pairwpEst[2][1=\varobjidx,2=\objidx]{\wh g_{#1#2}}
\def\paircand{g}

\newcommand{\obj}[1][]{
	\ifthenelse{\equal{#1}{}}
	{\ensuremath{X}}
	{\ensuremath{X(#1)}}
}
\newcommand{\objsmpl}[2][]{
	\ifthenelse{\equal{#1}{}}
	{\ensuremath{X_{#2}}}
	{\ensuremath{X_{#2}(#1)}}
}
\newcommand\obji[1][]{
	\ifthenelse{\equal{#1}{}}
	{\ensuremath{X_{\objidx}}}
	{\ensuremath{X_{\objidx}(#1)}}
}

\newcommand{\wobj}[1][]{
	\ifthenelse{\equal{#1}{}}
	{\ensuremath{Y}}
	{\ensuremath{Y(#1)}}
}
\newcommand{\wobjsmpl}[2][]{
	\ifthenelse{\equal{#1}{}}
	{\ensuremath{Y_{#2}}}
	{\ensuremath{Y_{#2}(#1)}}
}
\newcommand\wobji[1][]{\wobjsmpl[#1]{\objidx}}
\newcommand\varwobji[1][]{\wobjsmpl[#1]{\varobjidx}}

\newcommand{\wobjsmplEst}[2][]{
	\ifthenelse{\equal{#1}{}}
	{\ensuremath{\wh{Y}_{#2}}}
	{\ensuremath{\wh{Y}_{#2}(#1)}}
}
\newcommand\wobjiEst[1][]{\wobjsmplEst[#1]{\objidx}}
\newcommand\varwobjiEst[1][]{\wobjsmplEst[#1]{\varobjidx}}
\def\wobjiLm{\wt{Y}_{\objidx}}

\def\cdn{\mid}

\def\ptb{\mc{P}} 
\def\ptbi{\ptb_{\objidx}}
\def\ptbij{\ptb_{\objidx\dptidx}}

\def\obsobj{Z}
\newcommandx*\obsobji[1][1=\objidx, usedefault]{\obsobj_{#1}}
\newcommandx*\obsobjij[2][1=\objidx, 2=\dptidx, usedefault=@]{\obsobj_{#1#2}}

\def\dptidx{j}
\newcommand{\ndp}[1][]{
	\ifthenelse{ \equal{#1}{} }
	{\ensuremath{m}}
	{\ensuremath{m_{#1}}}
}
\def\ndpi{\ndp_{\objidx}}

\def\summi{\sum_{\dptidx = 1}^{\ndpi}}

\newcommand\mObj[1][\tm]{\mu(#1)} 
\def\mObjNull{\mu}

\def\rtm{T}
\newcommandx*\tmi[1][1=\objidx, usedefault]{\rtm_{#1}}
\newcommandx*\tmij[2][1=\objidx, 2=\dptidx, usedefault=@]{\rtm_{#1#2}}

\def\tdom{\mc T}
\def\maxtm{\tau}
\def\tm{t}
\def\vartm{s}
\def\forallt{\quad\text{for all } \tm\in\tdom}

\newcommand\tgridpt[1][\tgrididx]{t_{#1}}
\def\tgrididx{k}
\def\ntgrid{p}
\newcommand\lsvec[1][\paircand]{\theta_{#1}} 
\def\lsSp{\Theta}
\def\anyVec{\theta}
\def\fixVec{\theta_*}
\def\lsvecPw{\lsvec[\pairwp]}

\newcommand\lsvecEst[1][\pairwp]{\wh{\theta}_{#1}}
\def\lsbasis{A}

\def\lagrM{\msr{L}_{\mObjNull}} 
\def\obfnM{\msr{C}_{\mObjNull}} 
\def\constrM{I}
\newcommandx*\constrCoef[1][1=\tgrididx, usedefault]{\ensuremath{\zeta_{#1}}}
\def\wpSlope{\xi}
\def\pwM{2}
\def\pwc{\alpha}

\def\pnty{\lambda}
\def\obfnYest{\msr{C}_{\wh{Y},\pnty}}

\def\ker{K}
\def\bwAll{b}

\def\bwi{b_{\objidx}}
\def\bwvi{b_{\varobjidx}}
\newcommandx*\kerh[1][1=\bwi, usedefault]{\ker_{#1}}
\def\momidx{l}

\def \LS {{\rm LS}}
\def\wfspLS{\wfsp_{\LS}}

\def\prsp{\Omega} 
\def\sfield{\mathscr{F}}
\def\outcm{\omega} 

\def\prspPrcs{\prsp_{1}} 
\def\prspNoise{\prsp_{2}} 
\def\sfieldPrcs{\sfield_1}
\def\sfieldNoise{\sfield_2}
\def\probNoise{\prob_{\prspNoise}}
\def\probPrcs{\prob_{\prspPrcs}}

\def\outcmPrcs{\outcm_{1}}

\def\expectNoise{\expect_{\prspNoise}}

\ifx\unifPrspPrcs\undefined
\def\unifPrspPrcs{0}
\else
\fi

\if\unifPrspPrcs1

\def\varOutcmPrQ{[\outcmPrcs']}
\def\detmWobj{\outcmPrQ} 
\def\varDetmWobj{\varOutcmPrQ}
\def\detmSp{\prspPrQ}

\else
\def\detmWobj{\outcmPrcs} 
\def\detmSp{\prspPrcs}
\fi

\def\interior{^\circ}

\def\pdto{U}
\def\rpso{V}
\def\pdom{\tdom} 
\def\pdtj{\pdto_{\dptidx}}
\def\rpsj{\rpso_{\dptidx}}

\def\apdt{\tm} 
\def\vapdt{\vartm} 
\def\arps{\anyObj}
\def\varps{\varAnyObj} 

\newcommandx*\objCm[1][1=\tm,usedefault]{\ensuremath{\nu(#1)}} 
\def\objCmNull{\nu}
\newcommandx*\objLm[2][1=\tm,2=\bwo,usedefault=@]{\wt{\nu}_{#2}(#1)} 
\newcommandx*\objLmOrc[2][1=\tm,2=\ndpo,usedefault=@]{\wh{\nu}_{#2}(#1)} 
\def\objLmOrcNull{\wh\nu_{\ndpo}}

\newcommandx*\objfCm{L}
\newcommandx*\objfLm[1][1=\bwo, usedefault]{\wt{L}_{#1}}
\newcommandx*\objfLmOrc[1][1=\ndpo, usedefault=@]{\wh{L}_{#1}}

\newcommandx*\kmom[2][1=\momidx,2=\bwo, usedefault]{\ensuremath{\rho_{#1,#2}}}
\newcommandx*\kvar[1][1=\bwo, usedefault]{\ensuremath{\sigma^2_{#1}}}
\newcommandx*\kmomEst[2][1=\momidx,2=\ndpo, usedefault]{\wh{\rho}_{#1,#2}}
\newcommandx*\kvarEst[1][1=\ndpo,usedefault]{\ensuremath{\wh\sigma^2_{#1}}}
\newcommandx*\kermom[4][3=\tm, 4=\bwo, usedefault]{\mc{K}_{#1,#2}(#3,#4)} 

\def\wLoc#1#2#3{w(#1,#2,#3)}

\def\wLocEst#1#2#3{\wh w (#1,#2,#3)}

\def\wLocDfo#1{\wLoc{#1}{\apdt}{\bwo}}
\def\wLocEstDfo#1{\wLocEst{#1}{\apdt}{\bwo}}

\def\unifinPdom{\apdt\in\pdom}
\def\stinPdom{\unifTdom{\vapdt,\apdt\in\pdom}}
\def\twotinPdom{\unifTdom{\apdt_1,\apdt_2\in\pdom}}

\def\denPdt{f_{\pdto}}
\def\cdenPdt{f_{\pdto\cdn\rpso}}

\def\cdfRps{F_{\rpso}}
\def\ccdfRps{F_{\rpso\cdn\pdto}}

\def\dCL{\phi}
\def\disc{\psi}

\def\jointdtn{\mc{F}}
\def\ltwoNmPr#1{\|#1\|_{\jointdtn}}

\newcommand{\objPrcs}[1][]{
	\ifthenelse{\equal{#1}{}}
	{\ensuremath{Y}}
	{\ensuremath{Y(#1)}}
}
\newcommand\objPrcsEst[1][]{
	\ifthenelse{\equal{#1}{}}
	{\ensuremath{\wh{Y}}}
	{\ensuremath{\wh{Y}(#1)}}
}

\newcommand{\objPrcsSmpl}[2][]{
	\ifthenelse{\equal{#1}{}}
	{\ensuremath{Y_{#2}}}
	{\ensuremath{Y_{#2}(#1)}}
}

\newcommand{\objPrcsSmplEst}[2][]{
	\ifthenelse{\equal{#1}{}}
	{\ensuremath{\wh{Y}_{#2}}}
	{\ensuremath{\wh{Y}_{#2}(#1)}}
}

\def\objPrcso{Y_{\detmWobj}}
\def\objPrcsLm{\wt{Y}}
\def\tmj{\rtm_{\dptidx}}
\def\obsobjj{\obsobj_{\dptidx}}
\def\ptbj{\ptb_{\dptidx}}

\newcommandx*\objCmR[2][1=\tm,2=\detmWobj,usedefault=@]{Y_{#2}(#1)} 
\newcommandx*\objLmR[3][1=\tm,2=\detmWobj,3=\bwo,usedefault=@]{\wt{Y}_{#2,#3}(#1)} 
\newcommandx*\objLmOrcR[3][1=\tm,2=\detmWobj,3=\ndpo,usedefault=@]{\wh{Y}_{#2,#3}(#1)} 
\newcommandx*\varObjLmOrcR[3][1=\tm,2=\varDetmWobj,3=\ndpo,usedefault=@]{\wh{Y}_{#2,#3}(#1)} 

\newcommandx*\objfCmR[1][1=\detmWobj, usedefault]{M_{#1}}
\newcommandx*\objfLmR[2][1=\detmWobj,2=\bwo, usedefault=@]{\wt{M}_{#1,#2}}
\newcommandx*\objfLmOrcR[2][1=\detmWobj,2=\ndpo, usedefault=@]{\wh{M}_{#1,#2}}

\def\objLmRbw{\objLmR[@][@][\bwi]}
\def\objLmRvbw{\objLmR[@][@][\bwvi]}
\def\objLmOrcRm{\objLmOrcR[@][@][\ndpi]}

\def\objfLmRbw{\objfLmR[@][\bwi]}
\def\objfLmRvbw{\objfLmR[@][\bwvi]}

\newcommandx*\kmomR[2][1=\momidx,2=\bwo, usedefault]{\ensuremath{\varrho_{#1,#2}}}
\newcommandx*\kvarR[1][1=\bwo, usedefault]{\ensuremath{\varsigma^2_{#1}}}
\newcommandx*\kmomREst[2][1=\momidx,2=\ndpo, usedefault]{\wh{\varrho}_{#1,#2}}
\newcommandx*\kvarREst[1][1=\ndpo,usedefault]{\ensuremath{\wh\varsigma^2_{#1}}}

\def\wLocR#1#2#3{v(#1,#2,#3)}
\def\wLocREstNull{\wh v}
\def\wLocREst#1#2#3{\wh v (#1,#2,#3)}

\def\wLocRDfo#1{\wLocR{#1}{\tm}{\bwo}}
\def\wLocREstDfo#1{\wLocREst{#1}{\tm}{\bwo}}

\def\unifinWobj{\unifPrcs{\detmWobj\in\detmSp}}
\def\unifinTm{\unifTdom{\tm\in\tdom}}

\def\denRtm{f_{\rtm}}
\def\cdenRtm{f_{\rtm\cdn\obsobj_{\detmWobj}}}

\def\cdfObj{F_{\obsobj_{\detmWobj}}}

\def\dCLR{\phi_{\detmWobj}}

\def\jointdtnNoise{\mc{F}_{\detmWobj}}
\def\ltwoNmPrNoise#1{\|#1\|_{\jointdtnNoise}}

\def\smconst{\varepsilon}

\def\tmo{\rtm} 
\def\tmoj{\rtm_{\dptidx}} 

\def\obsobjo{\obsobj_{\detmWobj}}
\def\obsobjoj{\obsobj_{\detmWobj\dptidx}}

\def\bwo{b}
\def\kerho{\ker_{\bwo}}
\def\ndpo{\ndp}
\def\summo{\sum_{\dptidx = 1}^{\ndpo}}
\def\supi{\sup_{1\le\objidx\le\nobj}}
\def\vsupi{\sup_{1\le\varobjidx\le\vnobj}}
\def\supivi{\sup_{\substack{1\le\objidx\le\nobj,\ 1\le\varobjidx\le\vnobj}}}
\def\infi{\inf_{1\le\objidx\le\nobj}}

\def\covernum{N}
\def\pwCm{\beta_1}
\def\pwLm{\beta_2}
\def\pwLmUnif{\beta'_2}
\def\constCm{c_1}
\def\constLm{c_2}
\def\radCm{r_1}
\def\radLm{r_2}
\def\radEntropy{r}

\def\anyFctn{g}
\def\anySet{E}
\def\open{E}

\def\shpabeta{\xi_{\tm}}
\def\shpbbeta{\gamma_{\tm}}
\def\gaus{N}
\def\meantgaus{\xi_{\tm}}
\def\sdtgaus{\gamma_{\tm}}

\def\distort{a}
\newcommand\distortFctn[1][\distort]{\msr{T}_{#1}}
\def\distortTm{u}

\def\cdf{F}
\def\rarg{x}
\def\pushforward#1#2{#1\# #2}

\def\cdfStdGaus{\Phi}

\def\unif{\mathrm{Uniform}}
\def\pois{\mathrm{Poisson}}

\def \tmise {\ensuremath{\mathrm{TMISE}} }
\def \wmise {\ensuremath{\mathrm{WMISE}} }

\definecolor{dgray}{gray}{0.35}
\def\detail#1{#1}

\def\note#1{\textcolor{dgray}{#1}}
\def\note#1{\bco#1\fi}

\def\unifTdom{} 
\def\unifPrcs{}

\def\kurt{\vartheta_{\ndpo}} 
\def\radius{\delta}
\def\csetBR{R}
\def\setUnifR{B_{\csetBR,\ndpo}}

\def\c{^c} 

\def\setidx{k}
\def\setknt{\mc{D}_{\setidx,\apdt}}

\def\objfLmOvsLmt{J_{\apdt}}
\def\objfLmOvsLmtOne{J_{\apdt}^{(1)}}
\def\objfLmOvsLmtTwo{J_{\apdt}^{(2)}}

\def\objfLmORvsLmRtTwo{J_{\tm,\detmWobj}^{(2)}}

\def\gfctn#1{g_{#1}}
\def\gclass{\mc{G}_{\bwo,\radius}}
\def\genve{G_{\bwo,\radius}} 
\def\gdiff{D_{\bwo,\radius}}  

\def\pcon{\eta} 
\def\tpcon{\tilde{\pcon}}
\def\intL{\ell} 
\def\cn{a_{\ndpo}}

\def\metricTm{\metric[\tdom]}
\def\brktnum{N_{[]}}
\def\pwLarger{\upsilon}
\def\ch{q_{\bwo}}

\def\chbi{q_{\bwi}}
\def\chbo{q_{\bwo}}
\def\cbni{a_{\ndpi}}

\def\cbno{a_{\ndpo}}
\def\pwSumn{\gamma}

\def\lipm{C_{\mObjNull}}

\def\qntl{Q}

\renewcommand{\emph}{\text}

\usepackage{authblk}
\def\blind{0}

\usepackage{xspace}
\def\pfsec{the Supplementary Material\xspace}

\def\references{\bibliography{merged}}

\if0\blind{
\title{\vspace{-1cm}Uniform convergence of local Fr\'echet regression,\\ with applications to locating extrema and time warping\\ for metric space valued trajectories\thanks{Research supported by NSF grants DMS-1712864 and DMS-2014626.} \thanks{Data used in preparation of this article were obtained from the Alzheimer's Disease Neuroimaging Initiative (ADNI) database (\url{adni.loni.usc.edu}). As such, the investigators within the ADNI contributed to the design and implementation of ADNI and/or provided data but did not participate in analysis or writing of this report. A complete listing of ADNI investigators can be found at: \url{http://adni.loni.usc.edu/wp-content/uploads/how_to_apply/ADNI_Acknowledgement_List.pdf}}}
\author{Yaqing Chen}
\author{Hans-Georg M{\"u}ller}
\affil{Department of Statistics, University of California, Davis \protect\\ Davis, CA 95616, USA}
}\fi

\if1\blind{
\title{\vspace{-1cm}Uniform convergence of local Fr\'echet regression,\\ with applications to locating extrema and time warping\\ for metric space valued trajectories\thanks{Data used in preparation of this article were obtained from the Alzheimer's Disease Neuroimaging Initiative (ADNI) database (\url{adni.loni.usc.edu}). As such, the investigators within the ADNI contributed to the design and implementation of ADNI and/or provided data but did not participate in analysis or writing of this report. A complete listing of ADNI investigators can be found at: \url{http://adni.loni.usc.edu/wp-content/uploads/how_to_apply/ADNI_Acknowledgement_List.pdf}}}
\author{\vspace{-1cm}}
}\fi
\date{}

\newcommand\Title{}
\newcommand\Author{
}

\begin{document}

\maketitle

{\bc{\bf \sf ABSTRACT}\ec}
\no Local Fr\'echet regression is a nonparametric regression method for metric space valued responses and Euclidean predictors, which can be utilized to obtain estimates of smooth trajectories taking values in general metric spaces from noisy metric space valued random objects. We derive uniform rates of convergence, which so far have eluded theoretical analysis of this method, for both fixed and random target trajectories, where we utilize tools from empirical processes. These results are shown to be widely applicable in metric space valued data analysis. In addition to simulations, we provide two pertinent examples where these results are important: The consistent estimation of the location of properly defined extrema in metric space valued trajectories, which we illustrate with the problem of locating the age of minimum brain connectivity as obtained from fMRI data; Time warping for metric space valued trajectories, illustrated with yearly age-at-death distributions for different countries.


\vspace{0.7em}
\no {KEY WORDS}: Random objects, Metric space valued functional data, Rates of convergence, Smoothing, fMRI, Mortality distributions. 

\clearpage
\section{Introduction}

Non-Euclidean data, or random object data taking values in metric spaces, have become increasingly common in modern data analysis and data science  while there is a lack of principled and statistically justified methodology. 
Since such data are metric space valued, they generally do not lie in a vector space, which means that many classical notions of statistics such as the definition of sample or population mean as an average or  expected value do not apply anymore and need to be replaced by  barycenters or Fr\'echet means \citep{frec:48}, the mathematical and statistical properties of which have been studied for various metric spaces. These include finite-dimensional Riemannian manifolds, the space of symmetric positive definite matrices, Kendall's shape space, or the Wasserstein space of distributions \citep[][among others]{bhat:03,bhat:05,dryd:09,ague:11,huck:12,lego:17}; the last is not a Riemannian manifold \citep{ambr:04}. 

Another important topic is to study the relationship of such random objects with other variables, where regression analysis comes into play.
Nonparametric (local) regression techniques have been used for a long time for smoothing and  interpolation of Euclidean responses. While Nadaraya--Watson type methods have been proposed in the cases where data lie in finite-dimensional Riemannian manifolds \citep{pell:06,davi:07,stei:09,stei:10,yuan:12}, and also generic metric spaces \citep{hein:09}, 
local Fr\'echet regression \citep{pete:19}, can be viewed as a generalization of local linear regression for metric space valued responses. While pointwise asymptotic results for the corresponding estimates have been previously derived, uniform convergence is much more challenging.
Here we derive uniform rates of convergence for the local Fr\'echet regression estimates \rvone{of the fixed conditional Fr\'echet mean trajectory} using tools from empirical process theory; see Theorem~\ref{thm:unifFregRateOverTm}. 
We then  extend this result to the case where local Fr\'echet regression is applied to recover metric space valued random processes from discrete noisy observations; see Theorem~\ref{thm:unifFregRateRdmTgt}. 
While these results may be of interest in their own right,  our derivations are motivated by important applications of uniform convergence. 
These include the estimation of the location of suitably defined extrema in metric space valued functions as well as  time warping for metric space valued functional data. 

Estimation of modes or maximum locations has been well studied for regression functions in nonparametric regression for real-valued data \citep[e.g.,][]{devr:78,mull:89:3,beli:12} and densities of probability distributions \citep[e.g.,][]{parz:62,vieu:96,bala:09}. 
For object data in metric spaces, the location of extrema with regard to functionals of interest can be obtained based on the estimation of the complete conditional Fr\'echet mean trajectory through local Fr\'echet regression, where the consistency of the derived estimates of the  location of an extremum is guaranteed by the uniform convergence of local Fr\'echet regression under regularity  conditions. 

For real-valued functional data, a random function has two types of variation: amplitude variation and phase (or time) variation.  Confounding the two types of variation may seriously contaminate conventional statistical methods \citep{knei:92}. This issue has been addressed by time warping, also referred to as curve synchronization, registration or alignment. 
The prototypical method is dynamic time warping (DTW) \citep{sako:78} 
and various statistical approaches have been developed over the years  for real-valued functional data \citep[][among others]{knei:92,gass:95,wang:97,rams:98, gerv:04,jame:07}; see \citet{marr:15} for a recent review. 
Beyond classical functional data in $\hilbert$ Hilbert space, time synchronization has been investigated in engineering for non-Euclidean semi-metric spaces, also referred to as dissimilarity spaces \citep{fara:93}, where the DTW method and its variants have been adopted with applications including human motion recognition and video classification \citep[][among others]{gong:11,vu:12,trig:18}. We note that no theoretical results were provided in these works.
To our knowledge, no comprehensive studies exist of time warping for samples of metric space valued trajectories that include an investigation of statistical properties or asymptotic behavior. 

Statistical methods devised for real-valued functional data, or random elements of a Hilbert space, are usually not applicable to functional data taking values in a metric space \citep{huck:15}. 
Even extending functional data analysis methods without time warping to metric space valued trajectories is challenging \citep{dube:20}, due to the fact that in general metric spaces one cannot rely on an algebraic structure.
In this paper, we tackle the even more challenging task to extend  pairwise warping for real-valued curves  \citep{tang:08} to metric space valued functional data. 
Since random processes are usually not fully observed and only discrete and noisy measurements are available, local Fr\'echet regression needs to be used to obtain complete subject-specific trajectories. \rvone{The uniform convergence result in Theorem~\ref{thm:unifFregRateRdmTgt} for local Fr\'echet regression estimates of random processes is crucial} 
to derive the uniform consistency of estimates of the pairwise warping functions, which  form  the backbone of the proposed warping method; the uniform consistency provides the major justification for this approach.

The remainder of the paper is organized as follows. A  key result on the uniform rate of convergence for local Fr\'echet regression is presented  in Section~\ref{sec:locFregFix}, followed by a study of  
the case where the target of the local Fr\'echet regression is a random process rather than a fixed trajectory in Section~\ref{sec:locFregRdm}. 
We present two applications, where the  estimation of the location of extrema is based on Theorem~\ref{thm:unifFregRateRdmTgt} and presented in Section~\ref{sec:extremum}.  A second key application is the time synchronization for metric space valued functional data in  Section~\ref{sec:wp}, which is based on Theorem~\ref{thm:unifFregRateRdmTgt}. The proposed methods are shown to lead to consistent estimation of time warping functions in Theorem~\ref{thm:pai} and Corollary~\ref{cor:warp}. 
We then demonstrate the estimation of extrema locations with functional magnetic resonance imaging (fMRI) data from the Alzheimer's Disease Neuroimaging Initiative (ADNI) database in Section~\ref{sec:adni}, where we find the time of minimum brain connectivity, quantified by the Fiedler values of the brain network. The time warping for metric space valued functional data is illustrated with yearly age-at-death distribution data for different countries from the Human Mortality Database  in Section~\ref{sec:wpMort}. 
We also report the results of various \rvone{simulation studies for the proposed warping method for distribution-valued functional data in Section~\ref{sec:wpSimu} in \pfsec.}


\section{Uniform Rates of Convergence for Local Fr\'echet Regression}\label{sec:locFregFix}

Let $(\msp, \metric)$ be a totally bounded separable metric space, and $\tdom= [0,\maxtm]$ be a closed interval in $\real$. This will be assumed throughout the paper. 
Consider a random pair $(\pdto,\rpso)$ with a joint distribution on the product space $\pdom\times\msp$, where $\pdto$ is a real-valued predictor and $\rpso$ is a metric space valued response. 
Suppose $\{(\pdtj,\rpsj)\}_{\dptidx=1}^{\ndpo}$ are i.i.d. realizations of $(\pdto,\rpso)$. 
For any $\apdt\in\pdom$, the conditional Fr\'echet mean of $\rpso$ given $\pdto=\apdt$ is defined by 
\bgt\label{eq:objCm}
\objCm = \argmin_{\arps\in\msp} \objfCm(\arps,\apdt), \quad \objfCm(\arps,\apdt) = \expect[\metric^2(\rpso, \arps) \cdn \pdto = \apdt]. \egt 
We consider local Fr\'echet means \citep{pete:19}  
\bgt \label{eq:objLm}
\objLm = \argmin_{\arps\in\msp}\objfLm(\arps,\apdt), \quad \objfLm(\arps,\apdt) = \expect[\wLocDfo{\pdto} \metric^2(\rpso,\arps)], \egt
where $\wLocDfo{\vapdt} =  \kerho(\vapdt-\apdt)[\kmom[2](\apdt) - \kmom[1](\apdt)(\vapdt - \apdt)]/\kvar(\apdt)$,  $\kmom(\apdt) = \expect[\kerho(\pdto - \apdt)(\pdto - \apdt)^{\momidx}]$, for $\momidx=0,1,2$, $\kvar(\apdt) = \kmom[0](\apdt)\kmom[2](\apdt) - \kmom[1](\apdt)^2$, $\kerho(\cdot) = \ker(\cdot/\bwo)/\bwo$, $\ker$ is a smoothing kernel, and $\bwo = \bwo(\ndpo)>0$ is a bandwidth sequence. 
Local Fr\'echet regression estimates of $\objCm$ are given by 
\bgt\label{eq:objLmOrc}
\objLmOrc = \argmin_{\arps\in\msp} \objfLmOrc(\arps,\apdt),\quad \objfLmOrc(\arps,\apdt) = \ndpo\inv \summo \wLocEstDfo{\pdtj} \metric^2(\rpsj,\arps), \egt 
where \rvone{$\wLocEstDfo{\vartm} = \kerho(\vapdt-\apdt)[\kmomEst[2](\apdt) - \kmomEst[1](\apdt)(\vapdt - \apdt)]/\kvarEst(\apdt)$,  $\kmomEst(\apdt) =\ndpo\inv \summo[\kerho(\pdtj - \apdt)(\pdtj - \apdt)^{\momidx}]$, $\momidx=0,1,2$,  $\kvarEst(\apdt) = \kmomEst[0](\apdt)\kmomEst[2](\apdt) - \kmomEst[1](\apdt)^2$.}

Let $\pdom\interior = (0,\maxtm)$ be the interior of $\pdom$. We require the following assumptions to obtain uniform rates of convergence over $\unifinPdom$ for local Fr\'echet regression estimators in \eqref{eq:objLmOrc}. 
\ben[label = (K\arabic*), series = kernel, start = 0]
\item \label{ass:ker} 
The kernel $\ker$ is a probability density function, symmetric around zero and uniformly continuous on $\real$. Defining $\ker_{kl} = \int_{\real}\ker(\rarg)^k\rarg^l\diffop \rarg < \infty$, 
for $k,l\in\ntnum$, $\ker_{14}$ and $\ker_{26}$, are finite.
The derivative $\ker'$ exists and is bounded on the support of $\ker$, i.e., $\sup_{\ker(\rarg)>0}|\ker'(\rarg)| <\infty$; additionally, $\int_{\real}\rarg^2|\ker'(\rarg)|\sqrt{|\rarg\log|\rarg||}\diffop\rarg<\infty$.
\note{\\Note: These assumptions on the kernel $\ker$ are required so that the assumptions for the kernels $\ker(\rarg)\rarg^\momidx$ for $\momidx=0,1,2$ and $\ker(\rarg)|\rarg|$ are satisfied as per Silverman (1978) and Mack \& Silverman (1982). 
	Additionally, since we assume $\denRtm$ is bounded away from zero on the entire time domain $\pdom$, we do not need to assume the kernel $\ker$ has bounded support as in the Wgrad2g, where the boundedness of the support of $\ker$ is needed for the uniform equicontinuity of $\objfLm(\arps,\cdot)$ and to prove $\sup_{\apdt\in\pdom} \sup_{\metric(\arps,\objCm)<\radius} |(\objfLm-\objfCm)(\arps,\apdt)- (\objfLm-\objfCm)(\objCm,\apdt)|=\O(\bwo^2\radius)$.} 
\een
\ben[label = (R\arabic*), series = freg, start = 0]
\item \label{ass:jointdtn} 
The marginal density $\denPdt$ of $\pdto$ and the conditional densities $\cdenPdt(\cdot,\arps)$ of $\pdto$ given $\rpso = \arps$ exist and are continuous on $\pdom$ and twice continuously differentiable on $\pdom\interior$, the latter for all $\arps\in\msp$. 
The marginal density $\denPdt$ is bounded away from zero on $\pdom$, $\inf_{\apdt\in\pdom}\denPdt(\apdt)>0$. 
The second-order derivative $\denPdt''$ is bounded,  $\sup_{\apdt\in\pdom\interior}|\denPdt''(\apdt)|<\infty$. 
The second-order partial derivatives $(\partial^2\cdenPdt/\partial\apdt^2)(\cdot,\arps)$ are uniformly bounded, $\sup_{\apdt\in\pdom\interior,\, \arps\in\msp}|(\partial^2\cdenPdt/\partial\apdt^2)(\apdt,\arps)| < \infty$. 
\note{The boundedness and uniform boundedness of $\denPdt''$ and $(\partial^2\cdenPdt/\partial\apdt^2)(\cdot,\arps)$ are needed to show: (1) the uniform order of the remainder of $\expect[\kerho(\pdto - \apdt)(\pdto - \apdt)^{\momidx}]$ and $\expect[\kerho(\pdto - \apdt) (\pdto - \apdt)^{\momidx} \cdn \rpso = \arps]$; (2) the uniform equicontinuity of $\objfLm(\arps,\cdot)$; and (3) $\sup_{\apdt\in\pdom} \sup_{\metric(\arps,\objCm)<\radius} |(\objfLm-\objfCm)(\arps,\apdt)- (\objfLm-\objfCm)(\objCm,\apdt)|=\O(\bwo^2\radius)$. }Additionally, for any open set $\open\subset\msp$, $\prob(\rpso\in\open\cdn \pdto = \apdt)$ is continuous as a function of $\apdt$\note{{ }(this is needed in the proof of Theorem~3 of \citet{pete:19})}; 
for any $\apdt\in\pdom$, $\objfCm(\arps,\apdt)$ is equicontinuous, i.e., 
\bgt\label{eq:eqcont_objCm} 
\limsup_{\vapdt\ra\apdt} \sup_{\arps\in\msp} \left|\objfCm(\arps,\vapdt) - \objfCm(\arps,\apdt) \right| = 0. \egt 
\item \label{ass:minUnif} 
For all $\unifinPdom$, the minimizers $\objCm$, $\objLm$ and $\objLmOrc$ exist and are unique, the last $\prob$-almost surely. 
In addition, for any $\epsilon>0$, 
\bgt\nn
\inf_{\unifinPdom} 
\inf_{\metric(\objCm,\arps) > \epsilon} \bprts{\objfCm(\arps,\apdt) - \objfCm(\objCm,\apdt)} > 0,\\
\liminf_{\bwo\ra 0} \inf_{\unifinPdom} 
\inf_{\metric(\objLm,\arps) > \epsilon} \bprts{\objfLm(\arps,\apdt) - \objfLm(\objLm,\apdt)} > 0, \egt
and there exists $c = c(\epsilon)>0$ such that
\bgt\nn
\prob\left(\inf_{\unifinPdom} \inf_{\metric(\objLmOrc,\arps) > \epsilon} \bprts{\objfLmOrc(\arps,\apdt) - \objfLmOrc(\objLmOrc,\apdt)} \ge c\right) \ra 1. \egt
\item \label{ass:entropyUnif} 
Let $\ball{\objCm}{\radEntropy}\subset \msp$ be a ball of radius $\radEntropy$ centered at $\objCm$ and $\covernum(\epsilon, \ball{\objCm}{\radEntropy}, \metric)$ be its covering number using balls of radius $\epsilon$. Then 
\bgt\nn \int_0^1 \sup_{\unifinPdom} 
\sqrt{1+\log\covernum(\radEntropy\epsilon, \ball{\objCm}{\radEntropy}, \metric)} \diffop\epsilon = \O(1),\quad\text{as }\radEntropy\ra {0+}.\egt
\note{For convenience, we can modify the condition from ``$\dots = \O(1)$ as $\radEntropy\ra {0+}$'' to ``$\sup_{\radEntropy>0}\dots <\infty$''. In the latter case, we can set $\pcon=\radLm$ in \eqref{eq:tailprobUnif} (see the derivation of the bounds for the bracketing entropy integral before \eqref{eq:objfLmEvsLm2Unif} for more details).}
\item \label{ass:curvatureUnif} 
There exists $\radCm,\radLm>0$, $\constCm,\constLm>0$ and $\pwCm,\pwLm>1$ such that 
\bgt\nn
\inf_{\unifinPdom} 
\inf_{\metric(\arps,\objCm)<\radCm} \left[\objfCm(\arps,\apdt) - \objfCm(\objCm,\apdt)- \constCm \metric(\arps,\objCm)^{\pwCm}\right] \ge 0, \\
\liminf_{\bwo\ra 0} \inf_{\unifinPdom} \inf_{\metric(\arps,\objLm)<\radLm} 
\left[\objfLm(\arps,\apdt) - \objfLm(\objLm,\apdt) - \constLm \metric(\arps,\objLm)^{\pwLm}\right] \ge 0.
\egt
\een

Similar yet weaker assumptions have been made by \citet{pete:19} for pointwise rates of convergence for local Fr\'echet regression estimators.  
Assumption~\ref{ass:ker} is needed to apply  results of \citet{silv:78} and \citet{mack:82}, and \ref{ass:jointdtn} is a standard distributional assumption for  
local nonparametric regression. These assumptions  guarantee the asymptotic uniform equicontinuity of $\objfLm$ and control the behavior of $(\objfLm-\objfCm)$ around $\objCm$ uniformly over $\apdt\in\pdom$, whence we obtain the uniform rate for the bias part $\metric(\objCm,\objLm)$ and the uniform consistency of the stochastic part $\metric(\objLm,\objLmOrc)$ for the local Fr\'echet regression estimators. 
In particular, \eqref{eq:eqcont_objCm} guarantees the $\metric$-continuity of $\objCm$ in conjunction with \ref{ass:minUnif}. 
Assumption~\ref{ass:minUnif} is commonly used to establish the uniform consistency of M-estimators \citep{vand:96}. It ensures the uniform convergence of $\objfLm(\cdot,\apdt)$ to $\objfCm(\cdot,\apdt)$ and the weak convergence of the empirical process $\objfLmOrc(\cdot,\apdt)$ to $\objfLm(\cdot,\apdt)$, which, in conjunction with the assumption that the metric space $\msp$ is totally bounded, implies the pointwise convergence of the minimizers for any given $\apdt\in\pdom$; it 
also ensures that the (asymptotic) uniform equicontinuity of $\objfLm$ and $\objfLmOrc$ implies the (asymptotic) uniform equicontinuity of $\objLm[\cdot]$ and $\objLmOrc[\cdot]$, whence the uniform convergence of the minimizers follows as  the time domain $\pdom$ is compact.  
Assumptions~\ref{ass:entropyUnif} and \ref{ass:curvatureUnif} are adapted from empirical process theory to control the differences $(\objfLmOrc-\objfLm)$ and $(\objfLm-\objfCm)$ near the minimizers $\objLm$ and $\objCm$, respectively, which is necessary to obtain the convergence rates for the bias and stochastic parts. 


In the following, we discuss  assumptions \ref{ass:minUnif}--\ref{ass:curvatureUnif} in the context of some specific 
metric spaces. 

\begin{exm}\label{eg:wsp}
	Let $\msp$ be the set of probability distributions on a closed interval of $\real$ with finite second moments, endowed with the $\hilbert$-Wasserstein distance $\metric[W]$; specifically, for any two distributions $\arps_1,\arps_2\in\msp$,
	\bal\nn \metric[W](\arps_1,\arps_2) = \left(\int_0^1 \left(\qntl_{\arps_1}(\rarg) - \qntl_{\arps_2}(\rarg)\right)^2\diffop\rarg \right)\half = \metric[\hilbert]\left(\qntl_{\arps_1},\qntl_{\arps_2}\right),\eal
	where $\qntl_{\arps}$ is the quantile function for any given distribution $\arps\in\msp$. The Wasserstein space $(\msp,\metric[W])$ satisfies  \ref{ass:minUnif}--\ref{ass:curvatureUnif} with $\pwCm=\pwLm=2$.
\end{exm}

\begin{exm}\label{eg:cor}
	Let $\msp$ be the space of $\dcor$-dimensional correlation matrices, i.e., symmetric, positive semidefinite matrices in  $\real^{\dcor\times\dcor}$ with diagonal elements all equal to 1, endowed with the Frobenius metric $\metric[F]$. The space $(\msp,\metric[F])$ satisfies \ref{ass:minUnif}--\ref{ass:curvatureUnif} with $\pwCm=\pwLm=2$.
\end{exm}


For Examples~\ref{eg:wsp}--\ref{eg:cor}, we note that since the Wasserstein space and the space of correlation matrices are Hadamard spaces \citep[the former as per][]{kloe:10}, there exists a unique minimizer of $\objfCm(\cdot,\apdt)$, for any $\apdt\in\pdom$ \citep{stur:03}. Examples~\ref{eg:wsp}--\ref{eg:cor} 
follow from similar arguments as those in the proofs of Propositions~1--2 of \citet{pete:19}; we omit the details.   

We then obtain uniform rates of convergence over $\unifinPdom$ for local Fr\'echet regression estimators as follows. Proofs and auxiliary results are in \pfsec. 

\bthm\label{thm:unifFregRateOverTm}
Under  \ref{ass:ker},  \ref{ass:jointdtn}--\ref{ass:curvatureUnif} and if 
$\bwo\ra 0$, $\ndpo\bwo^2(-\log\bwo)\inv\gify$, as $\ndpo\gify$, for any $\smconst>0$,  it holds  for $\objCm$, $\objLm$, and $\objLmOrc$ as per \eqref{eq:objCm}--\eqref{eq:objLmOrc},  respectively,  that
\begin{gather}
\sup_{\unifinPdom} \metric\left(\objCm,\objLm\right) 
= \O\left(\bwo^{2/(\pwCm-1)}\right),\label{eq:unifRateCmvsLm}\\
\begin{aligned}
&\sup_{\unifinPdom} \metric\left(\objLm,\objLmOrc\right) \\
&\quad= \Op\left(\max\left\{(\ndpo\bwo^2)^{-1/[2(\pwLm-1)+\smconst]}, (\ndpo\bwo^2(-\log\bwo)\inv)^{-1/[2(\pwLm-1)]}\right\}\right).
\end{aligned} \label{eq:unifRateLmvsLmO}
\end{gather}
Furthermore, with $\bwo\sim\ndpo^{-(\pwCm-1)/(2\pwCm+4\pwLm-6+2\smconst)}$, it holds that 
\bal\label{eq:unifFregRateOverTm} 
\sup_{\unifinPdom} \metric\left(\objCm,\objLmOrc\right) = \Op\left(\ndpo^{-1/(\pwCm+2\pwLm-3+\smconst)}\right).\eal
\ethm

Theorem~\ref{thm:unifFregRateOverTm} is a novel and relevant result for local Fr\'echet regression in its own right; we expect it to be a useful and widely applicable tool for the study of metric space valued data. 
We note that although $\tdom$ is a closed interval, boundary effects do not pose a problem, similar to the situation for local polynomial regression with real-valued responses \citep[e.g.,][]{fan:96}. For the bias part in \eqref{eq:unifRateCmvsLm}, we obtain the same rate as for pointwise results. 
For the stochastic part, the proof is substantially more involved. 
When $\pwCm=\pwLm=2$ as in Examples~\ref{eg:wsp}--\ref{eg:cor}, 
the uniform convergence rate is found to  be arbitrarily close to $\Op(\ndpo^{-1/3})$. 

\section[Recovering Metric-space Valued Random Processes]{Recovering Metric-space Valued Random Processes\\ from Discrete Noisy Measurements}\label{sec:locFregRdm}

We consider a metric space valued random process $\objPrcs\colon \tdom\ra\msp$ that is  assumed to be $\metric$-continuous over $\tdom$. 
In practice, the process $\objPrcs$ is usually not fully observed; instead one observes noisy measurements at discrete time points. 
\rvone{Since a metric space in general is not a vector space and hence does not afford additive operations, it is not obvious how to express the deviation of noisy observations from the underlying process $\objPrcs$.} To address this issue,  we introduce a random perturbation map $\ptb\colon \msp\ra\msp$ such that 
\bgt\label{eq:ptbmean} \varAnyObj = \argmin_{\anyObj\in\msp} \expect\left[\metric^2(\ptb(\varAnyObj),\anyObj)\right],\quad\text{for all }\varAnyObj\in\msp.\egt
Consider a random pair $(\rtm, \obsobj)$ following a joint distribution on $\tdom\times\msp$, where $\rtm$ is the time of observation and $\obsobj$ is a noisy observation of the process $\objPrcs$ at a random time $\rtm$, given by
\bgt\label{eq:obsobj} \obsobj = \ptb(\objPrcs(\rtm)). \egt 
Then the conditional Fr\'echet mean of the observed object $\obsobj$ given the process $\objPrcs$ and time $\rtm$ is the process evaluated at that time, i.e., 
\bgt\label{eq:obsMean} \objPrcs[\rtm] = \argmin_{\anyObj\in\msp} \expect\left[\metric^2(\obsobj,\anyObj)\cdn \objPrcs, \rtm\right].\egt
Furthermore, we assume 
\ben[label = (P\arabic*), series = prcs]
\item \label{ass:indep} 
The time of observation $\rtm$ and the random perturbation map $\ptb$ are independent of the random process $\objPrcs$. 
\een
The analogue of assumption~\ref{ass:indep} in Euclidean regression is the standard assumption of independence between additive noise and underlying process. 

Suppose that available noisy observations of the process $\objPrcs$ are $\{(\tmj,\obsobjj)\}_{\dptidx = 1}^{\ndpo}$, where $\obsobjj = \ptbj(\objPrcs(\tmj))$, and $\{(\tmj,\ptbj)\}_{\dptidx = 1}^{\ndpo}$ are independent realizations of $(\rtm,\ptb)$. Hence,  $\{(\tmj,\obsobjj)\}_{\dptidx = 1}^{\ndpo}$ are conditionally independent realizations of $(\rtm,\obsobj)$ given the process $\objPrcs$. 
Local Fr\'echet regression can be utilized to estimate the  process trajectories $\objPrcs$ via  \eqref{eq:objLmOrc}, with trajectory estimates 
\bgt\label{eq:objPrcsEst}
\objPrcsEst[\tm] = \argmin_{\anyObj\in\msp} \frac{1}{\ndpo} \summo \wLocREst{\tmj}{\tm}{\bwo} \metric^2(\obsobjj,\anyObj),\forallt.\egt 
Here, $\wLocREst{\vartm}{\tm}{\bwo} = \kerh[\bwo](\vartm - \tm)[\kmomREst[2][\ndpo](\tm) - \kmomREst[1][\ndpo](\tm)(\vartm - \tm)]/\kvarREst[\ndpo](\tm)$, $\bwo = \bwo(\ndpo)>0$ is a bandwidth sequence, $\kmomREst[\momidx][\ndpo](\tm) = \ndpo\inv\summo\kerh[\bwo](\tmj - \tm)(\tmj - \tm)^{\momidx}$, $\momidx = 0,1,2$, and $\kvarREst[\ndpo](\tm) = \kmomREst[0][\ndpo](\tm)\kmomREst[2][\ndpo](\tm) - \kmomREst[1][\ndpo](\tm)^2$, $\kerh[\bwo](\cdot) = \ker(\cdot/\bwo)/\bwo$; $\ker$ is a kernel function.

We note that \rvone{while $\objLmOrcNull$ in \eqref{eq:objLmOrc} is a local Fr\'echet regression estimate of the fixed trajectory $\objCmNull$ as per \eqref{eq:objCm}, the target of the local Fr\'echet regression implemented as per \eqref{eq:objPrcsEst} is the random process $\objPrcs$.} 
We next extend the results in Section~\ref{sec:locFregFix} for local Fr\'echet regression with fixed targets to the case of such random targets, and obtain the uniform convergence rates for $\objPrcsEst$ over $\tdom$. 

Let $(\prsp,\sfield,\prob)$ be the probability space on which the observed data $(\tmj,\obsobjj)$ are defined, where $\prsp$ is the sample space, $\sfield$ is the $\sigma$-algebra of events, and $\prob\colon \sfield\ra[0,1]$ is the probability measure. 
As the random mechanisms  that generate the data as per~\ref{ass:indep} are independent, the probability space $(\prsp,\sfield,\prob)$ is a product space of two probability spaces, $(\prspPrcs,\sfieldPrcs,\probPrcs)$, where the metric space valued process $\objPrcs$ is defined, 
and $(\prspNoise,\sfieldNoise,\probNoise)$, where the observed times $\tmj$ and the random perturbation map $\ptbj$ associated with the noisy observations $\obsobjj$ are defined. 
Fixing an element $\detmWobj\in\detmSp$ corresponds to a realization of the metric space valued process $\objPrcs$. 
Given a fixed $\detmWobj\in\detmSp$, the observed pairs $\{(\tmj,\obsobjj)\}_{\dptidx=1}^{\ndpo}$ are independent in $(\prspNoise,\sfieldNoise,\probNoise)$ and $\rtm$ and $\tmj$ do not depend on $\detmWobj$. 
We  use $\objPrcso$, $\obsobjo$, $\obsobjoj$, $\tmo$, and $\tmoj$ to represent the corresponding quantities given $\detmWobj\in\detmSp$ in what follows, and also 
$\expectNoise$ for the expectation (integral) with respect to $\probNoise$.
For any fixed $\detmWobj\in\detmSp$,  $\{(\tmoj,\obsobjoj)\}_{\dptidx=1}^{\ndpo}$ are i.i.d. realizations of $(\tmo,\obsobjo)$. For any $\tm\in\tdom$, as per \eqref{eq:obsMean}, 
\bgt\label{eq:objCmR}
\objCmR = \argmin_{\anyObj\in\msp} \objfCmR(\anyObj,\tm),\, \objfCmR(\anyObj,\tm) = \expectNoise[\metric^2(\obsobjo, \anyObj) \cdn \tmo = \tm]. \egt 

The localized Fr\'echet mean \citep{pete:19} is
\bgt \label{eq:objLmR}
\objLmR = \argmin_{\anyObj\in\msp}\objfLmR(\anyObj,\tm),\,  \objfLmR(\anyObj,\tm) = \expectNoise[\wLocRDfo{\tmo} \metric^2(\obsobjo,\anyObj)]. \egt
Here, $\wLocRDfo{\vartm} =  \kerho(\vartm-\tm)[\kmomR[2](\tm) - \kmomR[1](\tm)(\vartm - \tm)]/\kvarR(\tm)$, where $\kmomR(\tm) = \expectNoise[\kerho(\tmo - \tm)(\tmo - \tm)^{\momidx}]$, for $\momidx=0,1,2$, and $\kvarR(\tm) = \kmomR[0](\tm)\kmomR[2](\tm) - \kmomR[1](\tm)^2$. 
The local Fr\'echet regression estimates $\objPrcsEst[\tm]$ in \eqref{eq:objPrcsEst} can be expressed as 
\bgt\label{eq:objLmOrcR}
\objLmOrcR = \argmin_{\anyObj\in\msp} \objfLmOrcR(\anyObj,\tm),\, \objfLmOrcR(\anyObj,\tm) = \ndpo\inv \summo \wLocREstDfo{\tmoj} \metric^2(\obsobjoj,\anyObj). \egt 

Let $\tdom\interior = (0,\maxtm)$ be the interior of the time domain $\tdom$. 
Considering an arbitrarily fixed $\detmWobj\in\detmSp$, for local Fr\'echet regression as described in \eqref{eq:objCmR}--\eqref{eq:objLmOrcR}, assumptions \ref{ass:jointdtn}--\ref{ass:curvatureUnif} can be adapted to obtain uniform rates of convergence of local Fr\'echet regression estimates $\objLmOrcR$ over $\unifinTm$. 
To obtain uniform rates of convergence of the local Fr\'echet regression estimate $\objPrcsEst$ of the random process $\objPrcs$ over $\tdom$, we need to deal with different $\detmWobj\in\detmSp$ simultaneously, for which we require the following stronger variants of assumptions \ref{ass:jointdtn}--\ref{ass:curvatureUnif}. 

\ben[label = (U\arabic*), series = fregStrg, start = 0]
\item \label{ass:jointdtnStrg} 
The marginal density $\denRtm$ of $\rtm$ and the conditional densities $\cdenRtm(\cdot,\anyObj)$ of $\rtm$ given $\obsobjo = \anyObj$ exist and are continuous on $\tdom$ and twice continuously differentiable on $\tdom\interior$, 
the latter for all $\anyObj\in\msp$ and $\detmWobj\in\detmSp$. 
The marginal density $\denRtm$ is bounded away from zero on $\tdom$, $\inf_{\tm\in\tdom}\denRtm(\tm)>0$. 
The second-order derivative $\denRtm''$ is bounded,  $\sup_{\tm\in\tdom\interior}|\denRtm''(\tm)|<\infty$. 
The second-order partial derivatives $(\partial^2\cdenRtm/\partial\tm^2)(\cdot,\anyObj)$ are uniformly bounded, $\sup_{\unifinWobj,\, \tm\in\tdom\interior,\, \anyObj\in\msp} |(\partial^2\cdenRtm/\partial\tm^2)(\tm,\anyObj)| < \infty$. 
\note{The boundedness and uniform boundedness of $\denRtm''$ and $(\partial^2\cdenRtm/\partial\tm^2)(\cdot,\anyObj)$ are needed to show: (1)~the uniform order of the remainder of $\expectNoise[\kerho(\tmo - \tm)(\tmo - \tm)^{\momidx}]$ and $\expectNoise[\kerho(\tmo - \tm) (\tmo - \tm)^{\momidx} \cdn \obsobjo = \anyObj]$; (2)~the uniform equicontinuity of $\objfLmR(\anyObj,\cdot)$; and  (3)~$\sup_{\tm\in\tdom} \sup_{\metric(\anyObj,\objCmR)<\radius} |(\objfLmR-\objfCmR)(\anyObj,\tm)- (\objfLmR-\objfCmR)(\objCmR,\tm)|=\O(\bwo^2\radius)$.\\
}Additionally, for any open set $\open\subset\msp$, $\probNoise(\obsobjo\in\open\cdn \rtm = \tm)$ is continuous as a function of $\tm$ for all $\detmWobj\in\detmSp$\note{{ }(this is needed in the proof of Theorem~3 of \citet{pete:19})}. 
\item \label{ass:minUnifStrg} 
For all $\unifinWobj$ and $\unifinTm$, the minimizers $\objCmR$, $\objLmR$ and $\objLmOrcR$ exist and are unique, the last $\probNoise$-almost surely. Additionally,  for any $\epsilon>0$, 
\bgt\nn
\inf_{\unifinWobj,\, \unifinTm} 
\inf_{\substack{\anyObj\in\msp\text{ s.t. }\\ \metric(\objCmR,\anyObj) > \epsilon}} \bprts{\objfCmR(\anyObj,\tm) - \objfCmR(\objCmR,\tm)} > 0,\\
\liminf_{\bwo\ra 0} \inf_{\unifinWobj,\, \unifinTm} 
\inf_{\substack{\anyObj\in\msp\text{ s.t. }\\ \metric(\objLmR,\anyObj) > \epsilon}} \bprts{\objfLmR(\anyObj,\tm) - \objfLmR(\objLmR,\tm)} > 0. \egt
\note{The following part is removed since we strengthen \ref{ass:entropyUnif} and \ref{ass:curvatureUnif} as follows so that the first term on the right hand side of \eqref{eq:tailprobUnif} is gone, whence we do not need to show the uniform consistency, where the following part is needed.\\ 
	and for all $\detmWobj\in\detmSp$, there exists $c = c(\detmWobj,\epsilon)>0$ such that \bgt\nn \probNoise\left(\inf_{\unifinTm} \inf_{\metric(\objLmOrcR,\anyObj) > \epsilon} \bprts{\objfLmOrcR(\anyObj,\tm) - \objfLmOrcR(\objLmOrcR,\tm)} \ge c\right) \ra 1. \egt}
\item \label{ass:entropyUnifStrg} 
Let $\ball{\objCmR}{\radEntropy}\subset \msp$ be a ball of radius $\radEntropy$ centered at $\objCmR$ and $\covernum(\epsilon, \ball{\objCmR}{\radEntropy}, \metric)$ be its covering number using balls of radius $\epsilon$. Then 
\bal\nn \sup_{\radEntropy>0} \sup_{\unifinWobj} \int_0^1 \sup_{\unifinTm} 
\sqrt{1+\log\covernum(\radEntropy\epsilon, \ball{\objCmR}{\radEntropy}, \metric)} \diffop\epsilon  < \infty.\eal
\item \label{ass:curvatureUnifStrg} 
There exist $\constCm,\constLm>0$, and $\pwCm,\pwLm>1$ such that for any $\radCm,\radLm>0$, 
\bgt\nn
\inf_{\unifinWobj,\, \unifinTm} 
\inf_{\substack{\anyObj\in\msp\text{ s.t. }\\ \metric(\anyObj,\objCmR)<\radCm}} \left[\objfCmR(\anyObj,\tm) - \objfCmR(\objCmR,\tm)- \constCm \metric\left(\anyObj,\objCmR\right)^{\pwCm}\right] \ge 0, \\
\liminf_{\bwo\ra 0} \inf_{\substack{\unifinWobj,\, \unifinTm}} \inf_{\substack{\anyObj\in\msp\text{ s.t. }\\ \metric(\anyObj,\objLmR)<\radLm}} 
\left[\objfLmR(\anyObj,\tm) - \objfLmR(\objLmR,\tm) -\constLm \metric(\anyObj,\objLmR)^{\pwLm}\right] \ge 0. \egt
\een

We note that the assumption of equicontinuity of $\objfCm$ as per~\eqref{eq:eqcont_objCm} to guarantee the $\metric$-continuity of $\objCmNull$ is not needed in this case, since the process $\objPrcs$ is assumed to be $\metric$-continuous. 
We then obtain the uniform convergence rates for $\objPrcsEst$ over $\tdom$ as follows. 
\bthm\label{thm:unifFregRateRdmTgt}
Under \ref{ass:indep}, 
\ref{ass:ker}, and \ref{ass:jointdtnStrg}--\ref{ass:curvatureUnifStrg}, for any $\smconst>0$, 
\bgt\label{eq:objPrcsBiasVarRate}
\sup_{\unifinTm} \metric\left(\objPrcs(\tm), \objPrcsLm(\tm)\right) = \O\left(\bwo^{2/(\pwCm-1)}\right);\\
\begin{aligned}
	&\sup_{\unifinTm} \metric\left(\objPrcsLm(\tm), \objPrcsEst(\tm)\right)\\ 
	&\quad= \Op\left(\max\left\{(\ndpo\bwo^2)^{-1/[2(\pwLm-1)+\smconst]}, (\ndpo\bwo^2(-\log\bwo)\inv)^{-1/[2(\pwLm-1)]}\right\}\right).\end{aligned}
\egt
Furthermore, if $\bwo\sim\ndpo^{-(\pwCm-1)/(2\pwCm+4\pwLm-6+2\smconst)}$, 
\bgt\label{eq:objPrcsEstRate}
\sup_{\tm\in\tdom} \metric\left(\objPrcs[\tm], \objPrcsEst[\tm]\right) =  \Op\left(\ndpo^{-1/(\pwCm + 2\pwLm-3+\smconst)}\right).\egt
\ethm

We note that Examples~\ref{eg:wsp} and \ref{eg:cor} indeed  satisfy \ref{ass:minUnifStrg}--\ref{ass:curvatureUnifStrg} with $\pwCm=\pwLm=2$, where the uniform convergence rate in \eqref{eq:objPrcsEstRate} can be arbitrarily close to $\Op(\ndpo^{-1/3})$. 
\rvone{We also note that the uniform convergence results for local Fr\'echet regression with fixed and random targets in Theorems~\ref{thm:unifFregRateOverTm} and \ref{thm:unifFregRateRdmTgt}, respectively, can be extended to the case of multivariate predictors at the expense of more tedious algebra similarly to multivariate nonparametric regression with scalar responses \citep{rupp:94}}. 

\section{Estimation of the Location of Extrema}\label{sec:extremum}

In this section, we consider the problem of estimating the locations of extrema (maxima and/or minima) of the conditional Fr\'echet mean trajectory $\objCm$ as per \eqref{eq:objCm} with regard to some functional of interest. Without loss of generality, we focus on the case of extrema that are minima. 
Consider a functional $\smap\colon \msp\ra\real$ that quantifies a property of interest of the objects situated in metric space $\msp$, whence the corresponding property of the conditional Fr\'echet mean $\objCm$ of $\rpso$ given $\pdto=\apdt$ as per \eqref{eq:objCm} is 
\bgt\nn\smrCm(\apdt) = \smap(\objCm),\quad\text{for all }\unifinPdom.\egt
Our goal is to find the location where   $\smrCm(\cdot)$ is minimized, 
\bgt\label{eq:pmCm} \pmCm = \argmin_{\unifinPdom} \smrCm(\apdt).\egt 
An estimate of the minimizer $\pmCm$ of $\smrCm(\cdot)$ is given by replacing $\objCm$ with its local Fr\'echet regression estimate, i.e., 
\bgt\label{eq:pmEst} \pmEst = \argmin_{\unifinPdom} \smrEst(\apdt),\quad \text{with }\smrEst(\apdt) = \smap(\objLmOrc).\egt
In addition, we assume 
\ben[label= (D\arabic*), series = extremum]
\item\label{ass:smapLips} 
There exists $\constSmap>0$ and $\pwSmap>1$ such that for all $\arps_1,\arps_2\in\msp$, $|\smap(\arps_1) - \smap(\arps_2)| \le \constSmap \metric(\arps_1,\arps_2)^{\pwSmap}$.
\item\label{ass:pmUniq} 
The minimizer $\pmCm$ exists and is unique. Additionally, for any $\epsilon>0$, $\inf_{|\apdt-\pmCm| > \epsilon} [\smrCm(\apdt) - \smrCm(\pmCm)] > 0$. 
\item\label{ass:smrCurv} There exists $\radSmr,\constSmr>0$ and $\pwSmr>1$ such that $\inf_{|\apdt-\pmCm| <\radSmr} [\smrCm(\apdt) - \smrCm(\pmCm) - \constSmr |\apdt-\pmCm|^{\pwSmr}] \ge 0$.
\een
Assumptions \ref{ass:smapLips} and \ref{ass:pmUniq} 
guarantee the consistency of the minimizer estimate $\pmEst$, and hence can be used to obtain the corresponding convergence rate in conjunction with \ref{ass:smrCurv}. 
An example scenario where \ref{ass:smapLips} holds with $\pwSmap=1$ will be given in Section~\ref{sec:adni}. 
For \ref{ass:pmUniq} and \ref{ass:smrCurv}, a sufficient condition is, for instance, that $\smrCm(\cdot)$ is twice continuously differentiable on $\pdom$ with unique minimizer $\pmCm$ and $\smrCm''(\pmCm) > 0$; specifically, $\pwSmr = 2$ in \ref{ass:smrCurv}. 

\rvone{Applying Theorem~\ref{thm:unifFregRateOverTm},} we obtain the following result of the minimum location  estimate $\pmEst$ based on local Fr\'echet regression.

\bcor\label{cor:extremum}
Under \ref{ass:ker},  \ref{ass:jointdtn}--\ref{ass:curvatureUnif}, and \ref{ass:smapLips}--\ref{ass:smrCurv}, 
for any $\smconst>0$ and for  $\bwo\sim \ndpo^{-(\pwCm-1)/(2\pwCm+4\pwLm-6+2\smconst)}$, it holds for the estimate $\pmEst$ in \eqref{eq:pmEst} of the minimizer $\pmCm$ in \eqref{eq:pmCm} that 
\bgt\label{eq:pmRate}
|\pmEst-\pmCm| = \Op\left(\ndpo^{-\pwSmap/[\pwSmr(\pwCm+2\pwLm-3+\smconst)]}\right).
\egt
\ecor
We will illustrate this approach with an application to the study of brain connectivity utilizing fMRI data  in Section~\ref{sec:adni}. 
\rvone{More generally, for an aggregation statistic determined by a functional $\smap^*$ such that $|\smap^*(\objLmOrcNull) - \smap^*(\objCmNull)|\le \constSmap^*\sup_{\unifinPdom} \metric(\objLmOrc,\objCm)^{\pwSmap^*}$, for some $\constSmap^*>0$ and $\pwSmap^*>1$, where $\smap^*(\objCmNull),\smap^*(\objLmOrcNull)\in\real$, analogous rates of convergence as in  \eqref{eq:pmRate} can be obtained for 
	$|\smap^*(\objLmOrcNull) - \smap^*(\objCmNull)|$.  Examples where such results are useful include the estimation of zero crossings  or more general level crossings and the estimation of intervals where $\smap^*(\objCmNull)$ exceeds a certain level.} 

\section{Time Warping for Metric-space Valued Functional Data}\label{sec:wp}

\subsection{Global Warping}\label{sec:wpGlo}

We consider the time warping problem for metric space valued random trajectories. 
With $\tdom=[0,\maxtm]$ being the time domain, consider a set of warping functions $\wfsp = \{\wpcand\colon \tdom\ra\tdom\mid \wpcand(0)=0,\, \wpcand(\maxtm)=\maxtm$, $\wpcand$ is continuous and strictly increasing on  $\tdom\}$. Note that for each function $\wpcand\in\wfsp$, $\wpcand(\cdot)/\maxtm$ is a strictly increasing cdf on $\tdom$. 
Suppose $\mObjNull\colon \tdom\ra\msp$ is a fixed metric space valued trajectory, and $\wp\in\wfsp$ is a random (global) warping function such that $\expect [\wp(\tm)] = \tm$, for all $\tm\in\tdom$. 
We consider the following model for the metric space valued random process $\wobj\colon \tdom\ra\msp$ in Section~\ref{sec:locFregRdm}, 
\bgt\label{eq:gloPop} 
\wobj(\tm) = \mObj[\wp\inv(\tm)],\forallt.\egt
where $\mObjNull$ is referred to as the mean trajectory, and the stochastic fluctuations of the random warping function $\wp$ around the identity function $\id$ determines the phase variation of the process $\wobj$. 
Considering a random pair $(\rtm, \obsobj)$ consisting of time of observation $T$  and process $Z$ which is observed with a perturbation that is determined by the map $\ptb$ satisfying \eqref{eq:ptbmean},  
suppose  $\{(\wpi,\wobji,\tmi,\ptbi,\obsobji)\}_{\objidx=1}^{\nobj}$ is a set of $\nobj$ independent realizations of the quintuple $(\wp,\wobj,\rtm,\ptb,\obsobj)$, 
where as per \eqref{eq:gloPop}, the metric space valued processes $\wobji$ are 
\bgt\label{eq:gloSam} 
\wobji(\tm) = \mObj[\wpi\inv(\tm)], 
\forallt,\egt
and the observed objects are  $\obsobji = \ptbi(\wobji(\tmi))$,  as per \eqref{eq:obsobj}. 

Furthermore, we make the following assumptions regarding the fixed mean trajectory $\mObjNull$ and random warping function $\wp\in\wfsp$.
\ben[label=(W\arabic*), resume = warp]
\item \label{ass:contCm}
The trajectory $\mObjNull$ is $\metric$-continuous, i.e., $\lim_{\Delta\ra 0} \metric(\mObj[\tm+\Delta],\mObj) = 0$, for any $\tm\in\tdom$. 
\item\label{ass:flat}
Defining a bivariate function $\metric[\mObjNull]\colon \tdom^2\ra\real$ as $\metric[\mObjNull](\vartm,\tm) = \metric(\mObj[\vartm],\mObj[\tm])$, $\metric[\mObjNull]$ is twice continuously differentiable with $\inf_{\vartm=\tm\in\tdom} |(\partial\metric[\mObjNull]/\partial\vartm)(\vartm,\tm)| >0$. 
For any $\tm_1,\tm_2\in\tdom$ with $\tm_1<\tm_2$, 
$\int_{\tm_1}^{\tm_2}[(\partial\metric[\mObjNull]/\partial\vartm) (\vartm,\tm)]^2\diffop\vartm >0$, for all $\tm\in\tdom$.
\item\label{ass:wpSlope}
The difference quotients of the global warping function $\wp$ are bounded from above and below, i.e., there exist constants $c,C\in(0,+\infty)$ with $c<C$ and $cC\le 1$ such that $c\le [\wp(\vartm) - \wp(\tm)] / (\vartm - \tm) \le C$, for all $\vartm,\tm\in\tdom$ with $\vartm<\tm$. \note{$cC\le 1$ ensures $c\le C\inv \le [\wp\inv(\vartm) - \wp\inv(\tm)]/(\vartm-\tm)\le c\inv\le C$.}
\een
Assumption \ref{ass:contCm} implies the $\metric$-continuity of the random process $\wobj$ in conjunction with the continuity of the warping function $\wp$;
\ref{ass:flat} excludes the possibility that any part of the trajectory $\mObjNull$ could be flat. This is necessary to ensure the uniqueness of the warping functions, 
and will be used to establish the uniform convergence of the proposed estimates for the pairwise warping functions; see  Section~\ref{sec:wpTheory}\note{{ }(the discrepancy between the true and estimated warping functions is controlled by that between the corresponding objective functions)}. Assumption 
\ref{ass:wpSlope} guarantees there are no plateaus or steep increases in the global warping function and its inverse. 

\subsection{Pairwise Warping}\label{sec:pairwp}
For any $\objidx,\varobjidx\in\{1,\dots,\nobj\}$ such that $\objidx\ne\varobjidx$, the random pairwise warping function $\pairwp\colon \tdom\ra\tdom$ is a temporal transformation from  $\wobjsmpl{\varobjidx}$ towards $\wobji$ defined by 
\bgt\nn
\pairwp(\tm) = \wpi[\varobjidx](\wpi\inv(\tm)),\forallt.\egt
We note that $\pairwp\in\wfsp$. 
Moreover, we assume that (warping) functions in $\wfsp$ can be parameterized by linear splines \citep[as per][]{tang:08}. 
Let $\tgridpt = \tgrididx \maxtm/(\ntgrid+1)$, for $\tgrididx=1,\dots,\ntgrid$, be $\ntgrid$ equidistant knots in $\tdom$, with $\tgridpt[0]=0$, and $\tgridpt[\ntgrid+1] = \maxtm$. 
For any function $\paircand\in\wfsp$, defining a coefficient vector $\lsvec = [\paircand(\tgridpt[1]), \dots, \paircand(\tgridpt[\ntgrid+1])]^\top$, the piecewise linear formulation of $\paircand$ can be expressed as
\bgt\label{eq:lspline}
\paircand(\tm) = \lsvec^\top \lsbasis(\tm),\forallt, \egt
where $\lsbasis(\tm) = [\lsbasis_1(\tm),\dots,\lsbasis_{\ntgrid+1}(\tm)]^\top$,
$\lsbasis_{\tgrididx}(\tm) = \lsbasis_{\tgrididx}^{(1)}(\tm) - \lsbasis_{\tgrididx+1}^{(2)}(\tm)$,
$\lsbasis_{\tgrididx}^{(1)}(\tm) = (\tm - \tgridpt[\tgrididx-1]) / (\tgridpt - \tgridpt[\tgrididx-1]) \cdot \mbf{1}_{[\tgridpt[\tgrididx-1],\tgridpt)}$, 
$\lsbasis_{\tgrididx}^{(2)}(\tm) = (\tm - \tgridpt) / (\tgridpt - \tgridpt[\tgrididx-1]) \cdot \mbf{1}_{[\tgridpt[\tgrididx-1], \tgridpt)}$, for $\tgrididx=1,\dots,\ntgrid+1$, and $\lsbasis_{\ntgrid+2}^{(2)} = 0$. 
Due to the definition of the warping function space $\wfsp$, the parameter space $\lsSp$ of the splines coefficient vector $\lsvec$ is
\bgt\label{eq:lsSp} \lsSp = \{\anyVec \in \real^{\ntgrid+1}: 0<\anyVec_1<\dotsb<\anyVec_{\ntgrid+1}=\maxtm\}. \egt

The corresponding family of warping functions is $\wfsp = \wfspLS = \{\paircand\in\wfsp: \paircand =\lsvec^\top \lsbasis \text{ with } \lsvec\in\lsSp\}$. 
We assume that the pairwise warping function $\pairwp$ can be represented by \eqref{eq:lspline}, i.e., 
\bgt\label{eq:lspairwp} 
\pairwp(\cdot) = \lsvecPw^\top \lsbasis(\cdot),\quad \text{with }\lsvecPw\in\lsSp.\egt

\subsection{Samples and Estimation}\label{sec:wpEst}
For each $\objidx=1,\dots,\nobj$, suppose available observations for the process $\wobji$ are $\{(\tmij,\obsobjij)\}_{\dptidx = 1}^{\ndpi}$, 
where $\obsobjij = \ptbij(\wobji(\tmij))$, and $\{(\tmij,\ptbij)\}_{\dptidx = 1}^{\ndpi}$ are $\ndpi$ independent realizations of $(\tmi,\ptbi)$. 
To estimate the warping functions $\wpi$, a first step is to estimate the processes $\wobji$ by local Fr\'echet regression. 
Specifically, as per \eqref{eq:objPrcsEst}, the estimated trajectories  are 
\bgt\label{eq:wobjEst}
\wobjiEst[\tm] = \argmin_{\anyObj\in\msp} \frac{1}{\ndpi} \summi \wLocREst{\tmij}{\tm}{\bwi} \metric^2(\obsobjij,\anyObj),\forallt,\egt  
where $\wLocREstNull$ is as defined after \eqref{eq:objPrcsEst} and $\bwi = \bwi(\ndpi)>0$ are bandwidth sequences. 

Our next step is to obtain an estimator for the pairwise warping functions $\pairwp$ as per \eqref{eq:lspairwp}, for any distinct $\varobjidx,\objidx\in\{1,\dots,\nobj\}$. This is equivalent to estimating the corresponding spline coefficients $\lsvecPw\in\lsSp$, 
which can be obtained by minimizing the integral of the squared distance between $\varwobjiEst$ with time shifted toward $\wobjiEst$ 
over the time domain $\tdom$, with a regularization penalty on the magnitude of warping. 
Specifically, an estimator for $\lsvecPw$ is 
\bgt\label{eq:obfnYEst} 
\lsvecEst = \argmin_{\anyVec\in\lsSp} \obfnYest(\anyVec;\varwobjiEst,\wobjiEst),\\
\text{with} \quad 
\obfnYest(\anyVec;\varwobjiEst,\wobjiEst)
=  \int_\tdom \left[\metric^2\left(\varwobjiEst[\anyVec^\top \lsbasis(\tm)], \wobjiEst[\tm]\right) + \pnty\left(\anyVec^\top \lsbasis(\tm) -\tm\right)^2 \right] \diffop\tm, \egt
whence we obtain an estimator $\pairwpEst$ of the pairwise warping functions 
\bgt\label{eq:pairwpEst}
\pairwpEst(\tm) = \lsvecEst^\top \lsbasis(\tm), \forallt. \egt
By the assumption $\expect[\wp(\tm)] = \tm$, we have $\expect[\pairwp(\tm)\cdn \wpi] = \expect[\wpi[\varobjidx](\wpi\inv(\tm)) \cdn \wpi] = \wpi\inv(\tm)$, for all $\tm\in\tdom$, which justifies estimating  the inverse global warping functions $\wpi\inv$ by
\bgt\label{eq:wpEst} \wpiEst\inv(\tm) = \nobj\inv \sum_{\varobjidx=1}^{\nobj} \pairwpEst(\tm),\forallt.\egt
Hence, estimators $\wpiEst$ for the global warping functions $\wpi$ can be obtained by inversion, with  estimated aligned trajectories  given by $\wobjiEst(\wpiEst(\tm))$, for $\tm\in\tdom$. 


\subsection{Asymptotic Results for Time Warping}\label{sec:wpTheory}

In order to obtain the convergence rate for the proposed estimates $\wpiEst$ for the warping functions as per \eqref{eq:wpEst} based on discrete and noisy observations $\{(\tmij,\obsobjij)\}_{\dptidx = 1}^{\ndpi}$, an initial step is to derive bounds for the difference between the actual metric space valued processes $\wobji$ and their estimates $\wobjiEst$ as per \eqref{eq:wobjEst}, obtained by local Fr\'echet regression. 
Specifically, a uniform rate of convergence over the time domain $\tdom$, beyond the pointwise results shown by \citet{pete:19}, is needed. 
Furthermore, the targets of the local Fr\'echet regression implemented here are random processes $\wobji$ rather than fixed trajectories as per~\eqref{eq:objCm}. 
Thus, Theorem~\ref{thm:unifFregRateRdmTgt}, where the targets are random processes, needs to be invoked. 
Subsequently, we derive the rate of convergence for the estimates for warping functions and time synchronized processes. 

For any distinct $\varobjidx,\objidx=1,\dots,\nobj$, define functions $\obfnM(\cdot; \wpi[\varobjidx],\wpi)\colon \real^{\ntgrid+1}\ra \real$, 
\bgt\label{eq:obfnM}
\obfnM(\anyVec; \wpi[\varobjidx],\wpi) =  \int_\tdom \metric^2\bprts{\mObj[{\wpi[\varobjidx]\inv[\anyVec^\top \lsbasis(\tm)]}], \mObj[\wpi\inv(\tm)]}\diffop\tm , \quad \anyVec\in\real^{\ntgrid+1}.\egt 
We  show in Lemma~\ref{lem:uniq} in \pfsec that for any distinct $\objidx,\varobjidx=1,\dots,\nobj$, the coefficient vector $\lsvecPw$ corresponding to the pairwise warping functions $\pairwp$ is the unique minimizer of $\obfnM(\anyVec; \wpi[\varobjidx],\wpi)$ under certain constraints. 

In order to deal with the estimation of $\nobj$ trajectories simultaneously, we make the following assumption on the bandwidths $\bwi$ and numbers of discrete observations per trajectory $\ndpi$. 
\ben[label = (W\arabic*), resume = warp]
\item \label{ass:nObsPerObj}
There exist sequences $\ndp = \ndp(\nobj)$ and $\bwAll = \bwAll(\nobj)$ such that (1)~$\infi \ndpi \ge \ndp$; (2)~$0<C_1<\infi\bwi/\bwAll \le \supi\bwi/\bwAll < C_2 <\infty$, for some constants $C_1$ and $C_2$; and (3)~$\ndp\gify$, $\bwAll\ra 0$, and $\ndp\bwAll^2(-\log\bwAll)\inv\gify$, as $\nobj\ra\infty$.
\een

We then derive an asymptotic bound for the discrepancy between the two objective functions $\obfnM$ and  $\obfnYest$, whence we obtain the convergence rates for the estimates of the coefficient vector $\lsvecPw$ and the corresponding pairwise warping function in conjunction with Theorem~\ref{thm:unifFregRateRdmTgt} as follows.

\bthm\label{thm:pai}
Under \ref{ass:indep}, \ref{ass:contCm}--\ref{ass:nObsPerObj}, \ref{ass:ker}, and \ref{ass:jointdtnStrg}--\ref{ass:curvatureUnifStrg}, 
for any $\smconst>0$, if \newline $\bwi\sim 
\ndpi^{-(\pwCm-1)/(2\pwCm+4\pwLm-6+2\smconst)}$ for all $\objidx=1,\dots,\nobj$, and if $\pnty\ra0$, as $\nobj\gify$, 
then, for any distinct $\varobjidx$ and $\objidx$, 
it holds for the constrained minimizer $\lsvecEst$ in \eqref{eq:obfnYEst} that
\bgt\label{eq:pairwpVecRate} 
\| \lsvecEst - \lsvecPw \|
= \O\left(\pnty^{1/\pwM}\right) + \Op\left(\ndp^{-1/[\pwM(\pwCm+2\pwLm-3+\smconst)]}\right), \egt
where $\ndp$ is defined in \ref{ass:nObsPerObj}. Furthermore,  for the corresponding estimate of the pairwise warping function $\pairwpEst$ in \eqref{eq:pairwpEst} it holds that
\bgt\label{eq:pairwpRate} 
\sup_{\tm\in\tdom} \left|\pairwpEst(\tm) - \pairwp(\tm)\right|
= \O\left(\pnty^{1/\pwM}\right) + \Op\left(\ndp^{-1/[\pwM(\pwCm+2\pwLm-3+\smconst)]}\right). \egt
\ethm

We next obtain  asymptotic results for local Fr\'echet regression estimates $\wobjiEst$ across trajectories $\objidx=1,\dots,\nobj$ in Corollary~\ref{cor:unifFregRateAveSubj}, which is used in conjunction with Theorem~\ref{thm:pai} to obtain the convergence rates for the estimates of  the warping functions $\wpiEst$, and the aligned trajectories $\wobjiEst(\wpiEst(\cdot))$ in Corollary~\ref{cor:warp}. 

\bcor\label{cor:unifFregRateAveSubj}
Under \ref{ass:indep}, \ref{ass:contCm}, \ref{ass:nObsPerObj}, \ref{ass:ker}, and \ref{ass:jointdtnStrg}--\ref{ass:curvatureUnifStrg}, 
for any $\smconst>0$, and $\smconst'\in(0,1)$, if $\bwi \sim \ndpi^{-(\pwCm-1)(1-\smconst') / (2\pwCm+4\pwLm-6+2\smconst)}$, and $\limsup_{\nobj\gify} \nobj \ndp^{-\smconst'(\pwLm-1)/[2(\pwLm-1+\smconst/2)]} < \infty$,  it holds that 
\begin{gather}
\sup_{\tm\in\tdom} \metric\left(\wobji[\tm], \wobjiEst[\tm]\right) = \Op\left(\ndp^{-(1-\smconst') / (\pwCm+2\pwLm-3+\smconst)}\right),\ \text{for all }\objidx = 1,\dots,\nobj;\label{eq:objPrcsiEstRateForWarp} \\
\sup_{\tm\in\tdom} \nobj\inv\sumn\metric\left(\wobji[\tm], \wobjiEst[\tm]\right)^{\pwc} = \Op\left(\ndp^{-\pwc(1-\smconst')/ (\pwCm+2\pwLm-3+\smconst)}\right), \label{eq:objPrcsiEstSumRate} 
\end{gather}
for any given $\pwc\in (0,1]$. 
\ecor

\bcor\label{cor:warp}
Under \ref{ass:indep}, \ref{ass:contCm}--\ref{ass:nObsPerObj}, \ref{ass:ker}, and \ref{ass:jointdtnStrg}--\ref{ass:curvatureUnifStrg}, 
for any $\smconst>0$ and $\smconst'\in(0,1)$, if $\bwi \sim \ndpi^{-(\pwCm-1)(1-\smconst') / (2\pwCm+4\pwLm-6+2\smconst)}$ for all $\objidx=1,\dots,\nobj$, $\limsup_{\nobj\gify} \nobj \ndp^{-\smconst'(\pwLm-1)/[2(\pwLm-1+\smconst/2)]} < \infty$, and if $\pnty\ra0$, as $\nobj\gify$, 
it holds for the estimated warping functions $\wpiEst$ that \note{(this result is not uniform over $\objidx$)}
\bgt\label{eq:wpRate}
\sup_{\tm\in\tdom} \left|\wpiEst(\tm) - \wpi(\tm)\right|
= \O\left(\pnty^{1/\pwM}\right) + \Op\left(\ndp^{-(1-\smconst')/[\pwM(\pwCm+2\pwLm-3+\smconst)]}\right) + \Op\left(\nobj\mhf\right).\egt
Furthermore, if the mean trajectory $\mObjNull$ is Lipschitz $\metric$-continuous, i.e., there exists $\lipm>0$, such that $\metric(\mObj[\tm_1],\mObj[\tm_2]) \le \lipm |\tm_1-\tm_2|$, for all $\tm_1,\tm_2\in\tdom$, then it holds for the estimates of the aligned trajectories $\wobjiEst(\wpiEst(\cdot))$ that 
\bal\label{eq:aliRate}
&\sup_{\unifinTm} \metric\left(\wobjiEst(\wpiEst(\tm)), \wobji(\wpi(\tm))\right) \\
&\quad= \O\left(\pnty^{1/\pwM}\right) + \Op\left(\ndp^{-(1-\smconst')/[\pwM(\pwCm+2\pwLm-3+\smconst)]}\right) + \Op\left(\nobj\mhf\right). \eal
\ecor

Defining $\gamma = (1-\smconst')\inv-1$, $\smconst'\in(0,1)$ entails $\gamma>0$. To discuss some more specific rates, 
under the assumptions of Corollary~\ref{cor:warp}, the minimum number of observations per trajectory $\ndp$ should be bounded below by a multiple of $\nobj^{2[1-(\gamma+1)\inv]\inv [1+\smconst/(2(\pwLm-1))]}$, which implies that  the rates in the second terms on the right hand sides of \eqref{eq:wpRate}--\eqref{eq:aliRate} are bounded  above by a multiple of $\nobj^{-\gamma\inv [(\pwLm-1+\smconst/2) / (\pwLm-1)] (\pwCm+2\pwLm-3+\smconst)\inv}$, where the latter  can be arbitrarily close to  $\nobj^{-1/[\gamma(\pwCm+2\pwLm-3)]}$. 
Consider $\pnty=\O(\nobj\inv)$. 
Then, if $\gamma \in(0,2(\pwCm+2\pwLm-3)\inv]$, the estimates for the warping functions $\wpi$ and mean trajectory $\mObjNull$ as per \eqref{eq:wpRate}--\eqref{eq:aliRate} converge with a rate of  $\nobj\mhf$. 
Otherwise, if $\gamma > 2(\pwCm+2\pwLm-3)\inv$, the rates in \eqref{eq:wpRate}--\eqref{eq:aliRate} can be arbitrarily close to $\nobj^{-1/[\gamma(\pwCm+2\pwLm-3)]}$. 
Taking $\pwCm=\pwLm=2$ as in Examples~\ref{eg:wsp}--\ref{eg:cor}, 
the estimates $\wpiEst$ and $\wobjiEst(\wpiEst(\cdot))$ achieve root-$\nobj$ rate, when $\gamma = 2(\pwCm+2\pwLm-3)\inv = 2/3$ and $\ndp\gtrsim \nobj^{5(1+\smconst/2)}$. When $\gamma>2/3$ and $\ndp\gtrsim \nobj^{(2+\smconst)/(1+(\gamma+1)\inv)}$, the rate becomes approximately $\nobj^{-1/(3\gamma)}$. 

\section{Data Illustrations}\label{sec:app}

\subsection{Age of Minimum Connectivity in Brain Networks: fMRI Data}\label{sec:adni}

Much work has been done in recent years  to investigate how normal aging affects functional connectivity in human brains, which reflects  spatial integration of brain activity based on resting-state functional magnetic resonance imaging (rs-fMRI) \citep{ferr:13,denn:14,zonn:19}. 
Fluctuations in regional brain activity are recorded by blood oxygen-level dependent (BOLD) signals while subjects  relax. This leads to   voxel-specific time series of activation strength. 
Patterns of subject-specific functional connectivity are frequently analyzed invoking a  spatial parcellation of the brain into a set of predefined regions  \citep{bull:09}. 
Connectivity  between pairwise brain regions in the parcellation is then usually quantified  by what is referred to in the field as temporal Pearson correlation of the fMRI time series of the corresponding regions in neuroimaging. 
When considering $r$ distinct brain regions, the temporal Pearson correlations then yield correlation matrices in $\real^{r\times r}$, where each row and column represent one brain region, and one such matrix is obtained for each of \rvone{$\ndpo$} subjects, where in the ADNI data that we analyze for each subject one fMRI scan is available. 

To study the relationship between age and functional connectivity, it is then natural to apply local Fr\'echet regression for the case where the  random objects that form the responses are situated  in the space of correlation matrices and age is a scalar predictor. 
Based on the correlation matrices, networks of connectivity across regions are constructed by standard procedures in neuroimaging \citep{rubi:10}; see also \citet{phil:15} and \citet{pete:16:3}. 
The resulting networks can then be converted to graph Laplacians, for which the second smallest eigenvalue is known as the Fiedler value, also referred to as algebraic connectivity \citep{fied:73}. The Fiedler value is a measure of the global connectivity of a graph that indicates how well connected a network is \citep{deha:12,phil:15,cai:19}. 
Based on the results obtained from local Fr\'echet regression, we can then express the Fiedler value as a function of age of a subject  and identify the age at which the resting human brain attains the minimum level of connectivity. This  is of interest to understand the aging brain and as brain connectivity has been reported  to mostly decrease during aging while also increases have been reported \citep{ferr:13}.

We investigated the dependence of brain connectivity on age for elderly cognitively normal people using the resting-state fMRI data obtained from the Alzheimer's Disease Neuroimaging Initiative (ADNI) database (\url{adni.loni.usc.edu}). 
The data used in our analysis consist of fMRI scans from $\ndpo=402$ clinically normal elderly subjects at ages ranging from 55.6 to 95.4 years old, where one randomly  selected scan is taken for those subjects where multiple scans are available. 

Our analysis focused on the inter-regional connectivity of $\dcor=10$ hubs 
\citep[][see Table~3]{buck:09}. 
Specifically, we considered spherical seed regions of diameter 8 mm centered at the seed voxels of these hubs. 
Preprocessing of the BOLD signals was implemented by adopting the standard procedures of head motion correction, slice-timing correction, coregistration, normalization, and spatial smoothing. 
Subsequently, average signals of voxels within each seed region were extracted, where linear detrending and band-pass filtering are performed to account for signal drift and global cerebral spinal fluid and white matter signals, including only frequencies between 0.01 and 0.1~Hz, respectively. 
These steps were performed in MATLAB using the Statistical Parametric Mapping (SPM12, \url{www.fil.ion.ucl.ac.uk/spm}) and Resting-State fMRI Data Analysis Toolkit V1.8 (REST1.8, \url{http://restfmri.net/forum/?q=rest}). 

Let $\{\sgnl_{\dptidx\hubidx\tsidx}\}_{\tsidx=1}^{\nts}$ be the signal time series of seed region $\hubidx$ of subject $\dptidx$ excluding the first four time points, which were discarded to eliminate nonequilibrium effects of magnetization, for $\hubidx=1,\dots,\dcor$, and $\dptidx=1,\dots,\ndpo$. 
For subject $\dptidx$, the correlation matrix calculated for analyzing connectivity in fMRI is
\bgt\label{eq:corr}
\csm = (\csij)_{1\le\hubidx,\vhubidx\le\dcor},\quad 
\csij = \frac{\sum_{\tsidx=1}^{\nts} (\sgnl_{\dptidx\hubidx\tsidx} - \bar{\sgnl}_{\dptidx\hubidx}) (\sgnl_{\dptidx\vhubidx\tsidx} - \bar{\sgnl}_{\dptidx\vhubidx})}{\left[\sum_{\tsidx=1}^{\nts} (\sgnl_{\dptidx\hubidx\tsidx} - \bar{\sgnl}_{\dptidx\hubidx})^2 \sum_{\tsidx=1}^{\nts} (\sgnl_{\dptidx\vhubidx\tsidx} - \bar{\sgnl}_{\dptidx\vhubidx})^2\right]\half}. \egt
In the local Fr\'echet regression, we used  age-at-scan as predictor, and the correlation matrices $\csm$ as response, taken to be elements in the space of correlation matrices of dimension $\dcor$ equipped with Frobenius metric, $(\msp,\metric[F])$, as in Example~\ref{eg:cor}. 

For any correlation matrix $\corm\in\msp$, the Fiedler value is the second smallest eigenvalue of the corresponding graph Laplacian matrix
\bgt\nn \lapf(\corm) = \dgrf(\corm) - \adjf(\corm).\egt 
Here, $\adjf(\corm) = (\corm-\idm_{\dcor})_+$ is the adjacency matrix obtained by applying a threshold and setting the diagonal elements to  zero, and $\dgrf(\corm) = \diag\{\adjf(\corm)\ones_{\dcor}\}$ is the (node) degree matrix, where $\idm_{\dcor} = \diag\{\ones_{\dcor}\}$, $\ones_{\dcor} = (1,\dots,1)^\top\in\real^{\dcor}$, and $\amat_+ = (\max\{\amij, 0\})_{1\le\hubidx,\vhubidx\le\dcor}$, for any $\amat\in\real^{\dcor\times\dcor}$. 
Then the Fiedler value corresponding to $\corm$ is given by a map $\smap\colon \msp\ra\real$, 
\bgt\nn\smap(\corm)=\lambda_{\dcor-1}(\lapf(\corm)),\egt 
that yields the $(\dcor-1)$th largest, i.e., second smallest eigenvalue of $\lapf(\corm)$, for any $\corm\in\msp$. 
Note that $\metric[F](\lapf(\corm_1),\lapf(\corm_2))^2 \le 3\metric[F](\corm_1,\corm_2)^2$. 
In view of the  Hoffman--Wielandt inequality \citep{hoff:53}, $\smap$ satisfies \ref{ass:smapLips} with $\constSmap=\sqrt{3}$ and $\pwSmap=1$. 
Applying local Fr\'echet regression with bandwidth $\bwo=13.26$,  chosen by leave-one-out cross validation, the Fiedler values for the local Fr\'echet regression estimates $\objLmOrc$ as per \eqref{eq:objLmOrc} of the conditional mean correlation matrix at age  $\tm$ are 
\bgt\label{eq:fdlEst} \smrEst(\tm) = \smap(\objLmOrc) = \lambda_{\dcor-1}(\lapf(\objLmOrc),\quad\text{for } \tm\in\tdom.\egt 



\begin{figure}[hbt!]
	\centering
	\includegraphics[width=0.45\textwidth]{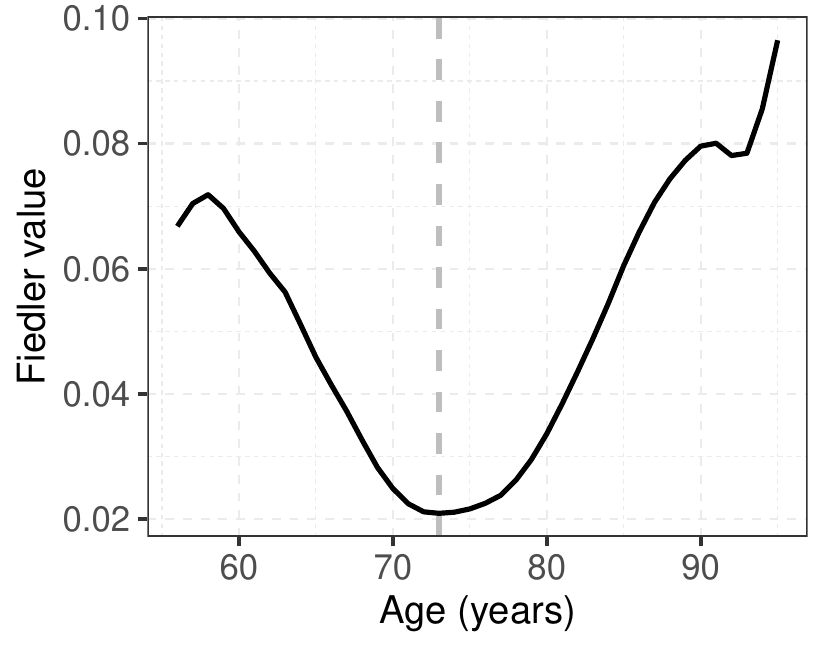}
	\caption{
		\rvone{Fiedler values as a function of age, corresponding to the local Fr\'echet regression estimate of the correlation matrix valued conditional Fr\'echet mean trajectory as per \eqref{eq:fdlEst}}, with the minimum attained at 73 years of age marked by a dashed line.}\label{fig:adnifdl}
\end{figure}

\rvone{Figure~\ref{fig:adnifdl} displays the trajectory $\smrEst$ of age-varying Fiedler values obtained for the local Fr\'echet regression estimate of the correlation matrix valued conditional Fr\'echet mean trajectory according to \eqref{eq:fdlEst},  based on the correlation matrices obtained from fMRI scans as per \eqref{eq:corr} for $\ndpo=402$ normal subjects in the ADNI data. 
	A convex pattern can be seen around the minimum of $\smrEst$, 
	which  is attained at 73 years of age.} 
While some studies have found that functional connectivity decreases during normal aging processes before 80 years of age \citep{ferr:13,meve:13}, we observe for these data that the decrease is reversed for older ages. 

\subsection{Time Warping for Distributional Trajectories: Human Mortality Data}\label{sec:wpMort}


There has been perpetual interest in understanding human longevity. One particular goal is to obtain a general pattern of how the distribution of age-at-death evolves over time.
Human mortality data for different countries  are available from the Human Mortality Database (\url{http://www.mortality.org/}). We consider the calendar time period from 1983 to 2013, for which the mortality data for 28 countries are available throughout. It is known that the mortality distributions generally shift to higher ages during this time interval, which reflects 
increasing longevity. It is then of interest to ascertain  which countries move faster and which move slower towards increased longevity, quantified by the rightward shift of the densities of age-at-death.  

To address this question, we apply the proposed time warping method in  the metric space of probability distributions with the Wasserstein metric, i.e., the Wasserstein space as per Example~\ref{eg:wsp}. In 1983 all countries start out with their warping functions taking values at the 
initial calendar year 1983, and in 2013 they all assume the value at the ending year 2013, so that the warping effect is considered between these two endpoints. 
A warping function below the identity function indicates that the country to which it belongs  
is on an accelerating course  towards enhanced longevity, while countries with warping functions above the identity are on a delayed course. 

\begin{figure}[hbt!]
	\centering
	\includegraphics[width=0.65\textwidth]{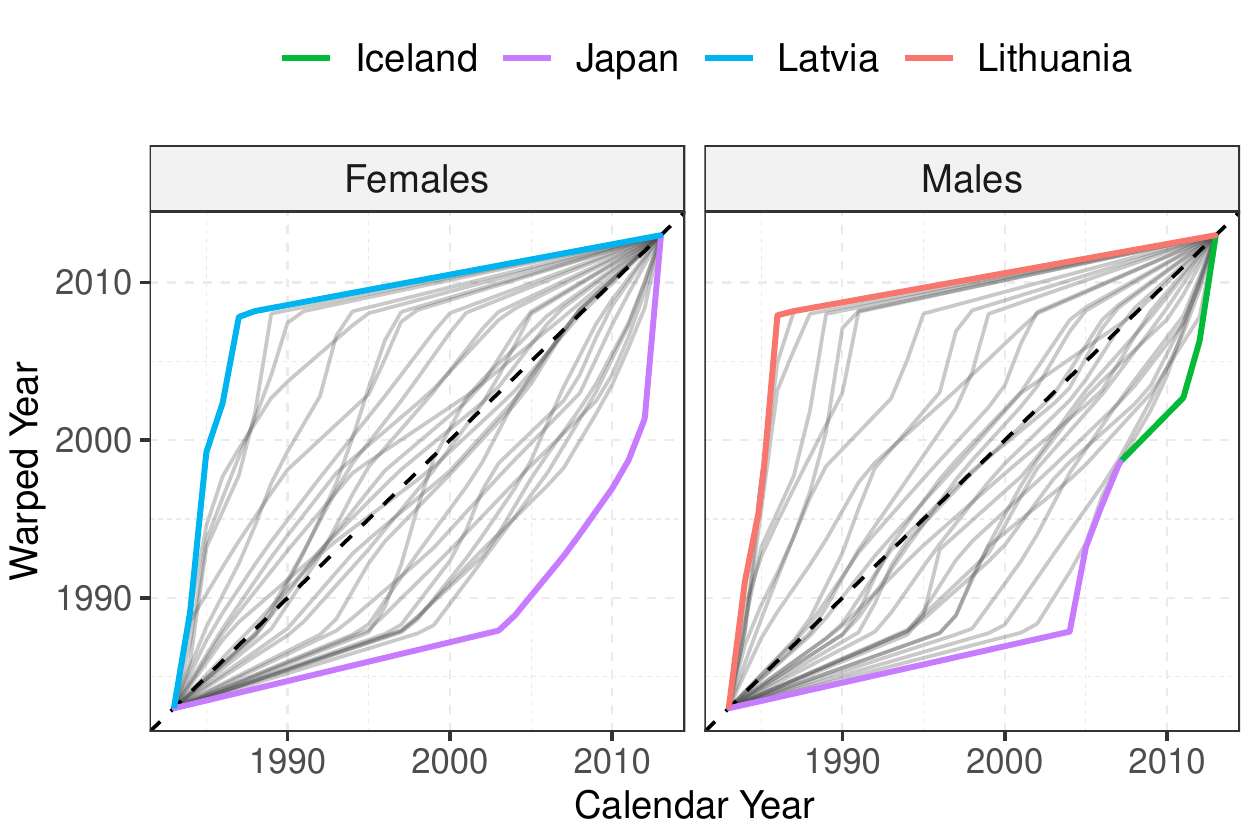}
	\caption{Estimated warping functions $\wpiEst$ as per \eqref{eq:wpEst} for the mortality distribution trajectories for females (left) and males (right) for each country in the sample (grey solid curves), where cross-sectional minimum and maximum warping functions are identified and  highlighted in different colors. The black dashed lines represent identity functions.}\label{fig:mortWp}
\end{figure}

Comparing the estimated warping functions across countries, we found that for males the enhancement in longevity of Japanese from 1983 to 2007 and for  Icelanders from 2008 to the 2013  accelerates  the fastest among all of the 28 countries between 1987 and 2013,  while males have the most delayed  increased  longevity for  Lithuania throughout the period  (Figure~\ref{fig:mortWp}). For females,  the  movement towards increased longevity is found to be fastest for Japanese women and slowest for Latvian women. The relative delay in increasing longevity  for 
Lithuania and Latvia, former Soviet republics, is likely due to  the aftermath of the breakup of the Soviet Union.

\begin{figure}[htbp] 
	\centering
	\includegraphics[width=0.8\textwidth]{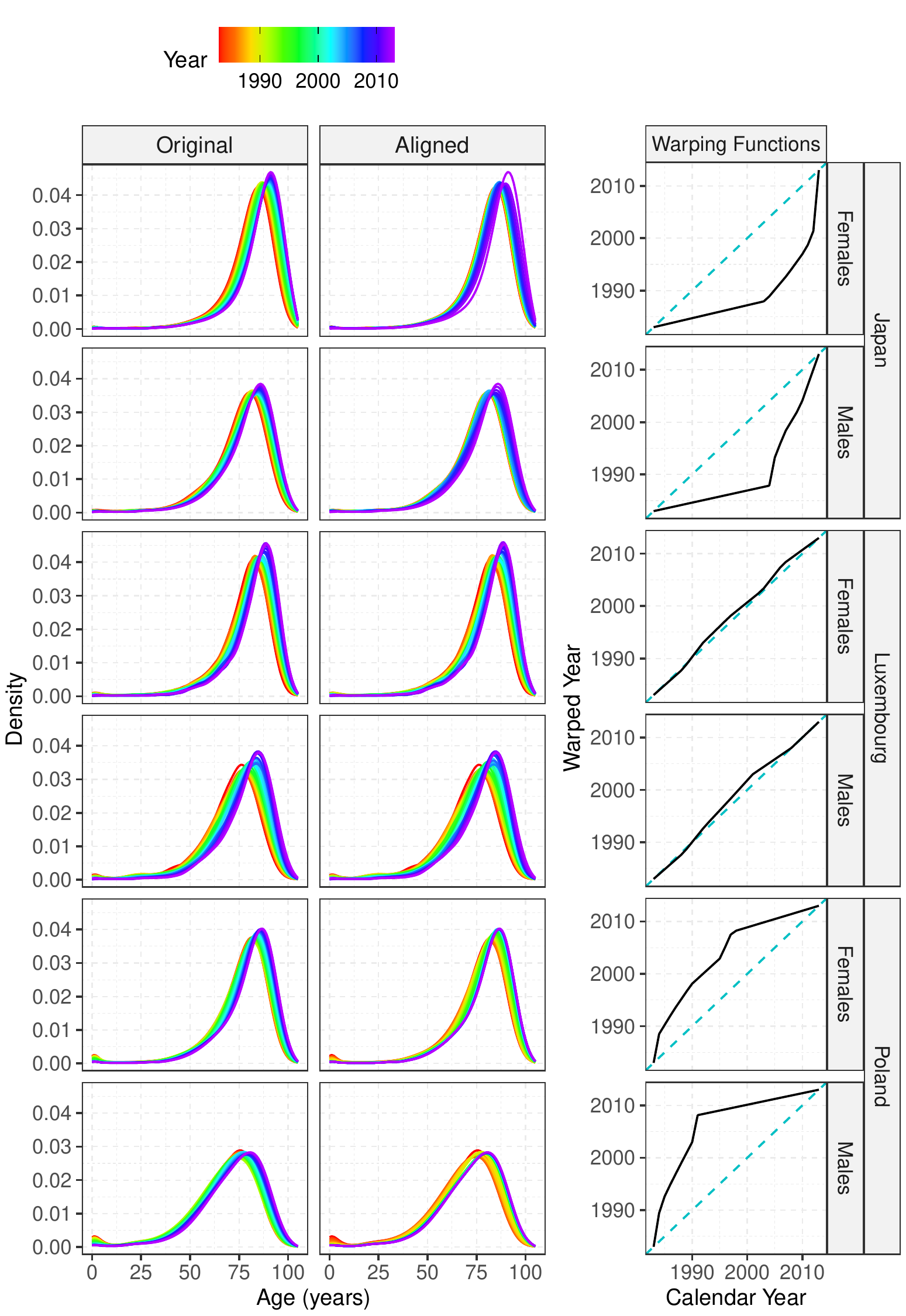}
	\caption{Density functions corresponding to the original (left) and aligned (middle) trajectories, $\wobjiEst[\cdot]$ and $\wobjiEst[\wpiEst(\cdot)]$, and the estimated warping functions $\wpiEst$ (right) during 1983--2013 for three countries, Japan (top), Luxembourg (middle), and Poland (bottom), for which the estimated warping functions for females and males are similar, where $\wobjiEst$ and $\wpiEst$ are as per \eqref{eq:wobjEst} and \eqref{eq:wpEst}. The blue dashed lines on the right panels represent identity functions.}\label{fig:mortSimi}
\end{figure}

\begin{figure}[hbt!]
	\centering
	\includegraphics[width=0.8\textwidth]{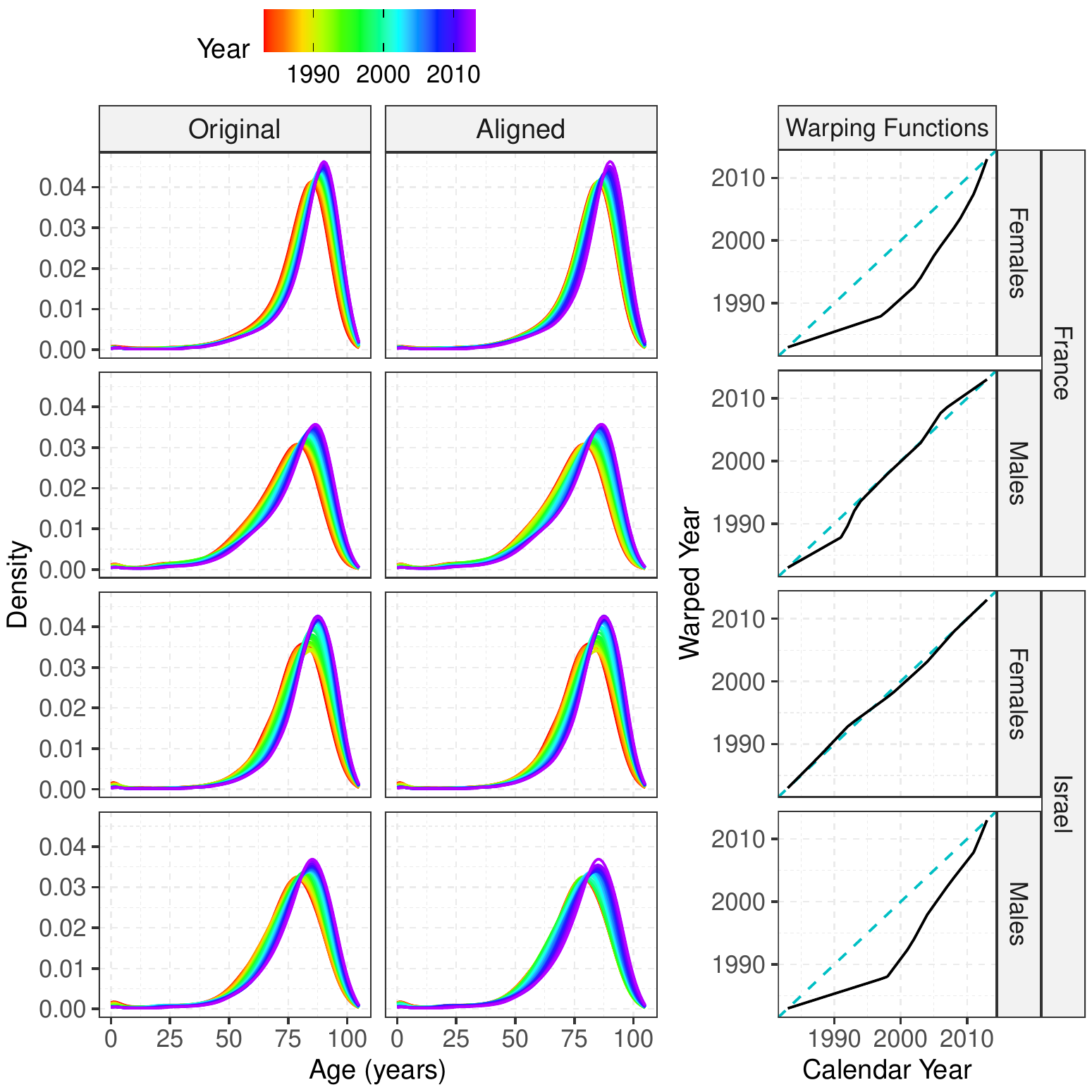}
	\caption{Density functions corresponding to the original (left) and aligned (middle) trajectories, $\wobjiEst[\cdot]$ and $\wobjiEst[\wpiEst(\cdot)]$, and the estimated warping functions $\wpiEst$ (right) during 1983--2013 for two countries, France (top) and Israel (bottom), for which the estimated warping functions for females and males differ,  where $\wobjiEst$ and $\wpiEst$ are as per \eqref{eq:wobjEst} and \eqref{eq:wpEst}. The blue dashed lines on the right panels represent identity functions.}\label{fig:mortDiff}
\end{figure}

The original and aligned trajectories along with the estimated warping functions for two selected groups of countries are demonstrated in Figures~\ref{fig:mortSimi} and \ref{fig:mortDiff}. The former group includes representative countries for which the pattern of the estimated warping functions are similar between females and males, while the  latter group consists of representative countries for which the estimated warping functions are mismatched between females and males.  

Among the countries shown in Figure~\ref{fig:mortSimi} with similar warping patterns between males and females, Luxembourg's warping functions are close to the identity and therefore its longevity increase represents the average increase across all countries, for both males and females. For  Japan, both male and female longevity are strongly accelerated compared to the other countries considered, in contrast to the situation  for 
Poland, where the increase in longevity for both males and females is much delayed relative to the average. 
In addition, countries shown in Figure~\ref{fig:mortDiff} exhibit an interesting gender heterogeneity. 
Both France and Israel show average longevity increase patterns for one gender, namely males in France and females in Israel, but not for the other gender, as  females in France and males in Israel exhibit accelerated longevity.

\references

\clearpage
\appendix
\renewcommand{\thesection}{}
\renewcommand{\thesubsection}{S.\arabic{subsection}}
\renewcommand{\theequation}{S.\arabic{equation}}
\renewcommand{\thethm}{S.\arabic{thm}}
\renewcommand{\thelem}{S.\arabic{lem}}
\setcounter{equation}{0}

\section{Supplementary Material: Theoretical Details}\label{app:proof}

\subsection{Proofs of the Uniform Convergence of Local Fr\'echet Regression with Fixed Targets as in Section~\ref{sec:locFregFix}}\label{app:pfUnifFreg}

\bpf[Proof of Theorem~\ref{thm:unifFregRateOverTm}]
For any $\smconst>0$, taking $\pwLmUnif =\pwLm+\smconst/2$, we will first show \eqref{eq:unifRateLmvsLmO}. 

We note that under \ref{ass:ker} and \ref{ass:jointdtn}, as $\bwo\ra 0$, it holds for $\momidx = 0,1,2$ that 
\bal\label{eq:kmomRate}
&\kmom[\momidx](\apdt) = \expect\left[\kerho(\pdto - \apdt)(\pdto - \apdt)^{\momidx}\right] = \bwo^{\momidx} \left[\denPdt(\apdt) \kermom{1}{\momidx} + \bwo\denPdt'(\apdt) \kermom{1}{\momidx+1} + \O(\bwo^2)\right],
\eal
where $\kermom{k}{l} = \int_{\{(\rarg-\apdt)/\bwo:\ \rarg\in\tdom\}}
\ker(\rarg)^k \rarg^l \diffop\rarg$, 
for $k,l\in\ntnum$, and the $\O$ terms are all uniform across $\unifinPdom$. These results are well known \citep{fan:96} and we omit the proofs here. 

Let $\kurt = (\ndpo\bwo)\mhf(-\log\bwo)\half$. Under \ref{ass:ker} and \ref{ass:jointdtn}, it follows from \eqref{eq:kmomRate} and similar arguments to the proof of Theorem~B of \citet{silv:78} that
\bgt\label{eq:kmomUnif}
\sup_{\unifinPdom}|\kmom[0](\apdt)| = \O(1),\quad
\sup_{\unifinPdom}|\kmom[1]^+(\apdt)| = \O(\bwo),\quad \sup_{\unifinPdom}|\kmom[2](\apdt)| = \O(\bwo^2),\\
\sup_{\unifinPdom} |\kmomEst(\apdt) - \kmom(\apdt)| = \Op(\kurt\bwo^{\momidx}),\ \momidx = 0,1,2,\quad
\sup_{\unifinPdom}|\kmomEst[1]^+(\apdt)| = \O(\bwo) + \Op(\kurt\bwo),
\egt
where $\kmom[1]^+(\apdt) = \expect[\kerho(\pdto- \apdt)|\pdto - \apdt|]$, and $\kmomEst[1]^+(\apdt) = \ndpo\inv\summo [\kerho(\pdtj - \apdt)|\pdtj - \apdt|]$\detail{, noting that $\bwo^{-\momidx}\kmomEst(\cdot)$ and $\bwo\inv\kmomEst[1]^+(\cdot)$ can all be viewed as kernel density estimators with kernels $\ker_{\momidx}(\rarg) = \ker(\rarg)\rarg^{\momidx}$ and $\ker_{1}^+(\rarg) = \ker(\rarg)|\rarg|$, respectively, as per \citet{silv:78}, 
and \eqref{eq:kmomUnif} implies $\sup_{\unifinPdom}|\kvarEst(\apdt) - \kvar(\apdt)| = \Op(\kurt\bwo^2)$}. 
Applying Taylor expansion yields 
\bgt\nn
\sup_{\unifinPdom} \left|[\kvarEst(\apdt)]\inv - [\kvar(\apdt)]\inv\right| = \Op(\kurt\bwo\mtwo),\\
\sup_{\unifinPdom} \left| \frac{\kmomEst[2](\apdt)}{\kvarEst(\apdt)} - \frac{\kmom[2](\apdt)}{\kvar(\apdt)} \right| = \Op(\kurt),\qaq \sup_{\unifinPdom} \left| \frac{\kmomEst[1](\apdt)}{\kvarEst(\apdt)} - \frac{\kmom[1](\apdt)}{\kvar(\apdt)} \right| = \Op(\kurt\bwo\inv).\egt
Hence, 
\bal\label{eq:wLocURt}
&\sup_{\unifinPdom}\ndpo\inv\summo |\wLocEstDfo{\pdtj} - \wLocDfo{\pdtj}| \\
&\quad\le \sup_{\unifinPdom}|\kmomEst[0](\apdt)| \sup_{\unifinPdom} \left| \frac{\kmomEst[2](\apdt)}{\kvarEst(\apdt)} - \frac{\kmom[2](\apdt)}{\kvar(\apdt)} \right| + \sup_{\unifinPdom}|\kmomEst[1]^+(\apdt)| \sup_{\unifinPdom} \left| \frac{\kmomEst[1](\apdt)}{\kvarEst(\apdt)} - \frac{\kmom[1](\apdt)}{\kvar(\apdt)} \right|\\
&\quad=\Op(\kurt). \eal
For any $\csetBR>0$, define a sequence of events 
\bal\label{eq:setUnifR} \setUnifR = \left\{\sup_{\unifinPdom}\ndpo\inv\summo |\wLocEstDfo{\pdtj} - \wLocDfo{\pdtj}| \le \csetBR \kurt \right\}. \eal

Given any $\pwLarger>1$, set $\cn = \min\{(\ndpo\bwo^2)^{\pwLm/[4(\pwLm-2+\pwLarger)]}, (\ndpo\bwo^2(-\log\bwo)\inv)^{\pwLm/[4(\pwLm-1)]} \}$; 
for some $\pcon\le \radLm$, set $\tpcon=\pcon^{\pwLm/2}$, 
with $\radLm$ and $\pwLm$ as per \ref{ass:curvatureUnif}. 
For any $\intL\in\ntnum_+$, considering $\ndpo$ large enough such that $\log_2(\tpcon\cn)>\intL$, 
\bal\label{eq:tailprobUnif}
&\prob\left(\cn \sup_{\unifinPdom} \metric\left(\objLm,\objLmOrc\right)^{\pwLm/2} > 2^{\intL}\right)\\
&\le \prob\left(\sup_{\unifinPdom} \metric\left(\objLm,\objLmOrc\right) > \pcon/2 \right) + \prob(\setUnifR\c)\\
&\quad\ + \sum_{\substack{\setidx > \intL\\ 2^{\setidx}\le\tpcon\cn}} \prob\left( \left\{ 2^{\setidx-1} \le \cn \sup_{\unifinPdom} \metric\left(\objLm,\objLmOrc\right)^{\pwLm/2} < 2^{\setidx} \right\} \cap \setUnifR\right).\eal
Note that \eqref{eq:wLocURt} implies 
$\lim_{\csetBR\gify}\limsup_{\ndpo\gify}\prob(\setUnifR\c) = 0$. 
Regarding the first term on the right hand side of \eqref{eq:tailprobUnif}, we will next show that 
\bgt\label{eq:opUnifLmvsLmO} \sup_{\unifinPdom} \metric\left(\objLm,\objLmOrc\right) = \op(1).\egt 

Given any $\unifinPdom$, $\metric(\objLm,\objLmOrc) = \op(1)$ follows from similar arguments to the proof of Lemma~2 in Section~S.3 of the Supplementary Material of \citet{pete:19}\note{{ }(noting that total boundedness of $\msp$ is needed here)}. 
By Theorems~1.5.4, 1.5.7 and 1.3.6 of \citet{vand:96}\detail{{ }and the total boundedness of $\pdom$}, it suffices to show that for any $\epsilon>0$, as $\radius\ra 0$, 
\[\limsup_{\ndpo\gify} \prob\left(\sup_{\stinPdom,\ |\vapdt-\apdt|<\radius} \left|\metric\left(\objLm[\vapdt],\objLmOrc[\vapdt]\right) - \metric\left(\objLm,\objLmOrc\right)\right| > 2\epsilon\right) \ra 0.\] 
In conjunction with \ref{ass:minUnif} and the fact that $|\metric(\objLm[\vapdt],\objLmOrc[\vapdt]) - \metric(\objLm,\objLmOrc)| \le  \metric(\objLm[\vapdt],\objLm[\apdt]) + \metric(\objLmOrc[\vapdt],\objLmOrc[\apdt])$, it suffices to show that
\bgt\label{eq:eqcont_objfLm}
\limsup_{\bwo\ra 0} \sup_{\stinPdom,\ |\vapdt-\apdt|<\radius} \sup_{\arps\in\msp} \left|\objfLm(\arps, \vapdt) - \objfLm(\arps, \apdt)\right| \ra 0,\quad\text{as }\radius\ra 0,\egt 
and that for any $\epsilon>0$, 
\bgt\label{eq:eqcont_objfLmOrc} 
\limsup_{\ndpo\gify} \prob\left(\sup_{\stinPdom,\ |\vapdt-\apdt|<\radius} \sup_{\arps\in\msp} \left|\objfLmOrc(\arps,\vapdt) - \objfLmOrc(\arps,\apdt)\right| > \epsilon \right) \ra 0,\quad\text{as }\radius\ra 0.\egt 

Noting that by \ref{ass:ker} and \ref{ass:jointdtn}, $\expect[\wLocDfo{\pdto}] = 1$ and $\expect[\wLocDfo{\pdto}(\pdto-\apdt)] = 0$, 
\bal\label{eq:objfCmvsLm}
\objfLm(\arps,\apdt)
&= \expect\left[\wLocDfo{\pdto} \objfCm(\arps,\pdto)\right]\\
&=  \expect\left[\wLocDfo{\pdto} \left(\objfCm(\arps,\apdt) + (\pdto-\apdt)\frac{\partial\objfCm}{\partial\apdt}(\arps,\apdt)\right)\right]\\
&\quad + \expect\left[\wLocDfo{\pdto} \left(\objfCm(\arps,\pdto) - \objfCm(\arps,\apdt) - (\pdto-\apdt)\frac{\partial\objfCm}{\partial\apdt}(\arps,\apdt)\right)\right]\\
&=\objfCm(\arps,\apdt) + \expect\left[\wLocDfo{\pdto} \left(\objfCm(\arps,\pdto) - \objfCm(\arps,\apdt) - (\pdto-\apdt)\frac{\partial\objfCm}{\partial\apdt}(\arps,\apdt)\right)\right].
\eal 
Defining a function $\dCL:\ \msp\times\pdom\ra\real$ as
\bal\label{eq:d2objfCm}
\dCL(\arps,\apdt) = \frac{\partial^2\objfCm}{\partial\apdt^2}(\arps,\apdt),\quad \arps\in\msp,\ \apdt\in\pdom,\eal 
it follows from \eqref{eq:objfCmvsLm}, \ref{ass:ker}, and \ref{ass:jointdtn}\note{{ }(in the warping framework we do not need to assume that $\ker$ has bounded support, since we assume $\denPdt$ is bounded away from zero on $\pdom$)} that 
\bal\nn \sup_{\arps\in\msp,\ \unifinPdom} \left| \objfLm(\arps,\apdt) - \objfCm(\arps,\apdt)\right| \le  \frac{1}{2}\bwo^2 \sup_{\unifinPdom} \expect\left[|\wLocDfo{\pdto}| \left(\frac{\pdto-\apdt}{\bwo}\right)^2\right] \sup_{\unifinPdom,\ \arps\in\msp}|\dCL(\arps,\apdt)|.\eal
Note that $\sup_{\unifinPdom} \expect[|\wLocDfo{\pdto}|(\pdto-\apdt)^2\bwo\mtwo] = \O(1)$. 
Furthermore, using similar arguments to the proof of Theorem~3 of \cite{pete:19}, \note{for any $\apdt$ such that $\denPdt(\apdt)>0$ }we obtain 
\bal\label{eq:objfCm}
\objfCm(\arps,\apdt)
= \int\metric^2(\varps,\arps) \diffop\ccdfRps(\apdt,\varps)
= \int\metric^2(\varps,\arps) \frac{\cdenPdt(\apdt,\varps)}{\denPdt(\apdt)} \diffop\cdfRps(\varps),\eal
where $\cdfRps$ and $\ccdfRps$ are the marginal and conditional distribution of $\rpso$, the latter given $\pdto$. In conjunction with \ref{ass:ker}, \ref{ass:jointdtn}, and the dominated convergence theorem, \eqref{eq:objfCm} implies 
\bal\label{eq:d2objfCmEq}
\dCL(\arps,\apdt)
&= \int\metric^2(\varps,\arps) \frac{\partial^2}{\partial\apdt^2}\left[\frac{\cdenPdt(\apdt,\varps)}{\denPdt(\apdt)}\right] \diffop\cdfRps(\varps),\eal
whence by \ref{ass:jointdtn} and the boundedness of $\msp$, we obtain $\sup_{\arps\in\msp,\ \unifinPdom} \left|\dCL(\arps,\apdt)\right|<\infty$.
Thus, 
\bal\label{eq:supObjfLmvsCm}
\sup_{\arps\in\msp,\ \apdt\in\pdom} \left| \objfLm(\arps,\apdt) - \objfCm(\arps,\apdt)\right| = \O(\bwo^2),\quad\text{as }\bwo\ra 0.\eal
Moreover, by \eqref{eq:objfCm} and \ref{ass:jointdtn},
\bal\label{eq:objfLmDiffinTm}
&\sup_{\stinPdom,\ |\vapdt-\apdt|<\radius} \sup_{\arps\in\msp} \left|\objfLm(\arps, \vapdt) - \objfLm(\arps, \apdt)\right|\\
&\quad\le \sup_{\stinPdom,\ |\vapdt-\apdt|<\radius} \sup_{\arps\in\msp} \left|\objfCm(\arps, \vapdt) - \objfCm(\arps, \apdt)\right| + 2\sup_{\arps\in\msp,\ \apdt\in\pdom} \left| \objfLm(\arps,\apdt) - \objfCm(\arps,\apdt)\right|\\
&\quad= \O(\radius) + \O(\bwo^2),\eal
whence \eqref{eq:eqcont_objfLm} follows. 
For \eqref{eq:eqcont_objfLmOrc}, let $\disc:\ \msp\ra\real$ be a function defined as 
\bal\nn\disc(\arps) \coloneqq \sup_{\unifinPdom} \left|\ndpo\inv\summo \left[\wLocDfo{\pdtj} \metric^2(\rpsj,\arps)\right] - \expect\left[\wLocDfo{\pdto} \metric^2(\rpso,\arps)\right] \right|.\eal 
Then
\bal\label{eq:objfLmOrcDiffinTm}
\sup_{\arps\in\msp} \left|\objfLmOrc(\arps, \vapdt) - \objfLmOrc(\arps, \apdt)\right| 
&\le 2\diam(\msp)^2 \sup_{\unifinPdom} \ndpo\inv\summo\left|\wLocEstDfo{\pdtj} - \wLocDfo{\pdtj}\right|\\
&\quad+ \sup_{\arps\in\msp} \left|\objfLm(\arps, \vapdt) - \objfLm(\arps, \apdt)\right| 
+ 2 \sup_{\arps\in\msp}\disc(\arps).\eal
Regarding $\disc(\arps)$, 
\detail{
	since
	\bal\nn
	&\ndpo\inv\summo \left[\wLocDfo{\pdtj} \metric^2(\rpsj,\arps)\right] - \expect\left[\wLocDfo{\pdto} \metric^2(\rpso,\arps)\right] \\
	&= \frac{\kmom[2](\apdt)}{\kvar(\apdt)} \left\{\ndpo\inv\summo \left[\kerho(\pdtj-\apdt) \metric^2(\rpsj,\arps)\right]  - \expect\left[\kerho(\pdto-\apdt) \metric^2(\rpso,\arps)\right]\right\}\\
	&\quad\  
	\begin{aligned}
		- \frac{\bwo\kmom[1](\apdt)}{\kvar(\apdt)} &\left\{\ndpo\inv\summo \left[\kerho(\pdtj-\apdt) \left(\frac{\pdtj-\apdt}{\bwo}\right) \metric^2(\rpsj,\arps)\right]\right.\\
		&\ - \left.\expect\left[\kerho(\pdto-\apdt) \left(\frac{\pdto-\apdt}{\bwo}\right) \metric^2(\rpso,\arps)\right]\right\},\end{aligned}
	\eal
}under \ref{ass:ker} and \ref{ass:jointdtn}, it follows from similar arguments to the proof of Proposition~4 of \citet{mack:82} with kernels $\ker_{\momidx}(\rarg) = \ker(\rarg)\rarg^{\momidx}$, for $\momidx=0,1$, that $\disc(\arps)= \op(1)$, for any given $\arps\in\msp$. 
Furthermore, \detail{noting that 
	\bgt\nn
	\sup_{\unifinPdom} \ndpo\inv\summo \left|\wLocDfo{\pdtj}\right|
	\le \sup_{\unifinPdom}|\kmomEst[0](\apdt)| \sup_{\unifinPdom}\left|\frac{\kmom[2](\apdt)}{\kvar(\apdt)}\right| + \sup_{\unifinPdom}|\kmomEst[1]^+(\apdt)| \sup_{\unifinPdom}\left|\frac{\kmom[1](\apdt)}{\kvar(\apdt)}\right|,\\
	\sup_{\unifinPdom} \expect\left|\wLocDfo{\pdto}\right| 
	\le \sup_{\unifinPdom} \frac{\kmom[0](\apdt)\kmom[2](\apdt)}{\kvar(\apdt)} + \sup_{\unifinPdom} \frac{\kmom[1]^+(\apdt)|\kmom[1](\apdt)|}{\kvar(\apdt)}, 
	\egt
}by \eqref{eq:kmomRate} and \eqref{eq:kmomUnif}, 
\[\sup_{\unifinPdom} \ndpo\inv\summo \left|\wLocDfo{\pdtj}\right| = \O(1) + \Op(\kurt),\quad \sup_{\unifinPdom} \expect\left|\wLocDfo{\pdto}\right| =\O(1),\]
and hence for any $\arps_1,\arps_2\in\msp$, 
\bal\nn
\left|\disc(\arps_1) - \disc(\arps_2)\right| 
&\le 2\diam(\msp)\metric(\arps_1,\arps_2) \left( \sup_{\unifinPdom}\ndpo\inv\summo \left|\wLocDfo{\pdtj}\right| +  \sup_{\unifinPdom} \expect\left|\wLocDfo{\pdto}\right| \right)\\
&= \metric(\arps_1,\arps_2) \left[\O(1) + \Op(\kurt)\right].
\eal
In conjunction with the total boundedness of $\msp$,  $\sup_{\arps\in\msp}\disc(\arps) = \op(1)$ follows \citep[Theorems~1.5.4, 1.5.7, and 1.3.6 of][]{vand:96}. 
This implies \eqref{eq:eqcont_objfLmOrc} in conjunction with \eqref{eq:objfLmOrcDiffinTm}, \eqref{eq:wLocURt} and \eqref{eq:objfLmDiffinTm}. 
Thus, \eqref{eq:opUnifLmvsLmO} follows. 

We move on to the third term on the right hand side of \eqref{eq:tailprobUnif}. For $\setidx\in\pint$, define sets 
\bal\label{eq:setknt} \setknt = \{\arps\in\msp:\ 2^{\setidx - 1} \le \cn\metric(\arps,\objLm)^{\pwLm/2} < 2^{\setidx}\}.\eal 
We note that under \ref{ass:curvatureUnif}, \[\liminf_{\ndpo\gify} \inf_{\unifinPdom} \inf_{\arps\in\setknt} \left[\objfLm(\arps,\apdt) - \objfLm(\objLm,\apdt)\right] \ge \constLm 2^{2(\setidx - 1)} \cn\mtwo.\] 
Defining functions $\objfLmOvsLmt(\cdot) = \objfLmOrc(\cdot,\apdt) - \objfLm(\cdot,\apdt)$ on $\msp$, applying Markov's inequality, the third term on the right hand side of \eqref{eq:tailprobUnif} can be bounded (from above) by
\bal\label{eq:sumExpectUnif}
&\sum_{\substack{\setidx > \intL\\ 2^{\setidx}\le\tpcon\cn}} \prob\left( \left\{\sup_{\unifinPdom} \sup_{\arps\in\setknt} \left|\objfLmOvsLmt(\arps) - \objfLmOvsLmt(\objLm)\right| \ge \constLm 2^{2(\setidx - 1)} \cn\mtwo\right\} \cap \setUnifR\right)\\
&\quad\le \sum_{\substack{\setidx > \intL\\ 2^{\setidx}\le\tpcon\cn}} \constLm\inv 2^{-2(\setidx - 1)} \cn^2 \expect\left(\indicator{\setUnifR} \sup_{\unifinPdom} \sup_{\arps\in\setknt} \left|\objfLmOvsLmt(\arps) - \objfLmOvsLmt(\objLm)\right| \right), \eal
where $\indicator{\anySet}$ is the indicator for an event $\anySet$. 
For any $\arps\in\msp$, defining 
\bal\label{eq:objfLmOvsLmtDecom}
\objfLmOvsLmtOne(\arps) &= \ndpo\inv\summo\left[\wLocEstDfo{\pdtj} - \wLocDfo{\pdtj}\right] \metric^2(\rpsj,\arps), \\
\objfLmOvsLmtTwo(\arps) &= \ndpo\inv\summo \left[\wLocDfo{\pdtj}\metric^2(\rpsj,\arps)\right] - \expect\left[\wLocDfo{\pdto} \metric^2(\rpso,\arps)\right],\eal
whence $\objfLmOvsLmt = \objfLmOvsLmtOne + \objfLmOvsLmtTwo$. 

For $\objfLmOvsLmtOne$, note that 
\[ \left|\objfLmOvsLmtOne(\arps) - \objfLmOvsLmtOne(\objLm)\right|
\le 2\diam(\msp)\metric(\arps,\objLm)\ndpo\inv\summo |\wLocEstDfo{\pdtj} - \wLocDfo{\pdtj}|, \]
for all $\arps\in\msp$ and $\unifinPdom$, 
and hence given $\radius>0$, it holds on $\setUnifR$ that
\bgt\label{eq:objfLmEvsLm1Unif} \sup_{\unifinPdom} \sup_{\metric(\arps,\objLm) < \radius} \left|\objfLmOvsLmtOne(\arps) - \objfLmOvsLmtOne(\objLm)\right| \le 2\diam(\msp) \csetBR \radius \kurt. \egt
For $\objfLmOvsLmtTwo$, given any $\arps\in\msp$, $\unifinPdom$ and $\radius>0$, 
defining functions $\gfctn{\apdt,\arps}:\ \pdom \times \msp \ra \real$ by
\bgt\nn \gfctn{\apdt,\arps}(\vapdt, \varps) = \wLocDfo{\vapdt} \left[\metric^2(\varps,\arps) - \metric^2(\varps,\objLm)\right],\quad \vapdt\in\pdom, \varps\in\msp,\egt
and a function class
\bgt\nn \gclass = \left\{\gfctn{\apdt,\arps}:\ \metric(\arps,\objLm) < \radius,\ \unifinPdom\right\}. \egt
For any $\twotinPdom$ and $\arps_l\in\ball{\objLm[\apdt_l]}{\radius}$ for $l=1,2$, 
\als{
	&\left|\gfctn{\apdt_1,\arps_1}(\vapdt,\varps) - \gfctn{\apdt_2,\arps_2}(\vapdt,\varps)\right| \\
	&\quad\le 2\diam(\msp) \left(\metric(\arps_1,\arps_2) + \metric(\objLm[\apdt_1],\objLm[\apdt_2])\right) \sup_{\unifinPdom} |\wLocDfo{\vapdt}| \\
	&\qquad + 2\diam(\msp)\radius \sup_{\unifTdom{\vapdt\in\pdom}} \left|\wLoc{\vapdt}{\apdt_1}{\bwo} - \wLoc{\vapdt}{\apdt_2}{\bwo}\right|.
}
By \eqref{eq:eqcont_objfLm} and \ref{ass:minUnif}, \bgt\label{eq:eqcont_objLm} \limsup_{\bwo\ra 0}\sup_{\stinPdom,\ |\vapdt-\apdt|<\radius} \metric(\objLm[\vapdt],\objLm)\ra 0, \quad\text{as }\radius\ra 0.\egt 
For small $|\apdt_1-\apdt_2|$\detail{{ }such that $\metric(\objLm[\apdt_1],\objLm[\apdt_2]) < \radLm$}, by \ref{ass:curvatureUnif}, it holds that as $\bwo\ra 0$, 
\bal\nn 
2\constLm \metric(\objLm[\apdt_1],\objLm[\apdt_2])^{\pwLm} 
&\le 2\sup_{\arps\in\msp} \left|\objfLm(\arps,\apdt_1) - \objfLm(\arps,\apdt_2)\right|\\ 
&\le 2\diam(\msp)^2 \sup_{\unifTdom{\vapdt\in\pdom}}|\wLoc{\vapdt}{\apdt_1}{\bwo} - \wLoc{\vapdt}{\apdt_2}{\bwo}|.\eal
Noting that $\sup_{\unifTdom{\vapdt\in\pdom}}|\wLoc{\vapdt}{\apdt_1}{\bwo} - \wLoc{\vapdt}{\apdt_2}{\bwo}| =|\apdt_1-\apdt_2|\O(\bwo\mtwo)$ by \ref{ass:ker} and \ref{ass:jointdtn}\note{{ }(since $\ker'$ and $\ker$ are assumed to be bounded, regardless of whether $\ker$ has bounded support or not)}, 
there exists a constant $C>0$ such that for small $|\apdt_1-\apdt_2|$, 
\[\left| \gfctn{\apdt_1,\arps_1}(\vapdt,\varps) - \gfctn{\apdt_2,\arps_2}(\vapdt,\varps) \right| \le C \left[\metric(\arps_1,\arps_2) + \metricTm(\apdt_1,\apdt_2) \right] \gdiff(\vapdt),\]
where $\metricTm:\pdom\times\pdom\ra[0,+\infty)$ is a metric on $\pdom$ defined as 
\bal\nn \metricTm(\apdt_1,\apdt_2) 
= \max\left\{\bwo\mtwo|\apdt_1-\apdt_2|,\ (\bwo\mtwo|\apdt_1-\apdt_2|)^{1/\pwLm}\right\}, \quad \twotinPdom,
\eal
which can be verified that is indeed a metric, and $\gdiff:\ \pdom\times\msp\ra\real$ is a function defined as 
\[ \gdiff(\vapdt,\arps) =  \sup_{\unifinPdom} |\wLocDfo{\vapdt}| + \radius,\quad \vapdt\in\pdom. \]
An envelope function $\genve:\ \pdom\times\msp\ra\real$ for the function class $\gclass$ is
\[\genve(\vapdt,\arps) = 2\diam(\msp)\radius \sup_{\unifinPdom} |\wLocDfo{\vapdt}|,\quad \vapdt\in\pdom.\]
Denoting the joint distribution of $(\pdto,\rpso)$ by $\jointdtn$, the $\hilbert[\jointdtn]$ norm $\ltwoNmPr{\cdot}$ is given by $\ltwoNmPr{\anyFctn} = [\expect(\anyFctn(\pdto, \rpso)^2)]\half$, for any function $\anyFctn:\ \pdom\times\msp \ra\real$. 
The envelope function entails $\ltwoNmPr{\genve} = \O(\radius\bwo\inv)$, by \ref{ass:ker} and \ref{ass:jointdtn}. 
Furthermore, by Theorem~2.7.11 of \citet{vand:96}, for $\epsilon>0$, the $\epsilon\ltwoNmPr{\genve}$ bracketing number of the function class $\gclass$ can be bounded as 
\detail{ 
	\bal\nn
	&\brktnum\left( \epsilon\ltwoNmPr{\genve}, \gclass, \ltwoNmPr{\cdot} \right) \\
	&= \brktnum\left( 2\frac{\epsilon\ltwoNmPr{\genve}}{2\ltwoNmPr{\gdiff}} \ltwoNmPr{\gdiff}, \gclass, \ltwoNmPr{\cdot} \right)\\ 
	&\le \covernum\left( \frac{\epsilon\ltwoNmPr{\genve}}{2\ltwoNmPr{\gdiff}}, \{(\apdt,\arps):\ \arps\in\ball{\objLm}{\radius},\ \unifinPdom\}, \metric[\pdom\times\msp] \right)\\
	&\le \covernum\left( \frac{\epsilon\ltwoNmPr{\genve}}{4\ltwoNmPr{\gdiff}}, \unifTdom{\pdom}, \metricTm \right)
	\cdot \sup_{\unifinPdom} \covernum\left( \frac{\epsilon\ltwoNmPr{\genve}}{4\ltwoNmPr{\gdiff}}, \ball{\objLm}{\radius},\metric \right), \eal
	where $\metric[\pdom\times\msp] ((\apdt_1,\arps_1),(\apdt_2,\arps_2)) = \metricTm(\apdt_1,\apdt_2) + \metric(\arps_1,\arps_2)$, for any $\twotinPdom$ and $\arps_1,\arps_2\in\msp$. Therefore, 
}
\bgt\label{eq:brkt2cover} \brktnum\left(\epsilon\ltwoNmPr{\genve}, \gclass, {\ltwoNmPr{\cdot}}\right)\le C_1(\epsilon\radius\bwo^2)^{-C_0} \sup_{\unifinPdom} \covernum\left(C_2\epsilon\radius,\ball{\objLm}{\radius},\metric\right), \egt 
where $C_0,C_1,C_2>0$ are constants\detail{{ }only depending on $\pwLm$}\detail{, noting that $\ltwoNmPr{\genve}/\ltwoNmPr{\gdiff} \sim\radius$}. 
In conjunction with \eqref{eq:unifRateCmvsLm} which will be shown later, \detail{for $\bwo$ sufficiently small, 
there exists a constant $C_3>C_2$ such that $\ball{\objLm}{\radius} \subset \ball{\objCm}{C_3\radius}$, for any $\radius>0$ and $\unifinPdom$. }Choose $\pcon$ in \eqref{eq:tailprobUnif} such that \ref{ass:entropyUnif} holds for all $\radEntropy\le C_3\pcon$. 
Observing that 
\bal\nn
&\int_0^1 \sup_{\unifinPdom} \sqrt{1 + \log\covernum\left(C_2\epsilon\radius,\ball{\objCm}{C_3\radius},\metric\right)} \diffop\epsilon \\
&\quad=\frac{C_3}{C_2} \int_0^{C_2/C_3} \sup_{\unifinPdom} \sqrt{1 + \log\covernum\left(\epsilon C_3\radius, \ball{\objCm}{C_3\radius}, \metric\right)} \diffop\epsilon \\
&\quad\le \frac{C_3}{C_2} \int_0^1 \sup_{\unifinPdom} \sqrt{1 + \log\covernum\left(\epsilon C_3\radius, \ball{\objCm}{C_3\radius}, \metric\right)} \diffop\epsilon, \eal 
\eqref{eq:brkt2cover} implies for any $\radius\le\pcon$, 
\bal\nn \int_0^1 &\sqrt{1+\log\brktnum\left(\epsilon\ltwoNmPr{\genve}, \gclass, \ltwoNmPr{\cdot}\right)}\diffop\epsilon\\
&\quad\le \int_0^1 \sup_{\unifinPdom} \sqrt{1 + \log\covernum\left(C_2\epsilon\radius,\ball{\objLm}{\radius},\metric\right)} \diffop\epsilon + \int_0^1\sqrt{-C_0\log(\epsilon\radius\bwo^2) + \log C_1}\diffop\epsilon\\
&\quad= \O\left(\mc{I} + \int_0^1\sqrt{-\log(\radius\bwo) - \log\epsilon}\diffop\epsilon\right)
= \O\left(\sqrt{-\log(\radius\bwo)}\right), \eal
with $\mc{I}$ being the integral in \ref{ass:entropyUnif}. 
By Theorem~2.14.2 of \citet{vand:96}, 
\bal \label{eq:objfLmEvsLm2Unif}
\expect\left( \sup_{\unifinPdom } \sup_{\metric(\arps,\objLm) < \radius} \left|\objfLmOvsLmtTwo(\arps) - \objfLmOvsLmtTwo(\objLm)\right|\right) 
&= \O\left(\radius\bwo\inv \sqrt{-\log(\radius\bwo)} \ndpo\mhf\right)\\
&= \O\left(\radius^{2-\pwLarger}(\ndpo\bwo^2)\mhf + \radius(\ndpo\bwo^2)\mhf \sqrt{-\log\bwo}\right). \eal

Combining \eqref{eq:objfLmEvsLm1Unif} and \eqref{eq:objfLmEvsLm2Unif}, it holds that 
\bal\label{eq:objfLmEvsLmUnif} \expect\left( \indicator{\setUnifR}  \sup_{\unifinPdom } \sup_{\metric(\arps,\objLm) < \radius} \left|\objfLmOvsLmt(\arps) - \objfLmOvsLmt(\objLm)\right| \right)
\le C (\ndpo\bwo^2)\mhf \left(\radius^{2-\pwLarger}+ \radius\sqrt{-\log\bwo}\right), \eal
where $C>0$ is a constant depending on $\csetBR$ and the entropy integral in \ref{ass:entropyUnif}. 
Note that on $\setknt$, it holds that $\metric(\arps,\objLm) < (2^\setidx \cn\inv)^{2/\pwLm}$. 
Hence, \eqref{eq:sumExpectUnif} can be bounded by
\bal\nn 
&C \sum_{\substack{\setidx > \intL\\ 2^{\setidx}\le\tpcon\cn}} 2^{-2(\setidx - 1)} \cn^2 (\ndpo\bwo^2)\mhf \left[(2^\setidx \cn\inv)^{2(2-\pwLarger)/\pwLm} + (2^\setidx \cn\inv)^{2/\pwLm} \sqrt{-\log\bwo}\right] \\
&\quad\le 4C \cn^{2(\pwLm-2+\pwLarger)/\pwLm} (\ndpo\bwo^2)\mhf \sum_{\setidx > \intL} 2^{-2\setidx(\pwLm-2+\pwLarger)/\pwLm}\\
&\qquad + 4C \cn^{2(\pwLm-1)/\pwLm} (\ndpo\bwo^2)\mhf \sqrt{-\log\bwo} \sum_{\setidx > \intL} 2^{-2\setidx(\pwLm-1)/\pwLm}, \eal
which converges to 0 as $\intL\gify$, since $\pwLm,\pwLarger>1$. 
Thus, 
\bal\nn 
\sup_{\unifinPdom} \metric\left(\objLm,\objLmOrc\right) &=\Op\left(\cn^{-2/\pwLm}\right), 
\eal
and \eqref{eq:unifRateLmvsLmO} follows.

Next, we will show \eqref{eq:unifRateCmvsLm}. 
By \eqref{eq:supObjfLmvsCm} and \ref{ass:minUnif}, $\metric(\objCm,\objLm) = \o(1)$, as $\bwo\ra 0$, for any $\unifinPdom$. 
By \eqref{eq:eqcont_objCm} and the compactness of $\pdom$, the conditional Fr\'echet mean trajectory $\objCmNull$ is $\metric$-continuous at any $\tm\in\pdom$ and hence uniformly $\metric$-continuous on $\pdom$. 
In conjunction with \eqref{eq:eqcont_objLm}, $\sup_{\unifinPdom} \metric(\objCm,\objLm) = \o(1)$ follows. 
Let $\gfctn{\arps,\apdt}:\ \pdom\ra\real$ be a function defined as $\gfctn{\arps,\apdt}(\vapdt) = \objfCm(\arps,\vapdt) - \objfCm(\objCm,\vapdt)$, for $\vapdt\in\pdom$. 
For any $\radius>0$, \eqref{eq:objfCmvsLm} and \eqref{eq:d2objfCmEq}, in conjunction with \ref{ass:ker}, \ref{ass:jointdtn}\note{{ }(in the warping framework we do not need to assume that $\ker$ has bounded support, since we assume $\denPdt$ is bounded away from zero on $\pdom$)}, and the boundedness of $\msp$,
\bal\nn
&\sup_{\unifinPdom} \sup_{\metric(\arps,\objCm)<\radius} \left|(\objfLm-\objfCm)(\arps,\apdt) - (\objfLm-\objfCm)(\objCm,\apdt)\right| \\
&\quad= \sup_{\unifinPdom} \sup_{\metric(\arps,\objCm)<\radius} \left|\expect\left[\wLocDfo{\pdto} \left(\gfctn{\arps,\apdt}(\pdto) - \gfctn{\arps,\apdt}(\apdt) - (\pdto-\apdt)\gfctn{\arps,\apdt}'(\apdt)\right)\right]\right|\\
&\quad\le \frac{1}{2}\bwo^2 \sup_{\unifinPdom} \expect\left[|\wLocDfo{\pdto}|\left(\frac{\pdto-\apdt}{\bwo}\right)^2\right]\cdot \sup_{\unifinPdom} \sup_{\metric(\arps,\objCm)<\radius}  \left|\dCL(\arps,\apdt) - \dCL(\objCm,\apdt)\right|\\
&\quad = \O(\bwo^2\radius),
\eal
with $\dCL(\arps,\apdt)$ defined as per \eqref{eq:d2objfCm}. 
Set $\ch = \bwo^{-\pwCm/(\pwCm-1)}$. 
Using similar arguments to the proof of \eqref{eq:unifRateLmvsLmO}, 
there exists a constant $C>0$ such that for small $\bwo$, 
\[ \indicator{\ch\sup_{\unifinPdom} \metric\left(\objCm,\objLm\right)^{\pwCm/2} > 2^\intL} 
\le C  \sum_{\setidx>\intL} \frac{\bwo^2(2^{\setidx}\ch\inv)^{2/\pwCm}} {2^{2(\setidx-1)} \ch\mtwo}
= 4C\sum_{\setidx>\intL} 2^{-2\setidx(\pwCm-1)/\pwCm}, \]
which converges to 0 as $\intL\gify$, and hence \eqref{eq:unifRateCmvsLm} follows. 

Lastly, we note that for any $\gamma\in(0,0.5)$, if $\bwo\sim\ndpo^{-\gamma}$, then 
\bal\nn \frac{(\ndpo\bwo^2(-\log\bwo)\inv)^{-1/[2(\pwLm-1)]}}{(\ndpo\bwo^2)^{-1/[2(\pwLmUnif-1)]}} \sim \frac{(\log\ndpo)^{1/[2(\pwLm-1)]}}{\ndpo^{(0.5-\gamma)\left[(\pwLm-1)\inv - (\pwLmUnif-1)\inv\right]}} \ra 0, \quad\text{as }\ndpo\gify.\eal
With $\bwo\sim\ndpo^{-(\pwCm-1)/(2\pwCm+4\pwLmUnif-6)}$, it holds that $\bwo^{2/(\pwCm-1)}\sim (\ndpo\bwo^2)^{-1/[2(\pwLmUnif-1)]} \sim \ndpo^{-1/(\pwCm+2\pwLmUnif-3)}$, whence \eqref{eq:unifFregRateOverTm} follows, which completes the proof.
\epf

\subsection{Proofs of the Uniform Convergence of Local Fr\'echet Regression with Random Targets as in Section~\ref{sec:locFregRdm}}


\bpf[Proof of Theorem~\ref{thm:unifFregRateRdmTgt}] 
Given any fixed $\smconst>0$, define 
\bgt\nn
\cbno = \min\left\{(\ndpo\bwo^2)^{\pwLm/[4(\pwLm-1+\smconst/2)]}, [\ndpo\bwo^2(-\log\bwo)\inv]^{\pwLm/[4(\pwLm-1)]}\right\}. \egt 
We will show for the bias and stochastic parts respectively that
\begin{gather}
\sup_{\unifinWobj} \sup_{\unifinTm} \metric\left(\objCmR,\objLmR\right) = \O\left(\bwo^{2/(\pwCm-1)}\right) , \label{eq:unifOverWobj_CmvsLm}\\
\limsup_{\ndpo\gify} \sup_{\unifinWobj} \probNoise\left(\cbno \sup_{\unifinTm} \metric\left(\objLmR,\objLmOrcR\right)^{\pwLm/2} > C\right) \ra 0, \quad\text{as }C\gify. \label{eq:unifOverWobj_LmvsLmOrc}
\end{gather}
Observing that 
\bal\nn
&\sup_{\unifinTm} \metric\left(\objPrcs(\tm), \objPrcsLm(\tm)\right) \le \sup_{\unifinWobj} \sup_{\unifinTm} \metric\left(\objCmR, \objLmR\right),\\
&\limsup_{\ndpo\gify} 
\prob\left( \sup_{\unifinTm} \metric\left(\objPrcsLm(\tm), \objPrcsEst(\tm)\right) > C \cbno^{-2/\pwLm}\right)\\
&\quad\le \limsup_{\ndpo\gify}\sup_{\unifinWobj}  \probNoise\left( \cbno \sup_{\unifinTm} \metric\left(\objLmR,\objLmOrcR\right)^{\pwLm/2} > C\right),\eal
\eqref{eq:objPrcsBiasVarRate} follows, which implies \eqref{eq:objPrcsEstRate} if $\bwo\sim\ndpo^{-(\pwCm-1)/(2\pwCm+4\pwLm-6+2\smconst)}$. 

For \eqref{eq:unifOverWobj_CmvsLm}, we note that 
for any given $\unifinWobj$, and $\unifinTm$, $\metric(\objCmR,\objLmR) = \o(1)$, as $\bwo\ra0$, by Theorem~\ref{thm:unifFregRateOverTm}. 
We note that by \ref{ass:ker} and \ref{ass:jointdtnStrg}, 
\bal\label{eq:wtmom2} \sup_{\unifinTm} \expectNoise\left[|\wLocR{\tmo}{\tm}{\bwo}| (\tmo-\tm)^2\bwo\mtwo\right] = \O(1),\quad\text{as }\nobj\gify.\eal
Defining 
\bgt\label{eq:d2objfCmR}
\dCLR(\anyObj,\tm) = \frac{\partial^2\objfCmR}{\partial\tm^2} (\anyObj,\tm),\egt
it holds following similar arguments to the proof of \eqref{eq:d2objfCmEq} that 
\bgt\label{eq:d2objfCmREq}
\dCLR(\anyObj,\tm)
= \int\metric^2(\varAnyObj,\anyObj) \frac{\partial^2}{\partial\tm^2}\left[\frac{\cdenRtm(\tm,\varAnyObj)}{\denRtm(\tm)}\right] \diffop\cdfObj(\varAnyObj).\egt
In conjunction with \ref{ass:jointdtnStrg} and the boundedness of $\msp$, 
\bgt\label{eq:2ndderivObjfCm}
\sup_{\unifinWobj,\ \unifinTm,\ \anyObj\in\msp} \left|\dCLR(\anyObj,\tm) \right| <\infty, \egt
whence we obtain 
\bal\nn
&\sup_{\unifinWobj,\ \unifinTm} \left|\objfLmR(\anyObj,\tm) - \objfCmR(\anyObj,\tm)\right| \\
&\quad\le \frac{1}{2}\bwo^2 \sup_{\unifinTm} \expectNoise\left[\left|\wLocRDfo{\tmo}\right| \left(\frac{\tmo-\tm}{\bwo}\right)^2\right] \sup_{\substack{\unifinWobj,\ \unifinTm,\ \anyObj\in\msp}} \left|\dCLR(\anyObj,\tm) \right|\\
&\quad=\O(\bwo^2).
\eal
This implies $\sup_{\unifinWobj,\ \unifinTm} \metric(\objCmR,\objLmR) = \o(1)$ under \ref{ass:minUnifStrg}. 
Furthermore, by \eqref{eq:wtmom2} and \eqref{eq:d2objfCmREq}, there exists a constant $C>0$ such that for $\ndpo$ large enough, 
\bal\label{eq:d2objfCmDiffR}
&\sup_{\unifinWobj,\ \unifinTm} \sup_{\metric(\anyObj,\objCmR)<\radius} \left|(\objfLmR-\objfCmR)(\anyObj,\tm) - (\objfLmR-\objfCmR)(\objCmR,\tm)\right| \\
&\quad\le \frac{1}{2} \bwo^2 \sup_{\unifinTm} \expectNoise\left[|\wLocRDfo{\tmo} |\left(\frac{\tmo-\tm}{\bwo}\right)^2\right]
\cdot \sup_{\substack{\unifinWobj\\ \unifinTm}} \sup_{\metric(\anyObj,\objCmR)<\radius}  \left|\dCLR(\anyObj,\tm) - \dCLR(\objCmR,\tm)\right|\\
&\quad \le C\bwo^2\radius. \eal
Using similar arguments to the proof of \eqref{eq:unifRateCmvsLm}, with $\chbo = \bwo^{-\pwCm/(\pwCm-1)}$, there exists a constant $C>0$ such that for large $\ndpo$, 
\bal\nn
&\indicator{\sup_{\unifinWobj}\sup_{\unifinTm} \metric\left(\objCmR,\objLmR\right)^{\pwCm/2} > 2^\intL \chbo\inv} 
\le C  \sum_{\setidx>\intL} \frac{\bwo^2(2^{\setidx}\chbo\inv)^{2/\pwCm}} {2^{2(\setidx-1)} \chbo\mtwo}
= 4C\sum_{\setidx>\intL} 2^{-2\setidx(\pwCm-1)/\pwCm}, \eal
which converges to zero as $\intL\gify$, whence \eqref{eq:unifOverWobj_CmvsLm} follows. 

Next, we will show \eqref{eq:unifOverWobj_LmvsLmOrc}. Let $\pcon$ be the minimum integer not less than $\log_2(\cbno\diam(\msp)^{\pwLm/2} + 1)$, and for any $\csetBR>0$, define sets 
\bal\label{eq:setUnifR_rdmTgt} \setUnifR = \left\{\sup_{\unifinPdom}\ndpo\inv\summo |\wLocREstDfo{\tmoj} - \wLocRDfo{\tmoj}| \le \csetBR \kurt \right\}. \eal 
For any $\intL\in\ntnum_+$, considering $\ndpo$ large enough such that $\pcon>\intL$, 
\bal\label{eq:tailprobUnifBoth}
&\sup_{\unifinWobj}  \probNoise\left( \cbno \sup_{\unifinTm} \metric\left(\objLmR,\objLmOrcR\right)^{\pwLm/2} > 2^{\intL}\right)
\le \probNoise(\setUnifR\c)\\
&\quad + \sum_{\substack{\intL< \setidx\le \pcon}} \sup_{\unifinWobj} \probNoise\left( \left\{ 2^{\setidx-1} \le \cbno \sup_{\unifinTm} \metric\left(\objLmR,\objLmOrcR\right)^{\pwLm/2} < 2^{\setidx} \right\} \cap \setUnifR\right).\eal 
Under \ref{ass:ker} and \ref{ass:jointdtnStrg}, it follows from similar arguments to the proof of Theorem~B of \citet{silv:78} that there exists $\csetBR>0$ with $\probNoise(\setUnifR\c) = 0$, for $\ndpo$ large enough. 
Regarding the convergence of the second term on the right hand side of \eqref{eq:tailprobUnifBoth}, using similar arguments to the proof of \eqref{eq:unifRateLmvsLmO}, 
it holds for $\objfLmORvsLmRtTwo(\anyObj) = \ndpo\inv\summo[\wLocR{\tmoj}{\tm}{\bwo} \metric^2(\obsobjoj,\anyObj)] - \expectNoise[\wLocR{\tmo}{\tm}{\bwo} \metric^2(\obsobjo,\anyObj)]$ that 
\bal\nn
\expect\left( \sup_{\unifinTm } \sup_{\metric(\anyObj,\objLmR) < \radius} \left|\objfLmORvsLmRtTwo(\anyObj) - \objfLmORvsLmRtTwo(\objLmR)\right|\right) 
&= \O\left(\radius^{2-\pwLarger}(\ndpo\bwo^2)\mhf + \radius(\ndpo\bwo^2)\mhf \sqrt{-\log\bwo}\right),\eal
where the $\O$ term is uniform over $\unifinWobj$, by \eqref{eq:unifOverWobj_CmvsLm}, \ref{ass:minUnifStrg} and  Theorem~2.14.2 of \citet{vand:96}\note{{ }and also the fact that $\ltwoNmPrNoise{\genve}$ and $\ltwoNmPrNoise{\gdiff}$ do not depend on $\detmWobj$, since $\genve(\vartm,\cdot)$ and $\gdiff(\vartm,\cdot)$ are constant functions for any fixed $\vartm\in\tdom$. Together with \eqref{eq:objfLmEvsLm1Unif}, this implies that the constant $C$ in \eqref{eq:objfLmEvsLmUnif} does not depend on $\detmWobj$}. 
Under \ref{ass:ker} and \ref{ass:jointdtnStrg}--\ref{ass:curvatureUnifStrg}, 
the second term on the right hand side of \eqref{eq:tailprobUnifBoth} can be bounded by 
\bal\nn
&C \sum_{\substack{\intL<\setidx\le\pcon}} 2^{-2(\setidx - 1)} \cbno^2 (\ndpo\bwo^2)\mhf \left[(2^\setidx \cbno\inv)^{2(1-\smconst/2)/\pwLm} + (2^\setidx \cbno\inv)^{2/\pwLm} \sqrt{-\log\bwo}\right]\\
&\quad\le 4C \cbno^{2(\pwLm-1+\smconst/2)/\pwLm} (\ndpo\bwo^2)\mhf \sum_{\setidx > \intL} 2^{-2\setidx(\pwLm-1+\smconst/2)/\pwLm}\\
&\qquad + 4C \cbno^{2(\pwLm-1)/\pwLm} (\ndpo\bwo^2)\mhf \sqrt{-\log\bwo} \sum_{\setidx > \intL} 2^{-2\setidx(\pwLm-1)/\pwLm}\\
&\quad\le 4C\sum_{\setidx > \intL} 2^{-2\setidx(\pwLm-1+\smconst/2)/\pwLm} + 4C \sum_{\setidx > \intL} 2^{-2\setidx(\pwLm-1)/\pwLm}, \eal
which converges to zero as $\intL\gify$, whence \eqref{eq:unifOverWobj_LmvsLmOrc} follows. 
\epf

\subsection{Proofs of Results in Section~\ref{sec:extremum}}\label{app:pfArgMax}

\bpf[Proof of Corollary~\ref{cor:extremum}]
By \eqref{eq:pmCm}, \eqref{eq:pmEst}, and \ref{ass:smapLips}, 
\bal\nn
\smrCm(\pmEst) - \smrCm(\pmCm)
&\le \smrCm(\pmEst) - \smrCm(\pmCm) + \smrEst(\pmCm) - \smrEst(\pmEst)\\
&\le 2\sup_{\unifinPdom} |\smrEst(\apdt) - \smrCm(\apdt)| 
\le 2\constSmap \sup_{\unifinPdom} \metric(\objLmOrc,\objCm)^{\pwSmap}. \eal
This implies $|\pmEst-\pmCm| = \op(1)$ in conjunction with \ref{ass:pmUniq} and Theorem~\ref{thm:unifFregRateOverTm}, whence  \eqref{eq:pmRate} follows under \ref{ass:smrCurv}. 
\epf

\subsection{Proofs of Results in Section~\ref{sec:wp}}\label{app:pfWp}

We will first present two auxiliary results (Lemmas~\ref{lem:muh} and \ref{lem:uniq}), where Lemma~\ref{lem:muh} will be needed in the proof of Lemma~\ref{lem:uniq}, and Lemma~\ref{lem:uniq} shows that the coefficient vector $\lsvecPw$ is the unique minimizer of $\obfnM$ given in \eqref{eq:obfnM} under certain constraints, which will be used to derive the rate of convergence for the M-estimator $\lsvecEst$ of the coefficient vector in Theorem~\ref{thm:pai}. 

\blem \label{lem:muh}
For any $\wpcand,\vwpcand\in\wfsp$ such that $\metric(\mObj[\wpcand(\tm)], \mObj[\vwpcand(\tm)]) = 0$, for all $\tm\in\tdom$, where $\mObjNull$ satisfies \ref{ass:contCm}--\ref{ass:flat}, it holds that
\[\wpcand(\tm) = \vwpcand(\tm), \forallt.\]
\elem
\bpf
Suppose there exists $\rarg_0\in(0,\maxtm)$ such that $\wpcand(\rarg_0) \neq \vwpcand(\rarg_0)$. Without loss of generality, we assume $\wpcand(\rarg_0) < \vwpcand(\rarg_0)$. 
Let $\tm_0 = \wpcand(\rarg_0)$. We define a sequence $\{\tm_k\}_{k=1}^\infty$ iteratively by 
\bgt\label{eq:seqt}
\tm_k = \vwpcand(\wpcand\inv(\tm_{k-1})),\quad\text{for }k=1,2,\dots\egt 
Then it can be shown by induction that $\{\tm_k\}_{k=1}^\infty$ is a strictly increasing sequence, whence there exists $\tm^*\in\tdom$ such that $\tm_k\uparrow\tm^*$ as $k\gify$, since $\{\tm_k\}_{k=1}^\infty\subset\tdom=[0,\maxtm]$. 
Due to the continuity of $\wpcand$ and $\vwpcand$, taking $k\gify$ on both sides of \eqref{eq:seqt} provides $\tm^* = \vwpcand(\wpcand\inv(\tm^*))$. 
Let $\rarg^* = \wpcand\inv(\tm^*)$, then $\tm^* = \wpcand(\rarg^*) = \vwpcand(\rarg^*)$. 
Furthermore, for all $k=1,2,\dotsc$, 
\bgt\label{eq:eqd}
\metric\left(\mObj[\tm_k], \mObj[\tm_{k-1}]\right) 
= \metric\left(\mObj[\vwpcand (\wpcand\inv(\tm_{k-1}))], \mObj[\wpcand(\wpcand\inv(\tm_{k-1})]\right)
= 0,\egt
since $\metric(\mObj[\wpcand(\tm)], \mObj[\vwpcand(\tm)]) = 0$, for all $\tm\in\tdom$. 
By \ref{ass:flat}, there exists $\vartm_0\in(\tm_0,\tm_1)$ such that 
\bgt\label{eq:notzero}
\metric\left(\mObj[\vartm_0],\mObj[\tm_0]\right) = \metric\left(\mObj[\vartm_0],\mObj[\tm_1]\right) >0. \egt
Similarly, we can iteratively define another sequence $\vartm_k = \vwpcand(\wpcand\inv(\vartm_{k-1}))$, for $k=1,2,\dots$, for which it also holds that $\vartm_k\uparrow \tm^*$ as $k\gify$ and $\metric(\mObj[\vartm_k], \mObj[\vartm_{k-1}])=0$, for all $k=1,2,\dots$ 
By \eqref{eq:eqd}, $\metric(\mObj[\tm_0], \mObj[\tm^*]) = \metric(\mObj[\tm_k], \mObj[\tm^*])$ for all $k=1,2,\dots$ Taking $k\gify$ yields $\metric(\mObj[\tm_0], \mObj[\tm^*]) = \lim_{k\ra\infty} \metric(\mObj[\tm_k], \mObj[\tm^*])=0$, by \ref{ass:contCm}. 
Similarly, it can be verified that $\metric(\mObj[\vartm_0], \mObj[\tm^*]) =0$, whence we obtain $\metric(\mObj[\vartm_0], \mObj[\tm_0]) = 0$, which contradicts \eqref{eq:notzero}.
\epf

\blem\label{lem:uniq}
Suppose \ref{ass:contCm}--\ref{ass:wpSlope} hold. For any $\objidx,\varobjidx=1,\dots,\nobj$ such that $\objidx\ne \varobjidx$, the coefficient vector $\lsvecPw$ corresponding to the pairwise warping function $\pairwp$ is the unique minimizer of the following constrained optimization problem 
\bgt\label{eq:optM} 
\min_{\anyVec\in\real^{\ntgrid+1}}\obfnM(\anyVec; \wpi[\varobjidx],\wpi),\\
\text{subject to }
\constrM_0(\anyVec) = \anyVec_{\ntgrid+1} - \maxtm = 0,\ \constrM_{\tgrididx}(\anyVec) = \anyVec_{\tgrididx-1} - \anyVec_{\tgrididx} + \wpSlope \le 0,\ \tgrididx=1,2,\dots,\ntgrid+1,\egt
where $\obfnM$ is as per \eqref{eq:obfnM}, $\anyVec = (\anyVec_1,\dots,\anyVec_{\ntgrid+1})^\top\in\real^{p+1}$, $\anyVec_0 = 0$, and $\wpSlope\in(0,cC\inv\maxtm/(\ntgrid+1))$ is a constant with $c$ and $C$ as per \ref{ass:wpSlope}. \note{Note: $\wpi[\varobjidx](\wpi\inv(\vartm)) - \wpi[\varobjidx](\wpi\inv(\tm)) \ge c(\wpi\inv(\vartm) - \wpi\inv(\tm)) \ge cC\inv (\vartm-\tm)$, for any $\vartm > \tm$.}
\elem
\bpf
Considering the fact that $\obfnM(\anyVec; \wpi[\varobjidx],\wpi) \ge 0$, for all $\anyVec\in\real^{\ntgrid+1}$, 
and that $\obfnM(\lsvecPw; \wpi, \wpi[\varobjidx]) = 0$, since $\wpi[\varobjidx]^{-1}(\lsvecPw^\top \lsbasis(\tm))) = \wpi[\varobjidx]^{-1}(\wpi[\varobjidx] (\wpi^{-1}(\tm))) = \wpi^{-1}(\tm)$, 
$\lsvecPw$ is a constrained minimizer of the optimization problem in \eqref{eq:optM} in conjunction with \ref{ass:wpSlope}; it suffices to show the uniqueness. 
Suppose $\fixVec$ is a constrained minimizer of $\obfnM(\cdot;\wpi[\varobjidx],\wpi)$. The Lagrangian function corresponding to \eqref{eq:optM} is 
$\lagrM(\anyVec;\wpi[\varobjidx],\wpi) = \obfnM(\anyVec;\wpi[\varobjidx],\wpi)  +\sum_{\tgrididx=0}^{\ntgrid+1} \constrCoef \constrM_{\tgrididx}(\anyVec)$, 
where $\constrCoef[0]\in\real$, and $\constrCoef\ge 0$ for $\tgrididx=1,\dots,\ntgrid+1$. 
By the Karush--Kuhn--Tucker condition \citep{karu:39,kuhn:51}, there exist $\constrCoef^*\ge 0$, $\tgrididx=0,1,\dots,\ntgrid+1$, such that $\nabla\obfnM(\fixVec;\wpi[\varobjidx],\wpi) + \sum_{\tgrididx = 0}^{\ntgrid +1} \constrCoef^* \nabla\constrM_{\tgrididx}(\fixVec) = 0$ and $\constrCoef^*\constrM_{\tgrididx}(\fixVec) = 0$ for $\tgrididx=1,\dots,\ntgrid+1$. 
By \ref{ass:wpSlope}\note{{ }($\constrM_{\tgrididx}(\fixVec)<0$, for all $\tgrididx=1,\dots,\ntgrid$, otherwise $\obfnM(\fixVec;\wpi[\varobjidx],\wpi)> 0$; and $\constrM_0(\fixVec)=0$)}, it holds that $\nabla\obfnM(\fixVec;\wpi[\varobjidx],\wpi) = 0$, 
which, in conjunction with \ref{ass:flat} and \ref{ass:wpSlope}, implies 
\bgt\nn \metric\bprts{\mObj[{\wpi[\varobjidx]\inv[\fixVec^\top \lsbasis(\tm)]}], \mObj[\wpi\inv(\tm)]} = 0,\egt
for almost everywhere $\tm\in\tdom$ and hence for all $\tm\in\tdom$ by \ref{ass:contCm} and the continuity of $\wpi$ and $\wpi[\varobjidx]$. 
Applying Lemma~\ref{lem:muh} yields
\bgt\nn \wpi[\varobjidx]\inv[\fixVec^\top \lsbasis(\tm)] = {\wpi\inv(\tm)},\forallt,\egt
and hence
\bgt\nn \fixVec^\top \lsbasis(\tm) = \wpi[\varobjidx]\circ \wpi\inv(\tm) = \lsvecPw^\top \lsbasis(\tm),\forallt. \egt
For any $\tgrididx=0,1,\dots,\ntgrid$ and $\tm\in[\tgridpt,\tgridpt[\tgrididx+1])$,
\bal\nn
\fixVec^\top \lsbasis(\tm) &= (\fixVec)_{\tgrididx+1} \frac{\tm-\tgridpt}{\tgridpt[\tgrididx+1] - \tgridpt} - (\fixVec)_{\tgrididx} \frac{\tm-\tgridpt[\tgrididx+1]}{\tgridpt[\tgrididx+1] - \tgridpt},\\
\lsvecPw^\top \lsbasis(\tm) &= (\lsvecPw)_{\tgrididx+1} \frac{\tm-\tgridpt}{\tgridpt[\tgrididx+1] - \tgridpt} - (\lsvecPw)_{\tgrididx} \frac{\tm-\tgridpt[\tgrididx+1]}{\tgridpt[\tgrididx+1] - \tgridpt}. \eal
If there exists $k_0\in\{0,1,\dots,\ntgrid\}$ such that $(\lsvecPw)_{k_0} \neq (\fixVec)_{k_0}$, then 
$(\lsvecPw)_{\tgrididx} \neq (\fixVec)_{\tgrididx}$, for all $\tgrididx= k_0,\dots,\ntgrid+1$, which is contradictory to $(\lsvecPw)_{\ntgrid+1} = (\fixVec)_{\ntgrid+1} = \maxtm$. Thus, $\fixVec = \lsvecPw$.
\epf

\bpf[Proof of Theorem \ref{thm:pai}]
For any $\objidx,\varobjidx=1,\dots,\nobj$ such that $\objidx\neq \varobjidx$, a 
Taylor expansion yields 
\bal\nn 
&\obfnM(\anyVec;\wpi[\varobjidx],\wpi) - \obfnM(\lsvecPw;\wpi[\varobjidx],\wpi) \\
&\quad= \frac{1}{2} (\anyVec-\lsvecPw)^\top \frac{\partial^2\obfnM}{\partial\anyVec\partial\anyVec^\top} (\lsvecPw;\wpi[\varobjidx],\wpi) (\anyVec-\lsvecPw)
+ \o\left(\|\anyVec-\lsvecPw\|^2\right)\\
&\quad= \frac{1}{2}  \int_{\tdom} \left[\frac{\partial^2\metric[\mObjNull]^2}{\partial\vartm^2}(\vartm,\wpi\inv(\tm)) \frac{1}{\wpi[\varobjidx]'(\vartm)^2}\right]_{\vartm = \wpi[\varobjidx]\inv(\lsvecPw^\top\lsbasis(\tm))} \left[(\anyVec-\lsvecPw)^\top \lsbasis(\tm)\right]^2 \diffop\tm 
+ \o\left(\|\anyVec-\lsvecPw\|^2\right)\\
&\quad\ge  \frac{1}{2} C\mtwo \left[\inf_{\vartm=\tm\in\tdom} \frac{\partial^2\metric[\mObjNull]^2}{\partial\vartm^2}(\vartm,\tm)\right] \int_{\tdom} \left[(\anyVec-\lsvecPw)^\top \lsbasis(\tm)\right]^2 \diffop\tm + \o\left(\|\anyVec-\lsvecPw\|^2\right), \eal
as $\|\anyVec-\lsvecPw\|\ra 0$, where $C$ is as per \ref{ass:wpSlope}, and $\inf_{\vartm=\tm\in\tdom} (\partial^2\metric[\mObjNull]/\partial\vartm^2)(\vartm,\tm) = \inf_{\vartm=\tm\in\tdom} 2[(\partial\metric[\mObjNull]/\partial\vartm)(\vartm,\tm)]^2 >0$ by \ref{ass:flat}. 
Noting that $\int_{\tdom}\lsbasis(\tm)\lsbasis(\tm)^\top\diffop\tm$ is positive definite, there exist $\delta_0>0$ such that 
\bal\nn \obfnM(\anyVec;\wpi[\varobjidx],\wpi) - \obfnM(\lsvecPw;\wpi[\varobjidx],\wpi) 
\ge \frac{1}{4} C\mtwo \left[\inf_{\vartm=\tm\in\tdom} \frac{\partial^2\metric[\mObjNull]^2}{\partial\vartm^2}(\vartm,\tm)\right] \lambda_{\min}^{\lsbasis} \|\anyVec-\lsvecPw\|^2, \quad\text{for all }\anyVec\in\ball{\lsvecPw}{\delta_0}, \eal
where $\lambda_{\min}^{\lsbasis}>0$ is the smallest eigenvalue of $\int_{\tdom}\lsbasis(\tm)\lsbasis(\tm)^\top\diffop\tm$, and $\ball{\lsvecPw}{\delta}$ is a ball of radius $\delta$ centered at $\lsvecPw$. 
Furthermore, by Lemma~\ref{lem:uniq} and the compactness of the feasible region $\lsSp_{\wpSlope} \coloneqq \{\anyVec\in\real^{\ntgrid+1}:\ \anyVec_{\tgrididx} - \anyVec_{\tgrididx-1}\ge \wpSlope,\ \tgrididx=1,\dots,\ntgrid+1,\ \anyVec_{\ntgrid+1}=\maxtm\}\subset \lsSp$ of the optimization problem in \eqref{eq:optM}, it holds for any $\delta>0$ that 
\bal\nn \inf_{\lsSp_{\wpSlope} \cap \ball{\lsvecPw}{\delta}\c} \obfnM(\anyVec;\wpi[\varobjidx],\wpi) - \obfnM(\lsvecPw;\wpi[\varobjidx],\wpi) > 0.\eal 
Observing that $\|\anyVec-\lsvecPw\|^2 \le \ntgrid\maxtm^2$, let \bal\nn C_0 = \min\left\{\frac{1}{4}C\mtwo \inf_{\vartm=\tm\in\tdom} \frac{\partial^2\metric[\mObjNull]^2}{\partial\vartm^2}(\vartm,\tm) \lambda_{\min}^{\lsbasis},\ (\ntgrid\maxtm^2)\inv \inf_{\lsSp_{\wpSlope} \cap \ball{\lsvecPw}{\delta_0}\c} \left[\obfnM(\anyVec;\wpi[\varobjidx],\wpi) - \obfnM(\lsvecPw;\wpi[\varobjidx],\wpi)\right] \right\},\eal
where we note that $C_0>0$ and 
\bgt\nn
\obfnM(\anyVec;\wpi[\varobjidx],\wpi) - \obfnM(\lsvecPw;\wpi[\varobjidx],\wpi) \ge C_0 \|\anyVec-\lsvecPw\|^{\pwM},\quad \text{for all }\anyVec\in \lsSp_{\wpSlope}.\egt
By \eqref{eq:obfnYEst}, $\lsvecEst$ minimizes $\obfnYest(\anyVec;\varwobjiEst,\wobjiEst)$ subject to the constraint $\anyVec\in\lsSp_{\wpSlope}$ for some $\wpSlope\in(0,cC\inv\maxtm/(\ntgrid+1))$, whence we obtain 
\bal\label{eq:lsvecEstvsTrue}
&\|\lsvecEst-\lsvecPw\|\\
&\quad\le C_0\mhf \left[\obfnM(\lsvecEst;\wpi[\varobjidx],\wpi) -  \obfnM(\lsvecPw;\wpi[\varobjidx],\wpi) + \obfnYest(\lsvecPw;\varwobjiEst,\wobjiEst) - \obfnYest(\lsvecEst;\varwobjiEst,\wobjiEst)\right]^{1/\pwM}\\
&\quad\le \sqrt{2} C_0\mhf \sup_{\anyVec\in\lsSp} \left|\obfnYest(\anyVec;\varwobjiEst,\wobjiEst) - \obfnM(\anyVec;\wpi[\varobjidx],\wpi)\right|^{1/\pwM}. \eal
Furthermore, noting that
\bal\nn
&\left|\obfnYest(\anyVec;\varwobjiEst,\wobjiEst) - \obfnM(\anyVec;\wpi[\varobjidx],\wpi)\right|\\
&\quad\le \int_\tdom \left| \metric^2\left(\wobjiEst[\tm], \varwobjiEst[\anyVec^\top \lsbasis(\tm)]\right) - \metric^2\left(\mObj[\wpi\inv(\tm)], \mObj[{\wpi[\varobjidx]\inv[\anyVec^\top \lsbasis(\tm)]}]\right)\right| \diffop\tm 
+ \pnty \int_\tdom \left(\anyVec^\top \lsbasis(\tm) -\tm\right)^2 \diffop\tm\\
&\quad\le 2\diam(\msp) \int_\tdom \left| \metric\left(\wobjiEst[\tm], \varwobjiEst[\anyVec^\top \lsbasis(\tm)]\right) -
\metric\left(\wobji[\tm], \varwobji[\anyVec^\top \lsbasis(\tm)]\right)\right| \diffop\tm
+ \frac{\maxtm^3}{3}\pnty\\
&\quad\le 2\diam(\msp) \int_\tdom \left[ \metric\left(\wobjiEst[\tm], \wobji[\tm]\right) + \metric\left( \varwobjiEst[\anyVec^\top \lsbasis(\tm)], \varwobji[\anyVec^\top \lsbasis(\tm)] \right) \right] \diffop\tm 
+ \frac{\maxtm^3}{3}\pnty, \eal 
\eqref{eq:lsvecEstvsTrue} can then be bounded as
\bal\nn
&\|\lsvecEst-\lsvecPw\|\\
&\quad\le \left(\frac{4\diam(\msp)\maxtm}{C_0}\right)^{1/\pwM} \left[ \sup_{\tm\in\tdom} \metric\left(\wobjiEst[\tm],\wobji[\tm]\right)^{1/\pwM} + \sup_{\tm\in\tdom} \metric\left(\varwobjiEst[\tm],\varwobji[\tm]\right)^{1/\pwM}\right]+ \left(\frac{2\maxtm^3}{3C_0}\right)^{1/\pwM} \pnty^{1/\pwM},\eal
whence \eqref{eq:pairwpVecRate} follows by Theorem~\ref{thm:unifFregRateRdmTgt}, and hence  \eqref{eq:pairwpRate} follows by observing that 
\bal\nn \sup_{\tm\in\tdom}|\pairwpEst(\tm) - \pairwp(\tm)| = \sup_{\tm\in\tdom}|(\lsvecEst - \lsvecPw)^\top \lsbasis(\tm)| \le \|\lsvecEst - \lsvecPw\|,\eal
since $\sup_{\tm\in\tdom}|\lsbasis_{\tgrididx}(\tm)| \le 1$, for all $\tgrididx=1,\dots,\ntgrid+1$. 
\epf

\bpf[Proof of Corollary~\ref{cor:unifFregRateAveSubj}]
With $\bwi\note{\ \sim \ndpi^{-(\pwCm-1)[1-4\pwSumn(\pwLm-1+\smconst/2)/\pwLm] / (2\pwCm+4\pwLm-6+2\smconst)}} \sim \ndpi^{-(1-\smconst')(\pwCm-1)/ (2\pwCm+4\pwLm-6+2\smconst)}$,  \eqref{eq:objPrcsiEstRateForWarp} follows from \eqref{eq:objPrcsBiasVarRate} in Theorem~\ref{thm:unifFregRateRdmTgt}. 
We only need to show \eqref{eq:objPrcsiEstSumRate}. 

Given any fixed $\smconst>0$ and $\smconst'\in(0,1)$, define $\pwSumn = \smconst'\pwLm/[4(\pwLm-1+\smconst/2)]$, and 
\bgt\nn
\cbni = \min\left\{(\ndpi\bwi^2)^{\pwLm/[4(\pwLm-1+\smconst/2)]}, [\ndpi\bwi^2(-\log\bwi)\inv]^{\pwLm/[4(\pwLm-1)]}\right\}.\egt 
We will show that for the bias part,
\bgt
\supi \sup_{\unifinWobj} \sup_{\unifinTm} \metric\left(\objCmR,\objLmRbw\right) = \O\left(\supi\bwi^{2/(\pwCm-1)}\right) = \O\left(\bwAll^{2/(\pwCm-1)}\right), \label{eq:unifOverWobjOverSubj_CmvsLm}
\egt
and for the stochastic part,
\bgt
\limsup_{\nobj\gify} \sumn \sup_{\unifinWobj} \probNoise\left( \cbni \ndpi^{-\pwSumn}\sup_{\tm\in\tdom} \metric\left(\objLmRbw,\objLmOrcRm\right)^{\pwLm/2} > C \right) \ra 0, \quad\text{as }C\gify. \label{eq:sumOverSubj_LmvsLmOrc}
\egt
For each $\objidx=1,\dots,\nobj$, and $\tm\in\tdom$, define $\wobjiLm(\tm):\ \detmSp\ra\msp$ as $\wobjiLm(\tm)(\detmWobj) = \objLmRbw$. 
Observing that 
\bal\nn
&\nobj\inv\sumn\sup_{\unifinTm} \metric\left(\wobji(\tm), \wobjiLm(\tm)\right)^{\pwc} \le \left[\supi\sup_{\unifinWobj} \sup_{\unifinTm} \metric\left(\objCmR, \objLmRbw\right)\right]^{\pwc},\\
&\limsup_{\nobj\gify} \prob\left( \nobj\inv \sumn \sup_{\tm\in\tdom} \metric\left(\wobjiLm(\tm),\wobjiEst(\tm)\right)^{\pwc}> C\supi(\cbni\ndpi^{-\pwSumn})^{-2\pwc/\pwLm} \right)\\
&\quad\le \limsup_{\nobj\gify} \sumn \prob\left( \sup_{\tm\in\tdom} \metric\left(\wobjiLm(\tm),\wobjiEst(\tm)\right) > C(\cbni\ndpi^{-\pwSumn})^{-2/\pwLm} \right)\\
&\quad\le \limsup_{\nobj\gify} \sumn \sup_{\unifinWobj} \probNoise\left( \cbni\ndpi^{-\pwSumn} \sup_{\tm\in\tdom} \metric\left(\objLmRbw,\objLmOrcRm\right)^{\pwLm/2} > C \right),\eal
\eqref{eq:objPrcsiEstSumRate} follows if $\bwi\note{\ \sim \ndpi^{-(\pwCm-1)[1-4\pwSumn(\pwLm-1+\smconst/2)/\pwLm] / (2\pwCm+4\pwLm-6+2\smconst)}} \sim \ndpi^{-(1-\smconst')(\pwCm-1)/ (2\pwCm+4\pwLm-6+2\smconst)}$. 
\note{Note that
	\bal\nn
	\supi(\cbni\ndpi^{-\pwSumn})^{-2/\pwLm} 
	&=\supi \ndpi^{-1/[2(\pwLm-1+\smconst/2)] + 2\pwSumn/\pwLm} \bwi^{-1/(\pwLm-1+\smconst/2)} \\
	&= \supi \ndpi^{-(1-\smconst')/[2(\pwLm-1+\smconst/2)]} \bwi^{-1/(\pwLm-1+\smconst/2)}\\
	&= \supi \ndpi^{-(1-\smconst')/[2(\pwLm-1+\smconst/2)]} \ndpi^{(1-\smconst')(\pwCm-1)/[2(\pwLm-1+\smconst/2)(\pwCm+2\pwLm-3+\smconst)]}\\
	&= \supi \ndpi^{(1-\smconst')/(\pwCm+2\pwLm-3+\smconst)}
	= \ndp^{(1-\smconst')/(\pwCm+2\pwLm-3+\smconst)}.\eal
}

For \eqref{eq:unifOverWobjOverSubj_CmvsLm}, 
we will first show $\supi\sup_{\unifinWobj,\ \unifinTm} \metric(\objCmR,\objLmRbw) = \o(1)$\note{, i.e., $\sup_{\unifinWobj,\ \unifinTm} \supi\metric(\objCmR,\objLmRbw) = \o(1)$}. 
By the Cauchy criterion for uniform convergence, it suffices to show 
\bgt\label{eq:cauchy} \sup_{\unifinWobj,\ \unifinTm}  \left|\supi\metric\left(\objCmR,\objLmRbw\right) - \vsupi\metric\left(\objCmR,\objLmRvbw\right)\right|\ra 0, \egt 
as $\nobj,\vnobj\gify$. 
We note that by \ref{ass:ker}, \ref{ass:jointdtnStrg}, and \ref{ass:nObsPerObj}, 
\bal\label{eq:wtmom2supi} \supi\sup_{\unifinTm} \expectNoise\left[|\wLocR{\tmo}{\tm}{\bwi}| (\tmo-\tm)^2\bwi\mtwo\right] = \O(1),\quad\text{as }\nobj\gify,\eal
whence in conjunction with \eqref{eq:2ndderivObjfCm} we obtain 
\bal\nn
&\supi\sup_{\substack{\unifinWobj,\ \unifinTm\\ \anyObj\in\msp}} \left|\objfLmR[@][\bwi](\anyObj,\tm) - \objfCmR(\anyObj,\tm)\right| \\&\quad
\le \frac{1}{2}\supi\bwi^2 \supi\sup_{\unifinTm} \expectNoise\left[\left|\wLocR{\tmo}{\tm}{\bwi}\right| \left(\frac{\tmo-\tm}{\bwi}\right)^2\right] \sup_{\substack{\unifinWobj,\ \unifinTm\\ \anyObj\in\msp}} \left|\dCLR(\anyObj,\tm) \right| \\&\quad
=\O\left(\bwAll(\nobj)^2\right),\eal
where $\dCLR$ is defined as per \eqref{eq:d2objfCmR}. 
Hence, \note{$\sup_{\unifinWobj,\ \unifinTm} [\objfLmRbw(\objLmRvbw,\tm) - \objfLmRbw(\objLmRbw,\tm) + \objfLmRvbw(\objLmRbw,\tm)- \newline \objfLmRvbw(\objLmRvbw,\tm)]\le 2\sup_{\unifinWobj,\ \unifinTm,\ \anyObj\in\msp} |\objfLmRbw(\anyObj,\tm) - \objfLmRvbw(\anyObj,\tm)|$ can be bounded by}
\bal\nn
&\supivi \sup_{\substack{\unifinWobj,\ \unifinTm\\ \anyObj\in\msp}} \left|\objfLmRbw(\anyObj,\tm) - \objfLmRvbw(\anyObj,\tm)\right| \\
&\quad\le  \supi \sup_{\substack{\unifinWobj,\ \unifinTm\\ \anyObj\in\msp}} \left|\objfLmRbw(\anyObj,\tm) - \objfCmR(\anyObj,\tm)\right| +  \vsupi \sup_{\substack{\unifinWobj,\ \unifinTm\\ \anyObj\in\msp}} \left|\objfLmRvbw(\anyObj,\tm) - \objfCmR(\anyObj,\tm)\right| \\
&\quad= \O\left(\bwAll(\nobj)^2\right) + \O\left(\bwAll(\vnobj)^2\right), \eal
as $\nobj,\vnobj\gify$. Observing that 
\bal\nn
&\sup_{\unifinWobj,\ \unifinTm}  \left|\supi\metric\left(\objCmR,\objLmRbw\right) - \vsupi\metric\left(\objCmR,\objLmRvbw\right)\right| \\
&\quad\le \sup_{\unifinWobj,\ \unifinTm} \supivi \left|\metric\left(\objCmR,\objLmRbw\right) - \metric\left(\objCmR,\objLmRvbw\right) \right|\\
&\quad\le \sup_{\unifinWobj,\ \unifinTm} \supivi \metric\left(\objLmRbw,\objLmRvbw\right),\eal
\eqref{eq:cauchy} follows in conjunction with \ref{ass:minUnifStrg}. 

Using similar arguments to the proof of \eqref{eq:unifRateCmvsLm}, with $\chbi = \bwi^{-\pwCm/(\pwCm-1)}$, by \eqref{eq:d2objfCmDiffR}, there exists a constant $C>0$ such that for large $\nobj$, 
\bal\nn
&\indicator{\supi \sup_{\unifinWobj}\sup_{\unifinTm} \metric\left(\objCmR,\objLmRbw\right)^{\pwCm/2} > 2^\intL \supi\chbi\inv} \\
&\quad\le \supi \indicator{\sup_{\unifinWobj}\sup_{\unifinTm} \metric\left(\objCmR,\objLmRbw\right)^{\pwCm/2} > 2^\intL \supi\chbi\inv} \\
&\quad\le \supi \indicator{\chbi \sup_{\unifinWobj}\sup_{\unifinTm} \metric\left(\objCmR,\objLmRbw\right)^{\pwCm/2} > 2^\intL}\\
&\quad\le \supi C  \sum_{\setidx>\intL} \frac{\bwi^2(2^{\setidx}\chbi\inv)^{2/\pwCm}} {2^{2(\setidx-1)} \chbi\mtwo}
= 4C\sum_{\setidx>\intL} 2^{-2\setidx(\pwCm-1)/\pwCm}, \eal
which converges to zero as $\intL\gify$, whence \eqref{eq:unifOverWobjOverSubj_CmvsLm} follows. 

Furthermore, by \ref{ass:nObsPerObj}, replacing $\cbno$ with $\cbni\ndpi^{-\pwSumn}$ in the proof of \eqref{eq:unifOverWobj_LmvsLmOrc} yields
\bal\nn
&\limsup_{\nobj\gify} \sumn \sup_{\unifinWobj} \probNoise\left( \cbni\ndpi^{-\pwSumn} \sup_{\tm\in\tdom} \metric\left(\objLmRbw,\objLmOrcRm\right)^{\pwLm/2} > 2^{\intL} \right)\\
&\quad\le 4C \sum_{\setidx > \intL} 2^{-2\setidx(\pwLm-1+\smconst/2)/\pwLm} \limsup_{\nobj\gify} \sumn \ndpi^{-2\pwSumn(\pwLm-1+\smconst/2)/\pwLm}\\
&\qquad + 4C \sum_{\setidx > \intL} 2^{-2\setidx(\pwLm-1)/\pwLm} \limsup_{\nobj\gify} \sumn \ndpi^{-2\pwSumn(\pwLm-1)/\pwLm}\\
&\quad\le 4C\sum_{\setidx > \intL} 2^{-2\setidx(\pwLm-1+\smconst/2)/\pwLm} \limsup_{\nobj\gify} \nobj \ndp^{-2\pwSumn(\pwLm-1+\smconst/2)/\pwLm}\\
&\qquad + 4C \sum_{\setidx > \intL} 2^{-2\setidx(\pwLm-1)/\pwLm} \limsup_{\nobj\gify} \nobj \ndp^{-2\pwSumn(\pwLm-1)/\pwLm}, \eal 
which converges to zero as $\intL\gify$, whence \eqref{eq:sumOverSubj_LmvsLmOrc} follows. 
\epf

\bpf[Proof of Corollary~\ref{cor:warp}]
By \eqref{eq:wpEst} and Theorem~\ref{thm:pai},
\bal\nn
\sup_{\tm\in\tdom} \left|\wpiEst^{-1}(\tm) - \wpi^{-1}(\tm)\right|
&\le \frac{1}{\nobj} \sum_{\varobjidx=1}^{\nobj} \sup_{\tm\in\tdom} \left|\pairwpEst(\tm) - \pairwp(\tm)\right| + \sup_{\tm\in\tdom} \left|\frac{1}{\nobj}\sum_{\varobjidx=1}^{\nobj} \pairwp(\tm) - \wpi\inv(\tm)\right|. \eal
By Theorem~2.7.5 of \citet{vand:96}, 
\bal\nn 
\sup_{\tm\in\tdom} \left|\frac{1}{\nobj} \sum_{\varobjidx=1}^{\nobj} \pairwp(\tm) - \wpi\inv(\tm)\right| 
= \sup_{\tm\in\tdom} \left|\frac{1}{\nobj} \sum_{\varobjidx=1}^{\nobj} \wpi[\varobjidx](\tm) - \tm\right| 
= \sup_{\tm\in\tdom} \left|\frac{1}{\nobj} \sum_{\varobjidx=1}^{\nobj} \wpi[\varobjidx](\tm) - \expect(\wpi[\varobjidx](\tm))\right| = \Op\left(\nobj^{-1/2}\right). \eal 
Observing that 
\bal\nn
&\frac{1}{\nobj} \sum_{\varobjidx=1}^{\nobj} \sup_{\tm\in\tdom} \left|\pairwpEst(\tm) - \pairwp(\tm)\right|
\le \frac{1}{\nobj} \sum_{\varobjidx=1}^{\nobj} \|\lsvecEst - \lsvecPw\|\\
&\quad\le \mathrm{const.} \left[\sup_{\tm\in\tdom} \metric\left(\wobjiEst[\tm],\wobji[\tm]\right)^{1/\pwM} + \nobj\inv \sum_{\varobjidx=1}^{\nobj} \sup_{\tm\in\tdom} \metric\left(\varwobjiEst[\tm],\varwobji[\tm]\right)^{1/\pwM} + \pnty^{1/\pwM}\right],\eal
it follows from Corollary~\ref{cor:unifFregRateAveSubj} that 
\bal\nn \sup_{\tm\in\tdom} \left|\wpiEst^{-1}(\tm) - \wpi^{-1}(\tm)\right| =  \O\left(\pnty^{1/\pwM}\right) + \Op\left(\ndp^{-(1-\smconst')/[\pwM(\pwCm+2\pwLm-3+\smconst)]}\right) + \Op\left(\nobj^{-1/2}\right).\eal
By \ref{ass:wpSlope}, \bal\nn \sup_{\tm\in\tdom} \left|\wpiEst(\tm) - \wpi(\tm)\right| &= \sup_{\tm\in\tdom} \left|\wpiEst(\wpiEst\inv(\tm)) - \wpi(\wpiEst\inv(\tm))\right|
= \sup_{\tm\in\tdom} \left|\tm - \wpi(\wpiEst\inv(\tm))\right| \\
&\le C \sup_{\tm\in\tdom} \left|\wpi\inv(\tm) - \wpi\inv(\wpi(\wpiEst\inv(\tm)))\right| = C\sup_{\tm\in\tdom} \left|\wpi\inv(\tm) - \wpiEst\inv(\tm)\right|,\eal
whence \eqref{eq:wpRate} follows.
Furthermore, observing that 
\bal\nn
\metric\left(\wobjiEst(\wpiEst(\tm)), \wobji(\wpi(\tm))\right) 
&\le \metric\left(\wobjiEst(\wpiEst(\tm)), \wobji(\wpiEst(\tm))\right) + \metric\left(\mObj[\wpi\inv(\wpiEst(\tm))], \mObj\right)\\
&\le \sup_{\unifinTm} \metric\left(\wobjiEst(\tm), \wobji(\tm)\right) + \lipm \sup_{\unifinTm} \left|\wpi\inv(\wpiEst(\tm)) - \tm\right|\\
&= \sup_{\unifinTm} \metric\left(\wobjiEst(\tm), \wobji(\tm)\right) + \lipm \sup_{\unifinTm} \left|\wpi\inv(\tm) - \wpiEst\inv(\tm)\right|, \eal
\eqref{eq:aliRate} also follows, which completes the proof.
\epf

\subsection{Simulation studies.}\label{sec:wpSimu}

In this section, we  compare the performance of the proposed warping method for metric valued functional data for different choices of the penalty parameter $\pnty$ and the number of knots $\ntgrid$.
Here, the time domain is $\tdom = [0,1]$, and the metric space $(\msp,\metric[W])$ considered is the Wasserstein space of continuous probability measures on $[0,1]$ with finite second moments endowed with the  $\hilbert$-Wasserstein distance as in Example~\ref{eg:wsp}. With sample size $\nobj =30$, two cases were implemented with fixed trajectories $\mObjNull$ in \eqref{eq:gloSam} as follows.
\ben[label = \text{Case} \arabic*:, itemindent=*, series = simu]
\item $\mObj[\tm] = \mathrm{Beta}(\shpabeta,\shpbbeta)$, where $\shpabeta = 1.1+10(\tm-0.4)^2$, and $\shpbbeta = 2.6 + 1.5\sin(2\pi \tm - \pi)$, for $\tm\in\tdom$. 
\item $\mObj = \gaus(\meantgaus, \sdtgaus^2)$ truncated on $[0,1]$, where $\meantgaus = 0.1+0.8 \tm$, and $\sdtgaus = 0.6 + 0.2 \sin(10\pi \tm)$, for $\tm\in\tdom$. 
Specifically, the corresponding distribution function is 
\bgt\nn \cdf_{\mObj[\tm]}(\rarg) = \frac{\cdfStdGaus((\rarg - \meantgaus)/\sdtgaus) - \cdfStdGaus(-\meantgaus/\sdtgaus)} {\cdfStdGaus((1 - \meantgaus)/\sdtgaus) - \cdfStdGaus(-\meantgaus/\sdtgaus)} \mbf{1}_{[0,1]}(\rarg) + \mbf{1}_{(1,+\infty)}(\rarg),\, \rarg\in\real,\egt
where $\cdfStdGaus$ is the distribution function of a standard Gaussian distribution. 
\een
We consider a family of perturbation/distortion functions $\{\distortFctn: \distort\in\integer\backslash\{0\}\}$, where $\distortFctn(\rarg) = \rarg - |\distort\pi|\inv\sin(\distort\pi\rarg)$, for $\rarg\in\real$. 
The warping functions $\wpi$ were generated through the distortion functions $\distortFctn$; specifically, $\wpi = \distortFctn[\distort_{\objidx 1}]\circ \distortFctn[\distort_{\objidx 2}]$, where 
$\distort_{\objidx l}$ are independent and identically distributed for $l=1,2$ and $\objidx = 1,\dots,\nobj$, such that $$\prob(\distort_{\objidx l} = -k) = \prob(\distort_{\objidx l} = k) = \prob(V_{2} = k)/[2(1-\prob(V_{2} = 0))],$$ for any $k\in\pint$, with $V_{2}\sim\pois(2)$. 
We note that this generation mechanism ensures $\wpi\in\wfsp$ and $\expect[\wpi(\tm)] = \tm$, for any $\tm\in\tdom$. 
With $\mObjNull$ and $\wpi$, the sample trajectories $\wobji$ were computed as per \eqref{eq:gloSam}. 

Set the number of discrete observations per trajectory $\ndpi= 30$, for all $\objidx=1,\dots,\nobj$. 
We sampled $\tmij\sim\unif(\tdom)$ independently, for $\dptidx = 1,\dots,\ndpi$, and $\objidx =1,\dots,\nobj$. 
Given a measurable function $\anyFctn\colon \real\ra\real$, 
a push-forward measure $\pushforward{\anyFctn}{\anyObj}$ is defined as $\pushforward{\anyFctn}{\anyObj}(\anySet) = \anyObj(\{\rarg: \anyFctn(\rarg)\in\anySet\})$, for any distribution $\anyObj\in\msp$ and set $\anySet\subset\real$. 
The observed distributions $\obsobjij$ were generated by adding perturbations to the trajectory evaluated at $\tmij$, $\wobj(\tmij)$, through push-forward measures; specifically  $\obsobjij=\pushforward{\distortFctn[\distortTm_{\objidx\dptidx}]}{(\wobji[\tmij])}$, where $\distortTm_{\objidx\dptidx}$ are independent and identically distributed following $\unif\{\pm 4\pi,\pm 5\pi, \dots, \pm 8\pi\}$, and are also independent of the observed times $\tmij$, $\dptidx = 1,\dots,\ndpi$, and $\objidx =1,\dots,\nobj$.

We applied the proposed pairwise warping method to the simulated data with Epanechnikov kernel and bandwidths $\bwi$ chosen by cross-validation in the presmoothing step as per \eqref{eq:wobjEst}, where the local Fr\'echet regression was implemented using the R package \texttt{frechet} \citep{frechet}.  We assessed the results through mean integrated squared errors (MISEs) for the estimated time-synchronized trajectories $\wobjiEst[\wpiEst(\cdot)]$ and the estimated warping functions $\wpiEst$ as per \eqref{eq:wobjEst} and \eqref{eq:wpEst}; specifically, 
\bal\label{eq:mise}
\tmise &= \frac{1}{\nobj}\sumn \int_\tdom \metric^2\left(\wobjiEst[\wpiEst(\tm)], \mObj\right) \diffop\tm, \\ 
\wmise &= \frac{1}{\nobj}\sumn \int_\tdom \left(\wpiEst(\tm) - \wpi(\tm)\right)^2 \diffop\tm.\eal

\begin{figure}[hbt!]
	\centering
	\includegraphics[width=\textwidth]{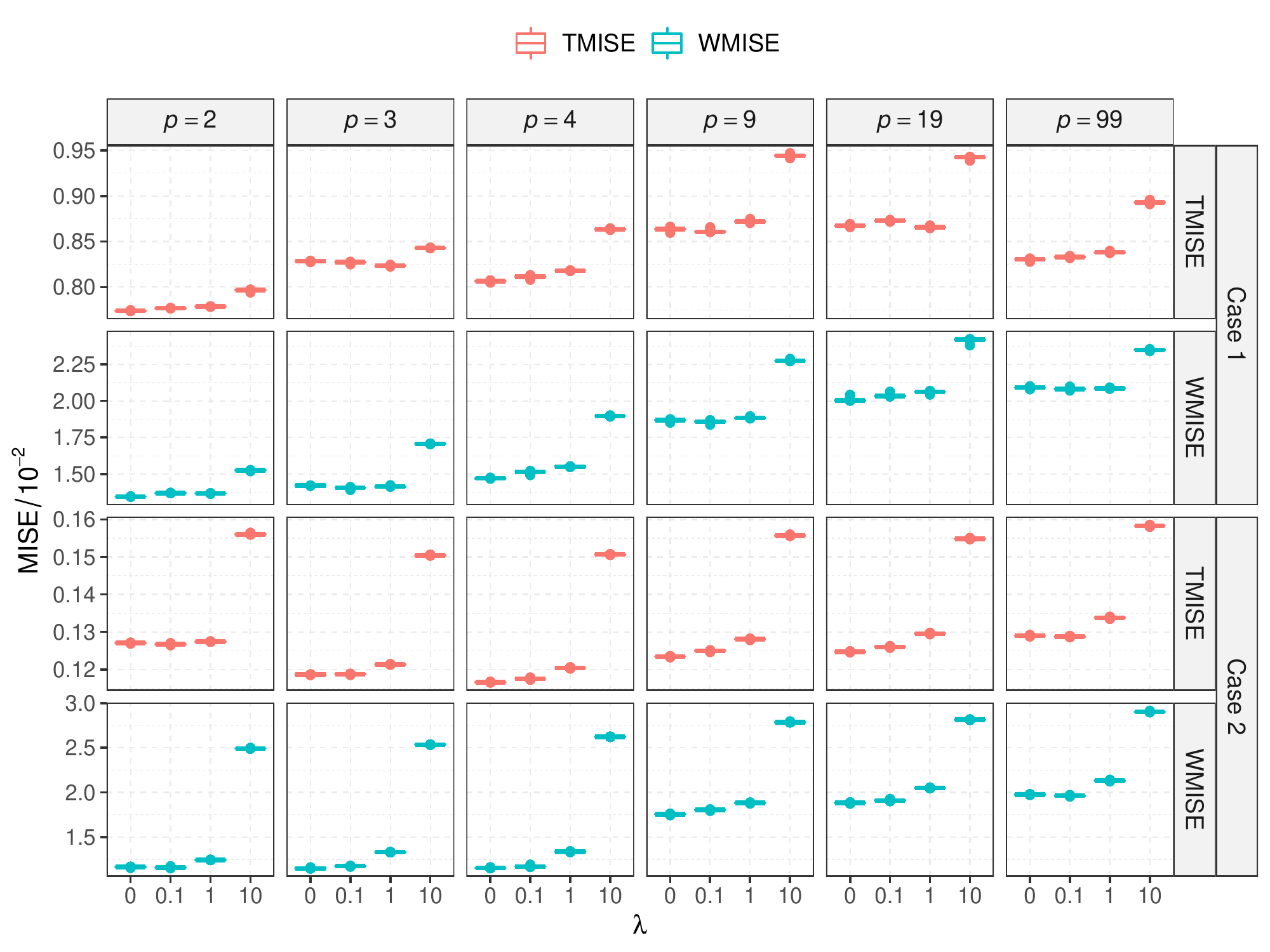}
	\caption{Summary of $\tmise$ (red) and $\wmise$ (blue) as per \eqref{eq:mise} out of 1000 Monte Carlo runs for Case~1 (top two rows) and Case~2 (bottom two rows).}\label{fig:MISE} 
\end{figure}

Since too many knots will result in shape distortion of the estimated warping function \citep{rams:98}, 1000 Monte Carlo runs were conducted  for $\ntgrid\in\{2, 3, 4, 9, 19, 99\}$, and $\pnty\in\{0\}\cup\{10^{l}: l=-1,0,1\}$. Results in terms of \tmise and \wmise for Case~1 and Case~2 are summarized in the boxplots in Figure~\ref{fig:MISE}, the top two rows show the results for Case~1 and the bottom two rows for Case~2. 
For both cases, for any given value of the number of knots $\ntgrid$, the proposed estimators perform almost equally well in terms of \tmise and \wmise with small values (no more than 1) of the penalty parameter $\pnty$, and the performance turns worse as $\pnty$ increases from 1 to 10. 
Furthermore, across different choices of the number of knots $\ntgrid$, the estimators achieve the minimum estimation errors with small $\ntgrid\in\{2,3,4\}$. 
\rvone{Thus, the  simulations indicate that the proposed method is not sensitive to the choice of $\ntgrid$ and $\pnty$ when $\ntgrid$ and $\pnty$ are relatively small, which is in agreement with findings in the literature \citep{rams:98,tang:08}.} 


\begin{thebibliography}{59}
\expandafter\ifx\csname natexlab\endcsname\relax\def\natexlab#1{#1}\fi
\expandafter\ifx\csname url\endcsname\relax
  \def\url#1{\texttt{#1}}\fi
\expandafter\ifx\csname urlprefix\endcsname\relax\def\urlprefix{URL }\fi

\bibitem[{Agueh and Carlier(2011)}]{ague:11}
\textsc{Agueh, M.} and \textsc{Carlier, G.} (2011).
\newblock Barycenters in the {W}asserstein space.
\newblock \textit{SIAM Journal on Mathematical Analysis} \textbf{43} 904--924.

\bibitem[{Ambrosio et~al.(2004)Ambrosio, Gigli and Savar{\'e}}]{ambr:04}
\textsc{Ambrosio, L.}, \textsc{Gigli, N.} and \textsc{Savar{\'e}, G.} (2004).
\newblock Gradient flows with metric and differentiable structures, and
  applications to the {W}asserstein space.
\newblock \textit{Atti Accad. Naz. Lincei Cl. Sci. Fis. Mat. Natur. Rend.
  Lincei (9) Mat. Appl} \textbf{15} 327--343.

\bibitem[{Balabdaoui et~al.(2009)Balabdaoui, Rufibach and Wellner}]{bala:09}
\textsc{Balabdaoui, F.}, \textsc{Rufibach, K.} and \textsc{Wellner, J.~A.}
  (2009).
\newblock Limit distribution theory for maximum likelihood estimation of a
  log-concave density.
\newblock \textit{The Annals of Statistics} \textbf{37} 1299--1331.

\bibitem[{Belitser et~al.(2012)Belitser, Ghosal and van Zanten}]{beli:12}
\textsc{Belitser, E.}, \textsc{Ghosal, S.} and \textsc{van Zanten, H.} (2012).
\newblock Optimal two-stage procedures for estimating location and size of the
  maximum of a multivariate regression function.
\newblock \textit{The Annals of Statistics} \textbf{40} 2850--2876.

\bibitem[{Bhattacharya and Patrangenaru(2003)}]{bhat:03}
\textsc{Bhattacharya, R.} and \textsc{Patrangenaru, V.} (2003).
\newblock Large sample theory of intrinsic and extrinsic sample means on
  manifolds - {I}.
\newblock \textit{The Annals of Statistics} \textbf{31} 1--29.

\bibitem[{Bhattacharya and Patrangenaru(2005)}]{bhat:05}
\textsc{Bhattacharya, R.} and \textsc{Patrangenaru, V.} (2005).
\newblock Large sample theory of intrinsic and extrinsic sample means on
  manifolds - {II}.
\newblock \textit{The Annals of Statistics} \textbf{33} 1225--1259.

\bibitem[{Buckner et~al.(2009)Buckner, Sepulcre, Talukdar, Krienen, Liu,
  Hedden, Andrews-Hanna, Sperling and Johnson}]{buck:09}
\textsc{Buckner, R.~L.}, \textsc{Sepulcre, J.}, \textsc{Talukdar, T.},
  \textsc{Krienen, F.~M.}, \textsc{Liu, H.}, \textsc{Hedden, T.},
  \textsc{Andrews-Hanna, J.~R.}, \textsc{Sperling, R.~A.} and \textsc{Johnson,
  K.~A.} (2009).
\newblock Cortical hubs revealed by intrinsic functional connectivity: mapping,
  assessment of stability, and relation to {A}lzheimer's disease.
\newblock \textit{Journal of Neuroscience} \textbf{29} 1860--1873.

\bibitem[{Bullmore and Sporns(2009)}]{bull:09}
\textsc{Bullmore, E.} and \textsc{Sporns, O.} (2009).
\newblock Complex brain networks: graph theoretical analysis of structural and
  functional systems.
\newblock \textit{Nature Reviews Neuroscience} \textbf{10} 186--198.

\bibitem[{Cai et~al.(2019)Cai, Liu, Mi, Garg, Trappe, McKeown and
  Wang}]{cai:19}
\textsc{Cai, J.}, \textsc{Liu, A.}, \textsc{Mi, T.}, \textsc{Garg, S.},
  \textsc{Trappe, W.}, \textsc{McKeown, M.~J.} and \textsc{Wang, Z.~J.} (2019).
\newblock Dynamic graph theoretical analysis of functional connectivity in
  {P}arkinson's disease: {T}he importance of {F}iedler value.
\newblock \textit{IEEE Journal of Biomedical and Health Informatics}
  \textbf{23} 1720--1729.

\bibitem[{Chen et~al.(2020)Chen, Gajardo, Fan, Zhong, Dubey, Han, Bhattacharjee
  and M\"uller}]{frechet}
\textsc{Chen, Y.}, \textsc{Gajardo, A.}, \textsc{Fan, J.}, \textsc{Zhong, Q.},
  \textsc{Dubey, P.}, \textsc{Han, K.}, \textsc{Bhattacharjee, S.} and
  \textsc{M\"uller, H.-G.} (2020).
\newblock \textit{{frechet: Statistical Analysis for Random Objects and
  Non-Euclidean Data}}.
\newblock R package version 0.1.0, available at
  \url{https://CRAN.R-project.org/package=frechet}.

\bibitem[{Davis et~al.(2007)Davis, Fletcher, Bullitt and Joshi}]{davi:07}
\textsc{Davis, B.~C.}, \textsc{Fletcher, P.~T.}, \textsc{Bullitt, E.} and
  \textsc{Joshi, S.} (2007).
\newblock Population shape regression from random design data.
\newblock In \textit{2007 IEEE 11th International Conference on Computer
  Vision}.

\bibitem[{de~Haan et~al.(2012)de~Haan, van~der Flier, Wang, Van~Mieghem,
  Scheltens and Stam}]{deha:12}
\textsc{de~Haan, W.}, \textsc{van~der Flier, W.~M.}, \textsc{Wang, H.},
  \textsc{Van~Mieghem, P.~F.}, \textsc{Scheltens, P.} and \textsc{Stam, C.~J.}
  (2012).
\newblock Disruption of functional brain networks in {A}lzheimer's disease:
  what can we learn from graph spectral analysis of resting-state
  magnetoencephalography?
\newblock \textit{Brain Connectivity} \textbf{2} 45--55.

\bibitem[{Dennis(2014)}]{denn:14}
\textsc{Dennis, P.~M., Emily L .and~Thompson} (2014).
\newblock Functional brain connectivity using {fMRI} in aging and {A}lzheimer's
  disease.
\newblock \textit{Neuropsychology Review} \textbf{24} 49--62.

\bibitem[{Devroye(1978)}]{devr:78}
\textsc{Devroye, L.} (1978).
\newblock The uniform convergence of nearest neighbor regression function
  estimators and their application in optimization.
\newblock \textit{IEEE Transactions on Information Theory} \textbf{24}
  142--151.

\bibitem[{Dryden et~al.(2009)Dryden, Koloydenko and Zhou}]{dryd:09}
\textsc{Dryden, I.~L.}, \textsc{Koloydenko, A.} and \textsc{Zhou, D.} (2009).
\newblock Non-{E}uclidean statistics for covariance matrices, with applications
  to diffusion tensor imaging.
\newblock \textit{The Annals of Applied Statistics} \textbf{3} 1102--1123.

\bibitem[{Dubey and M\"uller(2020)}]{dube:20}
\textsc{Dubey, P.} and \textsc{M\"uller, H.-G.} (2020).
\newblock Functional models for time-varying random objects.
\newblock \textit{Journal of the Royal Statistical Society: Series B}
  \textbf{82} 275--327.

\bibitem[{Fan and Gijbels(1996)}]{fan:96}
\textsc{Fan, J.} and \textsc{Gijbels, I.} (1996).
\newblock \textit{Local Polynomial Modelling and its Applications}.
\newblock Chapman \& Hall, London.

\bibitem[{Farag{\'o} et~al.(1993)Farag{\'o}, Linder and Lugosi}]{fara:93}
\textsc{Farag{\'o}, A.}, \textsc{Linder, T.} and \textsc{Lugosi, G.} (1993).
\newblock Fast nearest-neighbor search in dissimilarity spaces.
\newblock \textit{IEEE Transactions on Pattern Analysis and Machine
  Intelligence} \textbf{15} 957--962.

\bibitem[{Ferreira and Busatto(2013)}]{ferr:13}
\textsc{Ferreira, L.~K.} and \textsc{Busatto, G.~F.} (2013).
\newblock Resting-state functional connectivity in normal brain aging.
\newblock \textit{Neuroscience \& Biobehavioral Reviews} \textbf{37} 384--400.

\bibitem[{Fiedler(1973)}]{fied:73}
\textsc{Fiedler, M.} (1973).
\newblock Algebraic connectivity of graphs.
\newblock \textit{Czechoslovak Mathematical Journal} \textbf{23} 298--305.

\bibitem[{Fr{\'e}chet(1948)}]{frec:48}
\textsc{Fr{\'e}chet, M.} (1948).
\newblock Les {\'e}l{\'e}ments al{\'e}atoires de nature quelconque dans un
  espace distanci{\'e}.
\newblock In \textit{Annales de l'Institut Henri Poincar{\'e}}, vol.~10.
  215--310.

\bibitem[{Gasser and Kneip(1995)}]{gass:95}
\textsc{Gasser, T.} and \textsc{Kneip, A.} (1995).
\newblock Searching for structure in curve samples.
\newblock \textit{Journal of the American Statistical Association} \textbf{90}
  1179--1188.

\bibitem[{Gervini and Gasser(2004)}]{gerv:04}
\textsc{Gervini, D.} and \textsc{Gasser, T.} (2004).
\newblock Self-modeling warping functions.
\newblock \textit{Journal of the Royal Statistical Society: Series B}
  \textbf{66} 959--971.

\bibitem[{Gong and Medioni(2011)}]{gong:11}
\textsc{Gong, D.} and \textsc{Medioni, G.} (2011).
\newblock Dynamic manifold warping for view invariant action recognition.
\newblock In \textit{Proceedings of International Conference on Computer
  Vision}. 571--578.

\bibitem[{Hein(2009)}]{hein:09}
\textsc{Hein, M.} (2009).
\newblock Robust nonparametric regression with metric-space valued output.
\newblock In \textit{Advances in Neural Information Processing Systems}.

\bibitem[{Hoffman and Wielandt(1953)}]{hoff:53}
\textsc{Hoffman, A.~J.} and \textsc{Wielandt, H.~W.} (1953).
\newblock The variation of the spectrum of a normal matrix.
\newblock \textit{Duke Mathematical Journal} \textbf{20} 37--39.

\bibitem[{Huckemann(2012)}]{huck:12}
\textsc{Huckemann, S.~F.} (2012).
\newblock On the meaning of mean shape: manifold stability, locus and the two
  sample test.
\newblock \textit{Annals of the Institute of Statistical Mathematics}
  \textbf{64} 1227--1259.

\bibitem[{Huckemann(2015)}]{huck:15}
\textsc{Huckemann, S.~F.} (2015).
\newblock ({S}emi-)intrinsic statistical analysis on non-{E}uclidean spaces.
\newblock In \textit{Advances in Complex Data Modeling and Computational
  Methods in Statistics}. Springer, 103--118.

\bibitem[{James(2007)}]{jame:07}
\textsc{James, G.~M.} (2007).
\newblock Curve alignment by moments.
\newblock \textit{The Annals of Applied Statistics} \textbf{1} 480--501.

\bibitem[{Karush(1939)}]{karu:39}
\textsc{Karush, W.} (1939).
\newblock \textit{Minima of functions of several variables with inequalities as
  side constraints}.
\newblock Master's thesis, Department of Mathematics, University of Chicago.

\bibitem[{Kloeckner(2010)}]{kloe:10}
\textsc{Kloeckner, B.~R.} (2010).
\newblock A geometric study of {W}asserstein spaces: {E}uclidean spaces.
\newblock \textit{Annali della Scuola Normale Superiore di Pisa-Classe di
  Scienze} \textbf{9} 297--323.

\bibitem[{Kneip and Gasser(1992)}]{knei:92}
\textsc{Kneip, A.} and \textsc{Gasser, T.} (1992).
\newblock Statistical tools to analyze data representing a sample of curves.
\newblock \textit{The Annals of Statistics} \textbf{20} 1266--1305.

\bibitem[{Kuhn and Tucker(1951)}]{kuhn:51}
\textsc{Kuhn, H.} and \textsc{Tucker, A.} (1951).
\newblock Nonlinear programming.
\newblock In \textit{Proceedings of the Second Berkeley Symposium on
  Mathematical Statistics and Probability} (J.~Neyman, ed.). University of
  California Press, Berkeley, CA, 481--492.

\bibitem[{Le~Gouic and Loubes(2017)}]{lego:17}
\textsc{Le~Gouic, T.} and \textsc{Loubes, J.-M.} (2017).
\newblock Existence and consistency of {W}asserstein barycenters.
\newblock \textit{Probability Theory and Related Fields} \textbf{168} 901--917.

\bibitem[{Mack and Silverman(1982)}]{mack:82}
\textsc{Mack, Y.~P.} and \textsc{Silverman, B.~W.} (1982).
\newblock Weak and strong uniform consistency of kernel regression estimates.
\newblock \textit{Zeitschrift f\"{u}r Wahrscheinlichkeitstheorie und verwandte
  Gebiete} \textbf{61} 405--415.

\bibitem[{Marron et~al.(2015)Marron, Ramsay, Sangalli and Srivastava}]{marr:15}
\textsc{Marron, J.~S.}, \textsc{Ramsay, J.~O.}, \textsc{Sangalli, L.~M.} and
  \textsc{Srivastava, A.} (2015).
\newblock Functional data analysis of amplitude and phase variation.
\newblock \textit{Statistical Science} \textbf{30} 468--484.

\bibitem[{Mevel et~al.(2013)Mevel, Landeau, Fouquet, La~Joie, Villain,
  M{\'e}zenge, Perrotin, Eustache, Desgranges and Ch{\'e}telat}]{meve:13}
\textsc{Mevel, K.}, \textsc{Landeau, B.}, \textsc{Fouquet, M.},
  \textsc{La~Joie, R.}, \textsc{Villain, N.}, \textsc{M{\'e}zenge, F.},
  \textsc{Perrotin, A.}, \textsc{Eustache, F.}, \textsc{Desgranges, B.} and
  \textsc{Ch{\'e}telat, G.} (2013).
\newblock Age effect on the default mode network, inner thoughts, and cognitive
  abilities.
\newblock \textit{Neurobiology of Aging} \textbf{34} 1292--1301.

\bibitem[{M\"{u}ller(1989)}]{mull:89:3}
\textsc{M\"{u}ller, H.-G.} (1989).
\newblock Adaptive nonparametric peak estimation.
\newblock \textit{The Annals of Statistics} \textbf{17} 1053--1069.

\bibitem[{Parzen(1962)}]{parz:62}
\textsc{Parzen, E.} (1962).
\newblock On estimation of a probability density function and mode.
\newblock \textit{The Annals of Mathematical Statistics} \textbf{33}
  1065--1076.

\bibitem[{Pelletier(2006)}]{pell:06}
\textsc{Pelletier, B.} (2006).
\newblock Non-parametric regression estimation on closed {R}iemannian
  manifolds.
\newblock \textit{Journal of Nonparametric Statistics} \textbf{18} 57--67.

\bibitem[{Petersen and M{\"u}ller(2019)}]{pete:19}
\textsc{Petersen, A.} and \textsc{M{\"u}ller, H.-G.} (2019).
\newblock Fr\'echet regression for random objects with {E}uclidean predictors.
\newblock \textit{The Annals of Statistics} \textbf{47} 691--719.

\bibitem[{Petersen et~al.(2016)Petersen, Zhao, Carmichael and
  M{\"u}ller}]{pete:16:3}
\textsc{Petersen, A.}, \textsc{Zhao, J.}, \textsc{Carmichael, O.} and
  \textsc{M{\"u}ller, H.-G.} (2016).
\newblock Quantifying individual brain connectivity with functional principal
  component analysis for networks.
\newblock \textit{Brain Connectivity} \textbf{6}.

\bibitem[{Phillips et~al.(2015)Phillips, McGlaughlin, Ruth, Jager, Soldan and
  Initiative}]{phil:15}
\textsc{Phillips, D.~J.}, \textsc{McGlaughlin, A.}, \textsc{Ruth, D.},
  \textsc{Jager, L.~R.}, \textsc{Soldan, A.} and \textsc{Initiative, A. D.~N.}
  (2015).
\newblock Graph theoretic analysis of structural connectivity across the
  spectrum of {A}lzheimer's disease: the importance of graph creation methods.
\newblock \textit{NeuroImage: Clinical} \textbf{7} 377--390.

\bibitem[{Ramsay and Li(1998)}]{rams:98}
\textsc{Ramsay, J.~O.} and \textsc{Li, X.} (1998).
\newblock Curve registration.
\newblock \textit{Journal of the Royal Statistical Society: Series B}
  \textbf{60} 351--363.

\bibitem[{Rubinov and Sporns(2010)}]{rubi:10}
\textsc{Rubinov, M.} and \textsc{Sporns, O.} (2010).
\newblock Complex network measures of brain connectivity: uses and
  interpretations.
\newblock \textit{NeuroImage} \textbf{52} 1059--1069.

\bibitem[{Ruppert and Wand(1994)}]{rupp:94}
\textsc{Ruppert, D.} and \textsc{Wand, M.~P.} (1994).
\newblock Multivariate locally weighted least squares regression.
\newblock \textit{The Annals of Statistics} \textbf{22} 1346--1370.

\bibitem[{Sakoe and Chiba(1978)}]{sako:78}
\textsc{Sakoe, H.} and \textsc{Chiba, S.} (1978).
\newblock Dynamic programming algorithm optimization for spoken word
  recognition.
\newblock \textit{IEEE Transactions on Acoustics, Speech, and Signal
  Processing} \textbf{26} 43--49.

\bibitem[{Silverman(1978)}]{silv:78}
\textsc{Silverman, B.~W.} (1978).
\newblock Weak and strong uniform consistency of the kernel estimate of a
  density and its derivatives.
\newblock \textit{The Annals of Statistics} \textbf{6} 177--184.

\bibitem[{Steinke and Hein(2009)}]{stei:09}
\textsc{Steinke, F.} and \textsc{Hein, M.} (2009).
\newblock Non-parametric regression between manifolds.
\newblock In \textit{Advances in Neural Information Processing Systems}.
  1561--1568.

\bibitem[{Steinke et~al.(2010)Steinke, Hein and Sch{\"o}lkopf}]{stei:10}
\textsc{Steinke, F.}, \textsc{Hein, M.} and \textsc{Sch{\"o}lkopf, B.} (2010).
\newblock Nonparametric regression between general {R}iemannian manifolds.
\newblock \textit{SIAM Journal on Imaging Sciences} \textbf{3} 527--563.

\bibitem[{Sturm(2003)}]{stur:03}
\textsc{Sturm, K.-T.} (2003).
\newblock Probability measures on metric spaces of nonpositive curvature.
\newblock \textit{Heat Kernels and Analysis on Manifolds, Graphs, and Metric
  Spaces (Paris, 2002)} \textbf{338} 357--390.

\bibitem[{Tang and M{\"u}ller(2008)}]{tang:08}
\textsc{Tang, R.} and \textsc{M{\"u}ller, H.-G.} (2008).
\newblock Pairwise curve synchronization for functional data.
\newblock \textit{Biometrika} \textbf{95} 875--889.

\bibitem[{Trigeorgis et~al.(2018)Trigeorgis, Nicolaou, Schuller and
  Zafeiriou}]{trig:18}
\textsc{Trigeorgis, G.}, \textsc{Nicolaou, M.~A.}, \textsc{Schuller, B.~W.} and
  \textsc{Zafeiriou, S.} (2018).
\newblock Deep canonical time warping for simultaneous alignment and
  representation learning of sequences.
\newblock \textit{IEEE Transactions on Pattern Analysis and Machine
  Intelligence} \textbf{40} 1128--1138.

\bibitem[{van~der Vaart and Wellner(1996)}]{vand:96}
\textsc{van~der Vaart, A.~W.} and \textsc{Wellner, J.~A.} (1996).
\newblock \textit{Weak Convergence and Empirical Processes}.
\newblock Springer, New York.

\bibitem[{Vieu(1996)}]{vieu:96}
\textsc{Vieu, P.} (1996).
\newblock A note on density mode estimation.
\newblock \textit{Statistics \& Probability Letters} \textbf{26} 297--307.

\bibitem[{Vu et~al.(2012)Vu, Carey and Mahadevan}]{vu:12}
\textsc{Vu, H.~T.}, \textsc{Carey, C.} and \textsc{Mahadevan, S.} (2012).
\newblock Manifold warping: manifold alignment over time.
\newblock In \textit{Proceedings of the Twenty-Sixth AAAI Conference on
  Artificial Intelligence}. 1155--1161.

\bibitem[{Wang and Gasser(1997)}]{wang:97}
\textsc{Wang, K.} and \textsc{Gasser, T.} (1997).
\newblock Alignment of curves by dynamic time warping.
\newblock \textit{The Annals of Statistics} \textbf{25} 1251--1276.

\bibitem[{Yuan et~al.(2012)Yuan, Zhu, Lin and Marron}]{yuan:12}
\textsc{Yuan, Y.}, \textsc{Zhu, H.}, \textsc{Lin, W.} and \textsc{Marron, J.}
  (2012).
\newblock Local polynomial regression for symmetric positive definite matrices.
\newblock \textit{Journal of the Royal Statistical Society: Series B
  (Statistical Methodology)} \textbf{74} 697--719.

\bibitem[{Zonneveld et~al.(2019)Zonneveld, Pruim, Bos, Vrooman, Muetzel,
  Hofman, Rombouts, van~der Lugt, Niessen, Ikram and Vernooijab}]{zonn:19}
\textsc{Zonneveld, H.~I.}, \textsc{Pruim, R.~H.}, \textsc{Bos, D.},
  \textsc{Vrooman, H.~A.}, \textsc{Muetzel, R.~L.}, \textsc{Hofman, A.},
  \textsc{Rombouts, S.~A.}, \textsc{van~der Lugt, A.}, \textsc{Niessen, W.~J.},
  \textsc{Ikram, M.~A.} and \textsc{Vernooijab, M.~W.} (2019).
\newblock Patterns of functional connectivity in an aging population: {The
  Rotterdam Study}.
\newblock \textit{NeuroImage} \textbf{189} 432--444.

\end{thebibliography}
\end{document}